\documentclass[aps,pra,twocolumn,showpacs,superscriptaddress,longbibliography]{revtex4-1}
\usepackage{graphicx} 
\usepackage{amsmath}
\usepackage{graphicx,epstopdf}
\usepackage{gensymb}
\graphicspath{{../}}

\newcommand{\be}{\begin{equation}}
\newcommand{\ee}{\end{equation}}
\newcommand{\bea}{\begin{eqnarray}}
\newcommand{\eea}{\end{eqnarray}}
\newcommand{\bse}{\begin{subequations}}
\newcommand{\ese}{\end{subequations}}

\newcommand{\av}{\mathbf a}
\newcommand{\bv}{\mathbf b}
\newcommand{\cv}{\mathbf c}

\usepackage{color}
\usepackage[colorlinks,bookmarks=false,citecolor=darkblue,linkcolor=red,urlcolor=blue]{hyperref}

\definecolor{darkred}{rgb}{0.7,0.0,0.0}

\definecolor{darkblue}{rgb}{0,0.02,0.45}

\definecolor{darkgreen}{rgb}{0.02,0.45,0.0}

\definecolor{violet}{rgb}{0.8,0.2,0.6}

\begin{document}

\title{Two types of alternating spin-$\frac12$ chains \\ and their field-induced transitions in $\varepsilon$-LiVOPO$_4$}

\author{Prashanta K. Mukharjee}
\affiliation{School of Physics, Indian Institute of Science Education and Research, Thiruvananthapuram-695551, India}
\author{K. M. Ranjith}
\author{M. Baenitz}
\affiliation{Max Planck Institute for Chemical Physics of Solids, N$\ddot{o}$thnitzer Str. 40, 01187 Dresden, Germany}
\author{Y. Skourski}
\affiliation{Dresden High Magnetic Field Laboratory (HLD-EMFL), Helmholtz-Zentrum Dresden-Rossendorf, 01314 Dresden, Germany}

\author{A. A. Tsirlin}
\email{altsirlin@gmail.com}
\affiliation{Experimental Physics VI, Center for Electronic Correlations and Magnetism, Institute of Physics, University of Augsburg, 86135 Augsburg, Germany}

\author{R. Nath}
\email{rnath@iisertvm.ac.in}
\affiliation{School of Physics, Indian Institute of Science Education and Research, Thiruvananthapuram-695551, India}
\date{\today}

\begin{abstract}
Thermodynamic properties, $^{31}$P nuclear magnetic resonance (NMR) measurements, and density-functional band-structure calculations for $\varepsilon$-LiVOPO$_4$ are reported. This quantum magnet features a singlet ground state and comprises two types of alternating spin-$\frac12$ chains that manifest themselves by the double maxima in the susceptibility and magnetic specific heat, and by the two-step magnetization process with an intermediate $\frac12$-plateau. From thermodynamic data and band-structure calculations, we estimate the leading couplings of $J_1\simeq 20$~K and $J_2\simeq 60$~K and the alternation ratios of $\alpha_1=J_1'/J_1\simeq 0.6$ and $\alpha_2=J_2'/J_2\simeq 0.3$ within the two chains, respectively. The zero-field spin gap $\Delta_0/k_{\rm B}\simeq 7.3$~K probed by thermodynamic and NMR measurements is caused by the $J_1$-$J_1'$ spin chains and can be closed in the applied field of $\mu_{0}H_{\rm c1}\simeq 5.6$~T, giving rise to a field-induced long-range order. The NMR data reveal predominant three-dimensional spin-spin correlations at low temperatures. Field-induced magnetic ordering transition observed above $H_{c1}$ is attributed to the Bose-Einstein condensation of triplons in the sublattice formed by the $J_1$-$J_1'$ chains with weaker exchange couplings.
\end{abstract}

\maketitle

\section{Introduction}
Field-induced quantum phase transitions in magnets set a link between fermionic spin systems and lattice boson gas~\cite{Matsubara569,Giamarchi198,Zapf563}. In this context, spin-dimer compounds possessing a gap in the excitation spectrum are extensively studied~\cite{Giamarchi198,Zapf563}. Their triplet excitations (triplons) are equivalent to lattice bosons and can be tuned by applying magnetic field.
Gap closing in the applied field will usually lead to magnetic ordering that can be understood as Bose-Einstein condensation (BEC) of triplons~\cite{Mukhopadhyay177206,Sebastian617,Giamarchi198,Nikuni5868}. If interactions between the dimers are one-dimensional (1D) in nature, one further expects a non-Fermi-liquid-type Tomonaga-Luttinger Liquid (TLL) state realized at intermediate temperatures before the BEC state is reached~\cite{Klanjsek137207,Matsushita020408,Willenberg060407,Thielemann020408}.

The ground-state spin configuration in the applied field may also depend on the delicate balance between the kinetic energy and repulsive interaction of the triplons or $S_{\rm z} = +1$ bosons~\cite{Rice760}. The dominance of repulsive interaction would lead to the formation of superlattices, which result in magnetization plateaus~\cite{Narumi509,Shiramura1548}. This has been experimentally verified in the celebrated Shastry-Sutherland compound SrCu$_2$(BO$_3$)$_2$~\cite{Kodama395,Kageyama3168}. On the other hand, when the kinetic energy dominates over repulsive interactions, the triplons become delocalized, and the ground state is a superposition of singlet-triplet states, which can be approximated as a BEC of triplons. The phenomenon of BEC has been studied in detail for spin-dimer compounds TlCuCl$_{3}$\cite{Ruegg62}, BaCuSi$_{2}$O$_{6}$\cite{Jaime087203}, (Ba,Sr)$_{3}$Cr$_{2}$O$_{8}$~\cite{Aczel207203,Aczel100409}, etc. The transition from TLL as a 1D quantum critical state to the three-dimensional (3D) BEC state has often been observed in quasi-1D spin systems e.g. spin-$1/2$ ladders (C$_7$H$_{10}$N)$_2$CuBr$_4$~\cite{Jeong106404,Jeong167206,Moller020402} and (C$_5$H$_{12}$N)$_2$CuBr$_4$~\cite{Thielemann020408,Ruegg247202,Klanjsek137207} and spin-$1/2$ alternating spin chains Cu(NO$_3$)$_2\cdot$2.5D$_2$O and F$_5$PNN~\cite{Willenberg060407,Matsushita020408}. Thus, quasi-1D spin gap materials provide ample opportunities to tune the spin gap and study the field-induced quantum phase transitions.

The V$^{4+}$-based compounds offer an excellent playground to study gapped quantum magnets and related phenomena. Several of these compounds were already reported in the past in the context of spin gap physics~\mbox{\cite{Ueda2653,Yamauchi3729,Johnston219,Ghoshray214401}}. Recently, we studied magnetic properties of the $A$VO$X$O$_4$ series ($A$ = Na, Ag; $X$ = P, As), where all compounds were found to host alternating \mbox{spin-$\frac12$} chains not matching the structural chains of the VO$_6$ octahedra~\cite{Mukharjee144433,Arjun014421,Ahmed224423,Tsirlin144412}. In these systems, long-range superexchange couplings via two oxygen atoms play a central role. They are highly tunable and allow the variation of the zero-field spin gap $\Delta_{\rm 0}/k_{\rm B}$ from 21~K in NaVOAsO$_4$ to $\sim 2$~K in NaVOPO$_4$. External magnetic field closes the gap and leads to a field-induced magnetic transition, which is explained in terms of the triplon BEC~\cite{Mukharjee144433,Weickert104422}.

\begin{figure*}
	\includegraphics[scale=1]{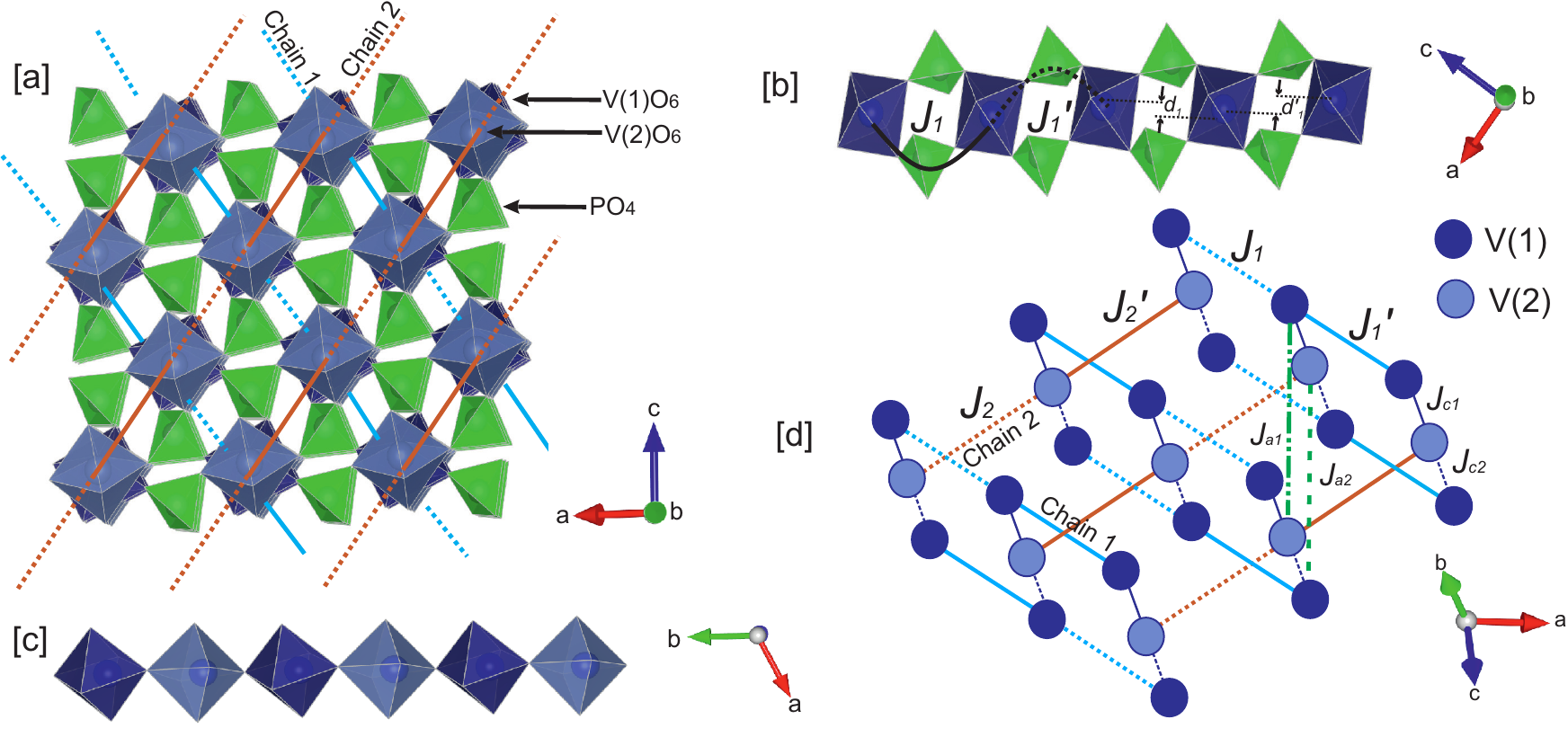}\\
	\caption{(a) Crystal structure of $\varepsilon$-LiVOPO$_4$ projected onto the $ac$-plane. The deep blue and light blue solid circles represent the V(1) and V(2) sites, respectively. Chain 1 with the couplings $J_1$ and $J_1'$ is formed by V(1) and chain 2 is formed by V(2) atoms with the couplings $J_2$ and $J_2'$ via the extended V-O$\ldots$O-V paths. These chains are nearly orthogonal to each other, whereas the structural chains are parallel comprising both V1 and V2 atoms at the same time. These chains run perpendicular to the $ac$-plane. (b) A segment of the chain 1 formed by the V(1)O$_6$ octahedra with the intrachain couplings $J_{1}$ and $J_{1}^{\prime}$. The distances $d_1$ and $d_1^{\prime}$ are the lateral displacements of the VO$_6$ octahedra in chain 1. (c) A section of the structural chain with the V(1)-O-V(2) paths along the $b$-axis. (d) An empirical/qualitative sketch of the spin model with all possible exchange interactions involving chain 1 ($J_1$, $J_1^{\prime}$) and chain 2 ($J_2, J_2^{\prime}$).}
	\label{Fig1}
\end{figure*}

Herein, we report ground-state properties of the chemically similar, but structurally different LiVOPO$_4$. Unlike the $A$VO$X$O$_4$ compounds, which are all monoclinic ($P2_1/c$), LiVOPO$_4$ crystallizes in several polymorphs with different symmetries and atomic arrangements~\cite{Harrison1751,Hidalgo8423}. We shall focus on the triclinic $\varepsilon$-LiVOPO$_4$ ($P\bar 1$) that can be seen as a distorted version of monoclinic $A$VO$X$O$_4$~\footnote{Triclinic crystal structure is derived from $\varepsilon$-VOPO$_4$. Alternatively, triclinic LiVOPO$_4$ is sometimes referred to as the $\alpha$-phase, because it was the earliest discovered LiVOPO$_4$ polymorph.} and has been actively studied in the context of battery research~\cite{Yang2008,Quackenbush2015,Lin2016,Shi2018,Chung2019}, but not as a quantum magnet. In its crystal structure, each of the Li, V, and P reside at two nonequivalent sites, whereas the O atoms have ten nonequivalent sites. The magnetic V$^{4+}$ ions form chains of the VO$_6$ octahedra with the alternation of V1 and V2, leading to two alternating V--V distances of 3.599 and 3.629\,\r A along these structural chains (Fig.~\ref{Fig1}).

Assuming that strongest magnetic interactions run along the structural chains, one expects the magnetic behavior of alternating spin-$\frac12$ chains that was indeed proposed by Onoda and Ikeda~\cite{Onoda053801} who reported, among others~\cite{Yang2008,Chung2019}, magnetic susceptibility of $\varepsilon$-LiVOPO$_4$. On the other hand, our recent results for the monoclinic $A$VO$X$O$_4$ compounds suggest that spin chains may not coincide with the structural chains, because leading interactions occur through the double bridges of the $X$O$_4$ tetrahedra. In this case, $\varepsilon$-LiVOPO$_4$ should feature two types of alternating spin-$\frac12$ chains, one formed by V(1) and the other one formed by V(2), each with different interactions and different spin gaps (Fig.~\ref{Fig1}). 
Below, we report experimental fingerprints of these two nonequivalent spin chains and thus directly confirm the proposed microscopic scenario. Moreover, we detect a field-induced phase transition, which is reminiscent of triplon BEC.

\begin{table}
	\caption{Atomic distances and bond angles along the superexchange paths involving P(1) and P(2) in the two alternating spin chains of $\varepsilon$-LiVOPO$_4$ at $300$~K to highlight the coupling of P atoms with V atoms.}
	\label{Cross Chain details}
	\begin{ruledtabular}
	\begin{tabular}{cll}
		Site & \multicolumn{1}{c}{Bond Length (\AA)}                                                                                                                                               & \multicolumn{1}{c}{Angle (deg.)}                                                                                                                                                                 \\ \hline
		P(1) & \begin{tabular}[c]{@{}l@{}}V(1)-O(1) = 1.96\\ O(1)-P(1) = 1.53\\ P(1)-O(2) = 1.54\\ O(2)-V(1) = 1.97\\ V(2)-O(7) = 1.98\\ O(7)-P(1) = 1.53\\ P(1)-O(8) = 1.53\\ O(8)-V(2) = 2.01\end{tabular} & \begin{tabular}[c]{@{}l@{}}$\angle$V(1)-O(1)-O(2) = 152.16\\ $\angle$V(1)-O(2)-O(1) = 109.37\\ $\angle$V(2)-O(8)-O(7) = 140.38\\ $\angle$V(2)-O(7)-O(8) = 124.64\\
		Average value $\simeq$ 131.63 \\ 
	    \end{tabular} \smallskip     \\ 
		P(2) & \begin{tabular}[c]{@{}l@{}}V(1)-O(3) = 1.96\\ O(3)-P(2) = 1.53\\ P(2)-O(4) = 1.55\\ O(4)-V(1) = 2.02\\ V(2)-O(9) = 1.98\\ O(9)-P(2) = 1.53\\ P(2)-O(10) = 1.52\\ O(10)-V(2) = 1.94\end{tabular} & \begin{tabular}[c]{@{}l@{}}$\angle$V(1)-O(4)-O(3) = 118.59\\ $\angle$V(1)-O(3)-O(4) = 149.06\\ $\angle$V(2)-O(9)-O(10) = 134.87\\$\angle$V(2)-O(10)-O(9) = 132.66\\
		 Average value $\simeq$ 133.79 \end{tabular} \\ 
	\end{tabular}
	\end{ruledtabular}
\end{table}

\section{Methods}
Polycrystalline sample of $\varepsilon$-LiVOPO$_4$  was prepared by the conventional solid-state reaction method from stoichiometric mixtures of LiPO$_3$ and VO$_2$ (Aldrich, 99.995\%). LiPO$_3$ was obtained by heating LiH$_2$PO$_4$.H$_{2}$O (Aldrich, 99.995\%) for 4~h at 400~$^{\circ}$C in air. The reactants were ground thoroughly, pelletized, and fired at 740~$^{\circ}$C for two days in flowing argon atmosphere with two intermediate grindings. Phase purity of the sample was confirmed by powder x-ray diffraction (XRD) recorded at room temperature using a PANalytical powder diffractometer (Cu\textit{K}$_{\alpha}$ radiation, $\lambda_{\rm avg}\simeq 1.5418$~{\AA}). Rietveld refinement of the acquired data was performed using \texttt{FULLPROF} software package~\cite{Carvajal55} taking the initial cell parameters from Ref.~[\onlinecite{Ateba1223}]. The low-temperature XRD data down to $15$~K were recorded using a low-temperature attachment (Oxford Phenix) to the x-ray diffractometer. 

Magnetization ($M$) was measured as a function of temperature (2~K~$\leq T \leq$~380~K) using the vibrating sample magnetometer (VSM) attachment to the Physical Property Measurement System [PPMS, Quantum Design]. A $^{3}$He attachment to the SQUID [MPMS-7T, Quantum Design] magnetometer was used for magnetization measurements in the low-temperature range ($0.5$~K $\leq T \leq$ $2$~K). Specific heat ($C_{\rm p}$) as a function of temperature was measured down to $0.35$ K using the thermal relaxation technique in PPMS under magnetic fields up to $14$~T. For $T \leq 2$~K, measurements were performed using an additional $^{3}$He attachment to PPMS. High-field magnetization was measured in pulsed magnetic field up to $60$~T at the Dresden High Magnetic Field Laboratory~\cite{Tsirlin132407,Skourski214420}.

The NMR experiments on the $^{31}$P nucleus (nuclear spin $I = 1/2$ and gyromagnetic ratio $\gamma/2\pi=17.235$~MHz/T) were carried out using pulsed NMR technique in the temperature range 1.6~K~$\leq T \leq$~230~K. The $^{31}$P NMR spectra as a function of temperature were obtained either by sweeping the field at a fixed frequency or by taking the Fourier transform (FT) of the echo signal, keeping the magnetic field fixed. The NMR shift $K(T)$ = [$H_{\rm ref}-H(T)$]/$H(T)$ was determined by measuring the resonant field $H(T)$ of the sample with respect to the standard H$_3$PO$_4$ sample (resonance frequency $H_{\rm ref}$). The $^{31}$P nuclear spin-lattice relaxation rate ($1/T_1$) was measured using the inversion recovery technique at different temperatures. Details of the positions of both P sites with respect to the magnetic V$^{4+}$ centers are given in Table~\ref{Cross Chain details}.

Density-functional (DFT) band-structure calculations were performed in the \texttt{FPLO} code~\cite{fplo} using experimental crystal structure from Ref.~[\onlinecite{lavrov1982}] and the Perdew-Burke-Ernzerhof (PBE) flavor of the exchange-correlation potential~\cite{pbe96}. Exchange couplings were obtained within superexchange theory or by a mapping procedure~\cite{Tsirlin014405} using total energies of collinear spin configurations calculated within DFT+$U$, where the on-site Coulomb repulsion $U_d=5$\,eV and Hund's coupling $J_d=1$\,eV~\cite{Nath024418,Tsirlin104436} were used to account for strong correlations in the V $3d$ shell. All calculations are performed for the $4\times 4\times 4$ $k$-mesh. To resolve all pertinent exchange couplings, the $(\av_t+\cv_t)\times\bv_t\times (\av_t-\cv_t)$, $\av_t\times 2\bv_t\times \cv_t$, and $(\av_t+\bv_t)\times 2\bv_t\times \cv_t$ supercells were used, where $\av_t$, $\bv_t$, and $\cv_t$ are lattice vectors of the triclinic unit cell given in Ref.~[\onlinecite{lavrov1982}]. 

\section{Results}
\subsection{X-ray Diffraction}
\begin{figure}
\includegraphics [width = \linewidth]{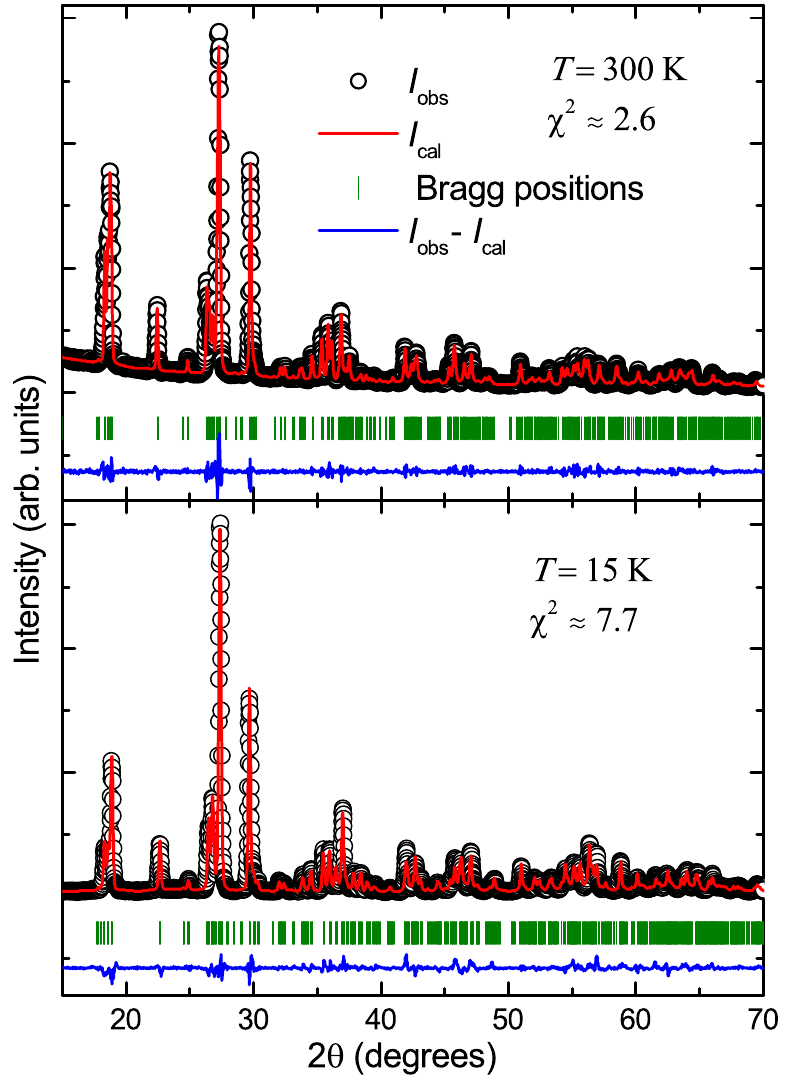}\\
\caption{X-ray powder diffraction patterns of $\varepsilon$-LiVOPO$_4$ at two different temperatures ($T = 300$~K and $15$~K). The solid lines denote the Rietveld refinement fit of the data. The Bragg-peak positions are indicated by green vertical bars and bottom blue line indicates the difference between the experimental and calculated intensities.}
\label{Fig2}
\end{figure}
\begin{figure}
\includegraphics [width = \linewidth]{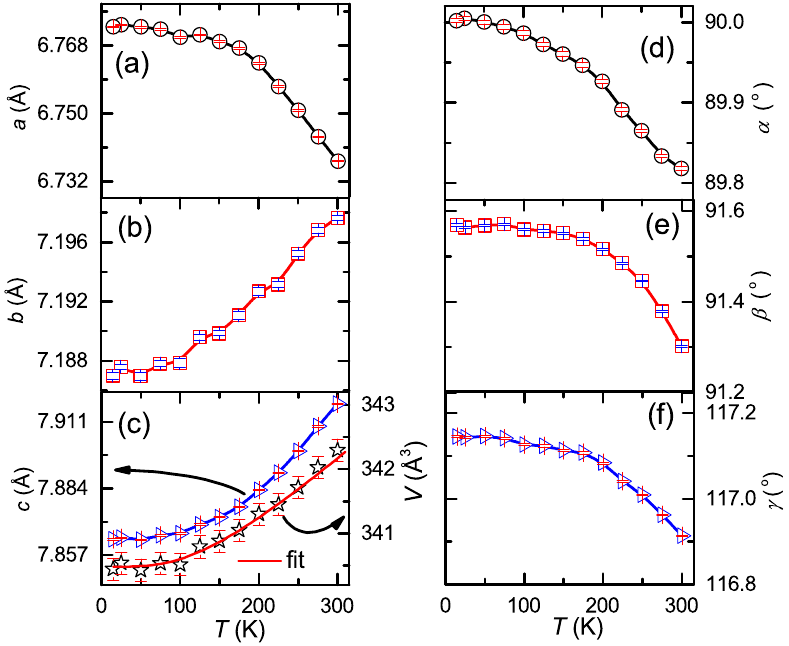}\\
\caption{Lattice parameters (a) $a$, (b) $b$, (c) $c$, (d) $\alpha$, (e) $\beta$, and (f) $\gamma$ vs temperature. The unit cell volume ($V_{\rm cell}$) is plotted as a function of temperature in right $y$-axis of (c) and the solid line represents the fit using Eq.~\eqref{VcellvsT}.}
\label{Fig3}
\end{figure}
In order to confirm the phase purity and to study the temperature variation of the crystal structure, the powder XRD patterns are analyzed for $15$~K $\leq$ $T$ $\leq$ $300$~K. The XRD patterns at two end temperatures, $T = 300$~K and 15~K, along with the refinement are shown in Fig.~\ref{Fig2}. At room temperature, all the peaks could be indexed based on the triclinic (space group: {\it $P\bar{1}$}) structure, implying phase purity of the sample. The refined lattice parameters [$a = 6.729(1)$~\AA, $b = 7.194(1)$~\AA, $c = 7.920(2)$~\AA, $\alpha = 89.82(2)^\circ$, $\beta = 91.2288(2)^\circ$, and $\gamma = 116.8799(2)^\circ$] at room temperature are in good agreement with the previous report~\cite{Ateba1223}. Identical XRD patterns with no extra peaks are observed in the whole measured temperature range, which excludes any structural phase transition or lattice deformation in $\varepsilon$-LiVOPO$_4$, unlike other spin gap compounds NaTiSi$_{2}$O$_{6}$ \cite{Isobe1423,Popovic036401}, CuGeO$_{3}$ \cite{Hirota736}, NaV$_{2}$O$_{5}$\cite{Fujii326}, and K-TCNQ \cite{Lepine3585}.

The variation of both lattice parameters and unit cell volume ($V_{\rm cell}$) as a function of temperature are shown in Fig.~\ref{Fig3}. The cell parameters $b$ and $c$ decrease systematically, while the rest of the parameters ($a$, $\alpha$, $\beta$, and $\gamma$) rise with decreasing temperature. However, the overall unit cell volume shrinks upon cooling. The temperature variation of $V_{\rm cell}$ was fitted by the equation~\cite{Kittel1986,Bag144436}
\begin{equation}
V_{\rm cell}(T)=\gamma U(T)/K_0+V_0,
\label{VcellvsT}
\end{equation}
where $V_{0}$ is the unit cell volume at $T$ = $0$~K, $K_{0}$ is the bulk modulus, and $\gamma$ is the Gr$\ddot{\rm u}$neisen parameter. The internal energy $U(T)$ can be expressed in terms of the Debye approximation as
\begin{equation}
U(T)=9pk_{\rm B}T\left(\frac{T}{\theta_{\rm D}}\right)^3\int_{0}^{\theta_{\rm D}/T}\dfrac{x^3}{e^x-1}dx.
\end{equation}
In the above, $p$ stands for the total number of atoms in the unit cell and $k_{\rm B}$ is the Boltzmann constant. The best fit of the data down to 15~K [see Fig.~\ref{Fig3}(c)] was obtained with the Debye temperature $\theta_{D} \simeq  530$~K, $\frac{\gamma}{K_{0}} \simeq 5.78 \times10^{-5}$~Pa$^{-1}$, and $V_{0} \simeq 340.5$~\AA$^{3}$.

\subsection{Magnetization}
\begin{figure}[h]
\includegraphics [width = \linewidth]{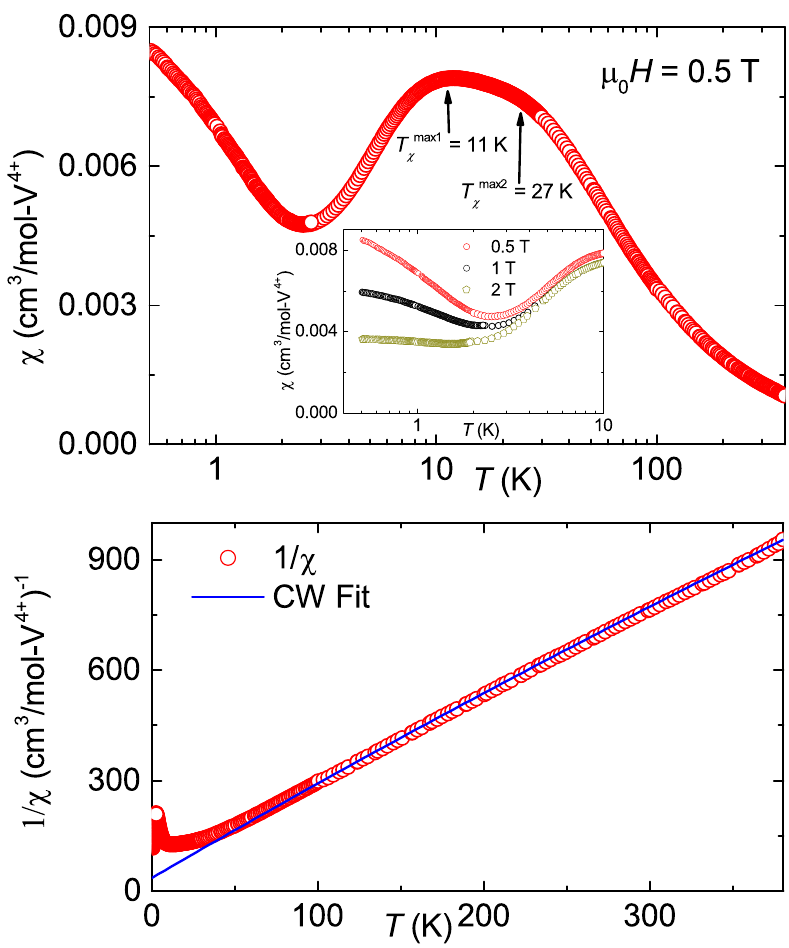}\\
\caption{Upper panel: $\chi(T)$ measured in $\mu_{0}H = 0.5 $~T. The two shoulders of the broad maximum are indicated by the vertical arrows. Inset: low-$T$ $\chi(T)$ measured in different applied fields. Lower panel: $1/\chi$ vs $T$ and the solid line is the CW fit using Eq.~\eqref{cw}.}
\label{Fig4}
\end{figure}
\vspace{0cm}
The temperature-dependent bulk magnetic susceptibility $\chi ~(\equiv M/H$) of $\varepsilon$-LiVOPO$_4$ measured in an applied field of $\mu_{0}H = 0.5$~T is shown in the upper panel of Fig.~\ref{Fig4}. It is consistent with earlier reports~\cite{Chung2019,Yang2008,Onoda053801} and exhibits a very broad maximum with two shoulders at around $T_{\chi}^{\rm max1} \simeq 11$~K and $T_{\chi}^{\rm max2}\simeq 27$~K. Such a broad maximum should not be mistaken with a magnetic transition~\cite{Chung2019} and mimics the short-range ordering of a low-dimensional quantum magnet. However, already the fact that two shoulders are observed in the susceptibility indicates the presence of two nonequivalent spin chains in $\varepsilon$-LiVOPO$_4$. Since peak position is related to the intrachain exchange coupling~\cite{Johnston9558}, we can estimate relative strengths of the interactions in the two chains, $\bar J_2/\bar J_1\simeq 2.45$, where $\bar J_1=(J_1+J_1')/2$ and $\bar J_2=(J_2+J_2')/2$ are average couplings in the chain 1 and chain 2, respectively. The susceptibility alone does not give information on which of the chains features stronger couplings, but our DFT calculations (Sec.~\ref{sec:model}) suggest that stronger interactions occur within chain 2. 

Below $T_{\chi}^{\rm max1}$, $\chi$ decreases rapidly suggesting the opening of a spin gap. However, below $2$~K, a large upturn is observed, which can be attributed to the effect of extrinsic paramagnetic contributions. As shown in the inset of the upper panel of Fig.~\ref{Fig4}, with the application of magnetic field this low-temperature Curie tail gets suppressed. Moreover, our powder XRD suggests high purity of the sample. Therefore, this low-temperature upturn in $\chi(T)$ may be due to the uncorrelated V$^{4+}$ free spins or chain-end effects that largely affect $\chi(T)$ at low temperatures~\cite{Eggert934}.

To extract the magnetic parameters, we first fitted the $1/\chi(T)$ data (see the lower panel of Fig.~\ref{Fig4}) above 150~K by the modified Curie-Weiss (CW) law,
\begin{equation}\label{cw}
\chi(T) = \chi_0 + \frac{C}{T - \theta_{\rm CW}},
\end{equation}
where $\chi_{\rm 0}$ represents the temperature-independent susceptibility, which includes the Van Vleck paramagnetic and core diamagnetic contributions, $C$ is the Curie constant, and $\theta_{\rm CW}$ is the CW temperature. The resulting fitting parameters are $\chi_{0} \simeq 7.76 \times 10^{-5}$~cm$^{3}$/mol-V$^{4+}$, $C \simeq 0.383$~cm$^{3}$K/mol-V$^{4+}$, and $\theta_{\rm CW} \simeq - 13.4$~K. Negative value of $\theta_{\rm CW}$ indicates that the dominant interactions between the V$^{4+}$ spins are antiferromagnetic (AFM) in nature. The effective moment was calculated by using the experimental value of $C$ in the relation $\mu_{\rm eff} = \sqrt{3k_{\rm B}C/N_{\rm A}}$, where $N_{\rm A}$ is the Avogadro's number. The calculated value of $\mu_{\rm eff} \simeq 1.72\mu_{\rm B}$/V$^{4+}$ is very close to the theoretical spin-only value [$\mu_{\rm eff} = g\sqrt{S(S+1)} \simeq 1.73\mu_{\rm B}$] for $S =1/2$, assuming the Land$\acute{e}$ $g$ factor $g = 2$. The experimental value of $\mu_{\rm eff}$ corresponds to a slightly reduced $g$ value of $g \simeq 1.98$ which is 
identical to the case of other V$^{4+}$-based compounds~\cite{Mukharjee144433,Tsirlin144412,Yogi024413}. The core diamagnetic susceptibility ($\chi_{\rm core}$) of $\varepsilon$-LiVOPO$_4$ was estimated to be $-6.8 \times 10^{-5}$~cm$^3$/mol by adding the $\chi_{\rm core}$ of the individual ions Li$^{1+}$, V$^{4+}$, P$^{5+}$, and O$^{2-}$~\cite{Selwood2013}. The Van Vleck paramagnetic susceptibility ($\chi_{\rm VV}$) of $\sim 14.6 \times 10^{-5}$~cm$^3$/mol is obtained by subtracting $\chi_{\rm core}$ from $\chi_{\rm 0}$.

\begin{figure}[h]
\includegraphics [width = \linewidth]{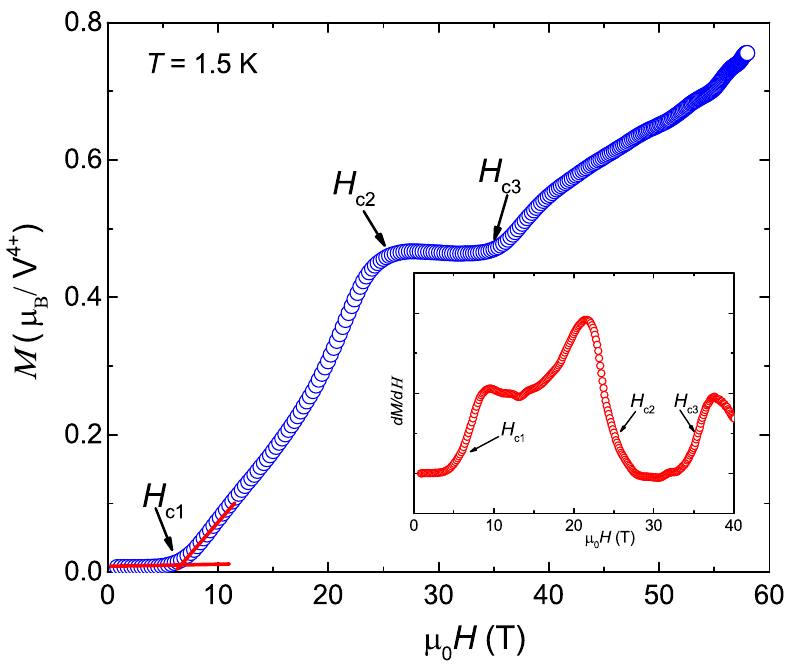}\\
\caption{Magnetization vs field measured at $T = 1.5$~K using pulsed magnetic field. Inset: $dM$/$dH$ vs $H$ to highlight the critical fields $H_{\rm c1}$, $H_{\rm c2}$, and $H_{\rm c3}$.}
\label{Fig5}
\end{figure}
Two alternating spin chains with different exchange parameters should also manifest themselves in the magnetization process. Indeed, experimental magnetization curve measured at the base temperature of $T = 1.5$~K (Fig.~\ref{Fig5}) shows a kink due to the gap closing at $\mu_{0}H_{\rm c1}\simeq 5.6$\,T that corresponds to the spin gap of $\Delta_{\rm 0}^{\rm M}/k_{\rm B}= g\mu_{0}\mu_{\rm B}H_{\rm c1}/k_{\rm B}\simeq 7.3$~K. At $\mu_{0}H_{\rm c2}\simeq 25$\,T, half of the spins are saturated, with the $\frac12$-plateau observed up to $\mu_{0}H_{\rm c3}\simeq 35$\,T. At even higher fields, the remaining spins are gradually polarized with nearly 80\,\% of the saturation magnetization reached at 58\,T, the highest field of our experiment. 

The two-step increase of the magnetization with the spin gap and intermediate $\frac12$-plateau is common for dimers built by two spin-1 ions~\cite{Samulon047202}. However, $\varepsilon$-LiVOPO$_4$ contains spins-$\frac12$, so this behavior should have a different origin. We infer that at $\mu_{0}H_{\rm c1}$ the spin gap corresponding to chain 1 is closed. This chain is fully polarized at $\mu_{0}H_{\rm c2}$, whereas at $\mu_{0}H_{\rm c3}$ the gap corresponding to chain 2 is closed, and the magnetization increases further. Such a scenario is confirmed by our quantitative analysis in Sec.~\ref{sec:model}.

\subsection{Specific heat}
\begin{figure}
\includegraphics[width = \linewidth]{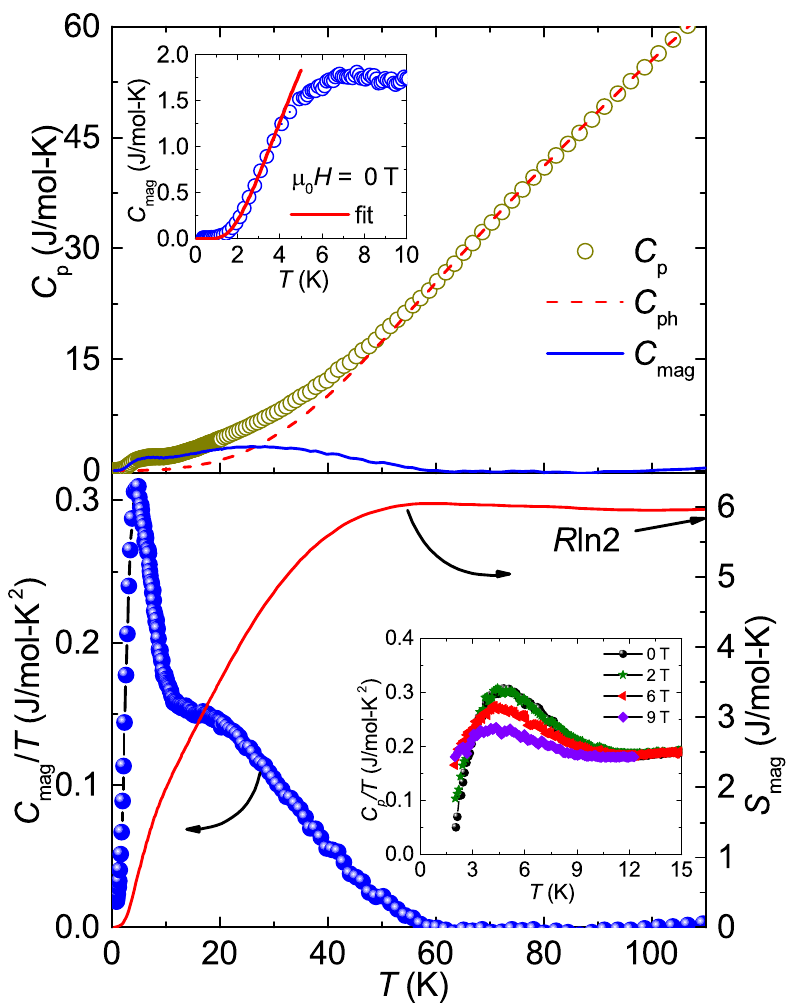}
\caption{Upper panel: specific heat $C_{\rm p}$ vs $T$ of $\varepsilon$-LiVOPO$_{4}$ in zero applied field. The dashed line stands for the phonon contribution to the specific heat ($C_{\rm ph}$) using Debye fit [Eq.~\eqref{Debye}]. The solid line is the magnetic contribution to the specific heat ($C_{\rm mag}$). Inset: $C_{\rm mag}$ vs $T$ in the low-temperature region. The solid line is an exponential fit as described in the text. Lower panel: $C_{\rm mag}/T$ and $S_{\rm mag}$ vs $T$ in the left and right $y$-axes, respectively. Inset: $C_{\rm p}/T$ vs $T$ in different magnetic fields in the low-temperature regime.}
\label{Fig6}
\end{figure}
The specific-heat ($C_{\rm p}$) data measured under zero field are shown in the upper panel of Fig.~\ref{Fig6}. No sharp anomaly/peak was noticed down to $T = 0.35$~K, thus ruling out the possibility of any magnetic or structural transition. A broad hump associated with low-dimensional short-range order is observed around $T_{C}^{\rm max}$ $\simeq 6$~K which moves weakly toward low temperatures with increasing magnetic field (see the inset of the lower panel of Fig.~\ref{Fig6}), reflecting the closing of the spin gap. The gap was estimated by fitting the data below 4~K with the activated behavior, $C_{\rm mag} \propto \exp (-\Delta_{0}^{\rm C}/k_{\rm B}T)$. The fit, as illustrated in the inset of the upper panel of Fig.~\ref{Fig6}, returns the zero-field spin gap of $\Delta_{0}^{\rm C}/k_{\rm B} \simeq 7.3$~K that matches nicely the value obtained from the high-field magnetization data.

Typically, in magnetic insulators, the high-temperature part of $C_{\rm p}$ is dominated by the phonon contribution, whereas the magnetic contribution becomes prominent at low temperatures. To estimate the phonon contribution ($C_{\rm ph}$), the experimental data at high temperatures (60~K~$\leq T \leq$~110~K) were fitted by a linear combination of three Debye functions~\cite{Nath064422}
\begin{equation}
\label{Debye}
C_{\rm ph}(T) = 9R\displaystyle\sum\limits_{\rm n=1}^{3} c_{\rm n} \left(\frac{T}{\theta_{\rm Dn}}\right)^3 \int_0^{\frac{\theta_{\rm Dn}}{T}} \frac{x^4e^x}{(e^x-1)^2} dx.
\end{equation} 
In the above, $R$ is the universal gas constant, the coefficients $c_{\rm n}$ stand for the groups of different atoms present in the crystal, and $\theta_{\rm Dn}$ are the corresponding Debye temperatures. The $C_{\rm ph}$ was extrapolated down to low temperatures and subtracted from the total specific heat to obtain the magnetic contribution to the specific heat ($C_{\rm mag}$). The obtained $C_{\rm mag}(T)$ is presented in the upper panel of Fig.~\ref{Fig6}. The accuracy of the above fitting procedure was further verified by calculating the magnetic entropy $S_{\rm mag}$ obtained by integrating $C_{\rm mag}/T$ (see the lower panel of Fig.~\ref{Fig6}). The value of $S_{\rm mag}$ is calculated to be $\sim 5.9$~J/mol K at $T \simeq 100$~K, which is close to $S_{\rm mag}=R\ln 2=5.76$~J/mol K expected for spin $\frac12$. 

As shown in the upper panel of Fig.~\ref{Fig6}, $C_{\rm mag}$ develops two broad maxima at $T_{C}^{\rm max1} \simeq 7$~K and $T_{C}^{\rm max2} \simeq 27$~K, similar to the two shoulders in the $\chi(T)$ data. The clear separation of these maxima indicates the different interaction strength in chains 1 and 2.

\begin{figure}
	\includegraphics[width = \linewidth]{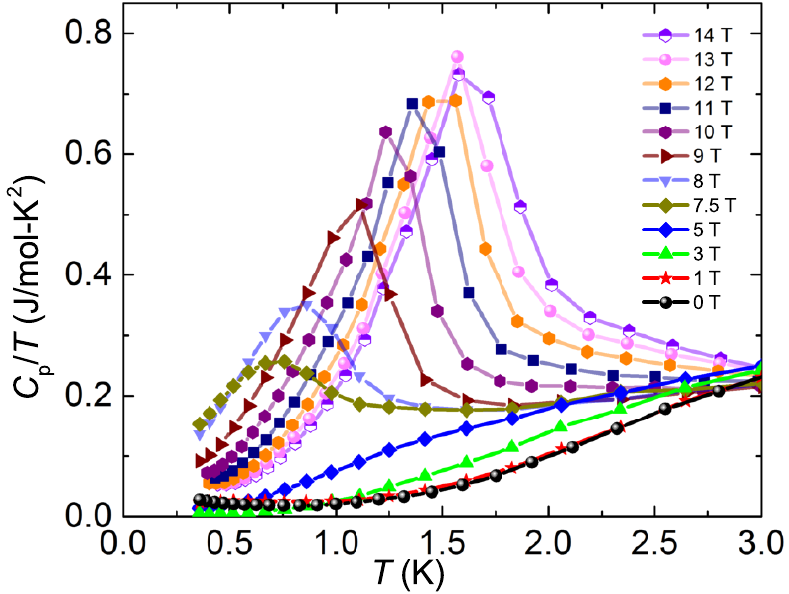}
	\caption{$C_{\rm p}/T$ vs $T$ measured in different applied magnetic fields up to 14~T.}
	\label{Fig7}
\end{figure}
With the spin gap closed around $\mu_{0}H_{\rm c1} \simeq 5.6$~T, the system may enter a long-range order (LRO) state. This state is indeed observed in the specific-heat data measured above $\mu_{0}H_{\rm c1}$. As shown in Fig.~\ref{Fig7}, no anomaly in $C_{\rm p}$ is found down to $T = 0.35$~K and up to $\mu_{0}H = 5$~T. However, for $\mu_{0}H > 7$~T a clear anomaly appears indicating the onset of a field-induced magnetic LRO ($T_{\rm N}$). This peak shifts toward higher temperature with increasing magnetic field.

\subsection{$^{31}$P NMR}
As mentioned previously, $\chi(T)$ does not show an exponential decrease at low temperatures anticipated for a gapped spin system, and may be influenced by extrinsic contributions. To access the intrinsic susceptibility of $\varepsilon$-LiVOPO$_4$, we performed NMR measurements on the $^{31}$P nuclei.

\subsubsection{NMR Shift}
\begin{figure}
	\includegraphics [width = \linewidth]{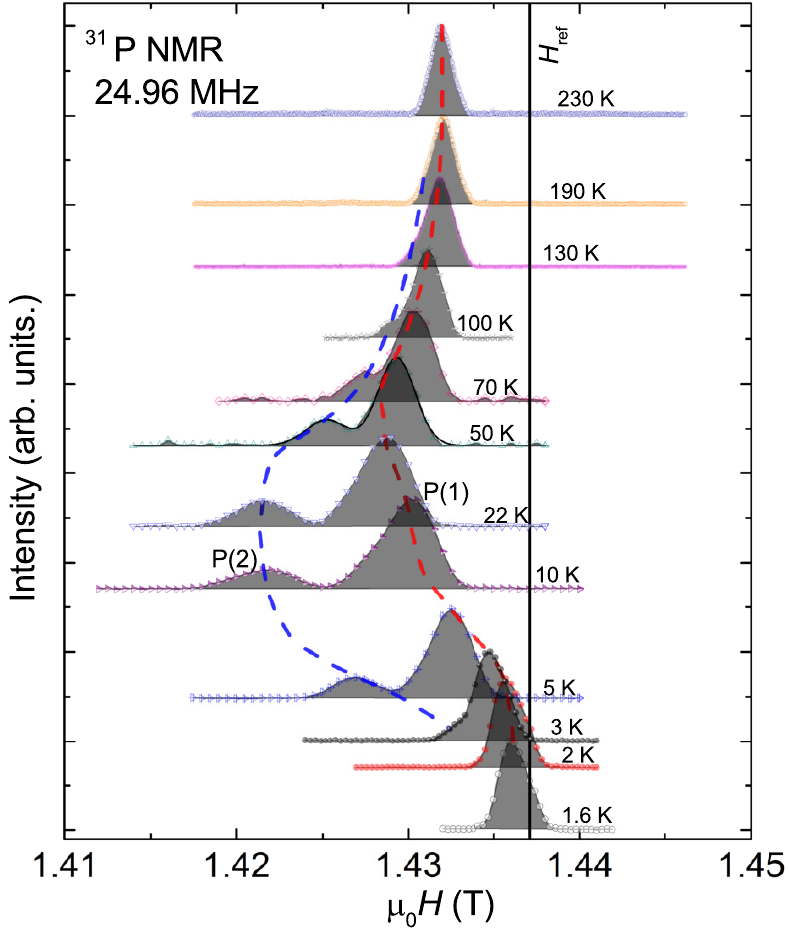}\\
	\caption{ Typical $^{31}$P field-sweep NMR spectra of $\varepsilon$-LiVOPO$_{4}$ measured at 24.96~MHz at different temperatures. The dashed lines track the shifting of NMR line positions. The vertical line corresponds to the $^{31}$P reference line ($H_{\rm ref}$) measured for H$_{3}$PO$_{4}$. The solid line is the fit of the spectra at $T = 50$~K using a double Gaussian function.}
	\label{Fig8}
\end{figure}
Figure~\ref{Fig8} presents the field-sweep $^{31}$P NMR spectra measured over a wide temperature range. At high temperatures, a narrow and symmetric spectral line typical for a $I = 1/2$ nucleus is observed. As the temperature is lowered, the line shape becomes asymmetric, followed by a complete splitting of the two spectral lines below about 120~K. This suggests the existence of two nonequivalent P sites, P(1) and P(2), with a different crystallographic environment, which is consistent with the structural data (Table~\ref{Cross Chain details}). Both lines shift with temperature and merge below about 4~K. The absence of drastic line broadening and/or line splitting down to 1.6~K, except for the one caused by the two nonequivalent P sites, rules out the occurrence of any structural and magnetic transitions. The line with a lower intensity shifts more strongly than the one with the higher intensity. The former can be assigned to P(2) and the latter to P(1), because the P(2)O$_4$ tetrahedra mediate stronger exchange interactions, $J_1$ in chain 1 and $J_2$ in chain 2, whereas the P(1)O$_4$ tetrahedra mediate weaker interactions $J_1'$ and $J_2'$, respectively (see Sec.~\ref{sec:model}). At $T = 1.6$~K, the position of the peak is very close to the zero shift value suggesting that the ground state of $\varepsilon$-LiVOPO$_{4}$ is nonmagnetic.

\begin{figure}
	\includegraphics[width = \linewidth]{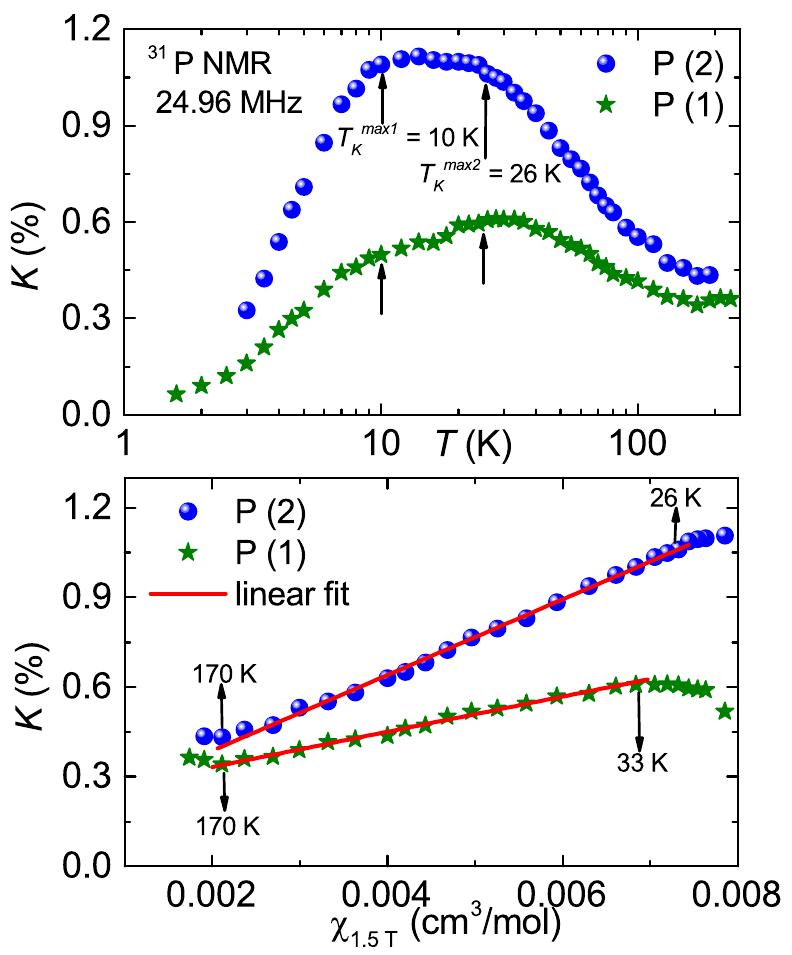}
	\caption{ Upper panel: temperature-dependent $^{31}$P NMR shift $K(T)$ measured at 24.96~MHz as a function of temperature for both P(1) and P(2) sites. Lower panel: $K$ vs $\chi$ (measured at 1.5~T) with temperature as an implicit parameter. The solid lines are the linear fits.}
	\label{Fig9}
\end{figure}

The temperature-dependent NMR shift $K(T)$ for both $^{31}$P sites was extracted by fitting each spectrum to a sum of two Gaussian functions. The results are shown in the upper panel of Fig.~\ref{Fig9}. One advantage of the NMR shift over the bulk $\chi(T)$ is that the Curie-Weiss term due to foreign phases and/or defects does not appear. The randomly distributed defects/impurities only broaden the NMR line but do not contribute to the NMR shift~\cite{Walstedt1974}. Therefore, $K(T)$ is more favorable than bulk $\chi(T)$ data for a reliable determination of magnetic parameters.

The $K(T)$'s corresponding to the P(1) and P(2) sites pass through a very broad maximum, similar to the $\chi(T)$ data. The overall temperature dependence is similar for P(1) and P(2), but the absolute values differ due to the different hyperfine couplings. The presence of several distinct P sites in the structure is reminiscent of the ambient-pressure polymorph of (VO)$_2$P$_2$O$_7$ with its two non-equivalent alternating spin-$\frac12$ chains. In that case, different phosphorous sites probe the behavior of different spin chains in the structure~\cite{Yamauchi3729,Kikuchi6731}. In contrast, each of the P sites in $\varepsilon$-LiVOPO$_4$ is coupled to both spin chains, so it is not possible to probe $K(T)$ separately. Two distinct shoulders are observed in $K(T)$ at $T_{K}^{\rm max1} \simeq 10$~K and $T_{K}^{\rm max2}\simeq 26$~K and closely resemble bulk magnetic susceptibility (Fig.~\ref{Fig4}). 

At low temperatures, both shifts decrease rapidly toward zero, suggesting the opening of a spin gap between the singlet ($S=0$) ground state and triplet ($S=1$) excited states. As NMR shift is insensitive to the impurities and defects, one can use it to accurately measure the intrinsic spin susceptibility. In the upper panel of Fig.~\ref{Fig9}, we show that for both P sites, $K(T)$ decreases toward zero, which is in contrast to the upturn observed in the low-temperature $\chi(T)$ data. This confirms the extrinsic nature of the low-temperature upturn observed in $\chi(T)$. In powder samples, the defects often break spin chains, with the unpaired spins at the ends of finite chains giving rise to the staggered magnetization, which also appears as the low-temperature Curie tail in $\chi(T)$.

The direct relation between $K(T)$ and spin susceptibility $\chi_{\rm spin}$ can be written as
\begin{equation}\label{K_chi}
K(T) = K_0 + \frac{A_{\rm hf}}{N_{A}\mu_{\rm B}}\chi_{\rm spin},
\end{equation}
where $K_0$ is the temperature-independent NMR shift and $A_{\rm hf}$ is the total hyperfine coupling constant between the $^{31}$P nuclei and V$^{4+}$ spins. The $A_{\rm hf}$ consists of the contributions due to transferred hyperfine coupling and nuclear dipolar coupling constants. Since both of the aforementioned couplings are temperature-independent, $K(T)$ is a direct measure of $\chi_{\rm spin}$. Using Eq.~\eqref{K_chi}, $A_{\rm hf}$ can be calculated from the slope of the linear $K$ vs $\chi$ plot with temperature as an implicit parameter. The lower panel of Fig.~\ref{Fig9} presents the $K$ vs $\chi$ plots for both P sites showing linear behavior at high temperatures. From the linear fit, the hyperfine coupling constants $A_{\rm hf}^{\rm P(1)} \simeq 3290$~Oe/$\mu_{\rm B}$ and $A_{\rm hf}^{\rm P(2)} \simeq 7068$~Oe/$\mu_{\rm B}$ are estimated for the P(1) and P(2) sites, respectively. Thus, the P(2) site is coupled with the V$^{4+}$ ions twice stronger than the P(1) site. These values are comparable to the $A_{\rm hf}$ values reported for other spin chains with similar interaction geometry~\cite{Mukharjee144433,Nath134451,Nath174436}.

\begin{figure}
\includegraphics[width = \linewidth]{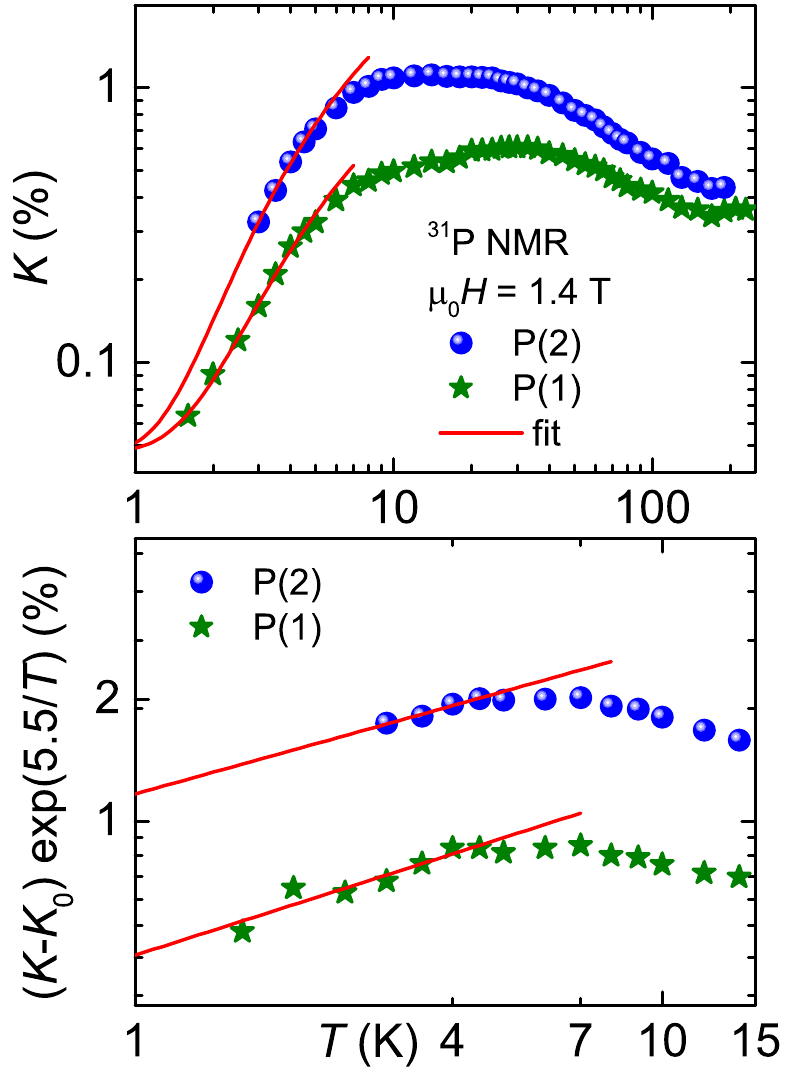}
\caption{Upper panel: $K$ vs $T$ for both P(1) and P(2) sites. The solid lines are the fits below 5~K by $K = K_{0} + mT^{(d/2-1)}e^{-\Delta/k_{\rm B}T}$, fixing $\Delta/k_{\rm B} \simeq 5.5$~K. Lower panel: $(K-K_{0})e^{5.5/T}$ vs $T$ is shown in the low-temperature regime. The solid lines are the fits with $d \simeq 2.83$ and 2.73 for the P(1) and P(2) sites, respectively.}
\label{Fig10}
\end{figure}

It is also possible to estimate the value of spin gap and the effective lattice dimensionality ($d$) by analyzing the low-temperature $K(T)$ data. Typically, the non-negligible interchain couplings are inevitably present in real materials and significantly influence the $K(T)$ data at low temperatures. For a $d$-dimensional system, the susceptibility at $k_{\rm B}T\ll\Delta$ can be approximated as~\cite{Taniguchi2758}
\begin{equation}\label{chi_d}
\chi_d \propto T^{(d/2)-1}\times e^{-\Delta/k_{\rm B}T}.
\end{equation}
Our NMR experiments were carried out in the magnetic field of $\mu_{0}H = 1.4$~T. Assuming a linear variation of $\Delta/k_{\rm B}$ with $H$, the zero-field spin gap $\Delta_{\rm 0}/k_{\rm B} \simeq 7.3$~K determined from the specific heat and magnetization is expected to be reduced to $\Delta_{\rm 1.4~T}/k_{\rm B} \simeq 5.5$~K at $\mu_{0}H = 1.4$~T. In the upper panel of Fig.~\ref{Fig10}, the $K(T)$ data below 5~K are fitted by $K = K_{0} + mT^{(d/2-1)}e^{-\Delta/k_{\rm B}T}$, fixing $\Delta/k_{\rm B} \simeq 5.5$~K where $m$ is a proportionality constant. 
In order to highlight the low-temperature linear regime and the fit, in the lower panel of Fig.~\ref{Fig10}, we have plotted $(K-K_{0})e^{5.5/T}$ vs $T$ in the log-log plot. The fit in this region returns $K_{0} \simeq 0.04694$\%, $m \simeq 0.4654$~\%/K$^{1/2}$, and $d \simeq 2.83$ for the P(1) site and $K_{0} \simeq 0.0462$\%, $m \simeq 1.1663$~\%/K$^{1/2}$, and $d \simeq 2.73$ for the P(2) site. The average value of $d \simeq 2.78$, which is very close to 3, suggests the dominant role of 3D spin-spin correlations at low temperatures. Indeed, we find rather strong interchain couplings from DFT (Sec.~\ref{sec:model}).

\subsubsection{Spin-lattice relaxation rate $1/T_1$}
\label{sec:T1}
\begin{figure}
\includegraphics[width = \linewidth]{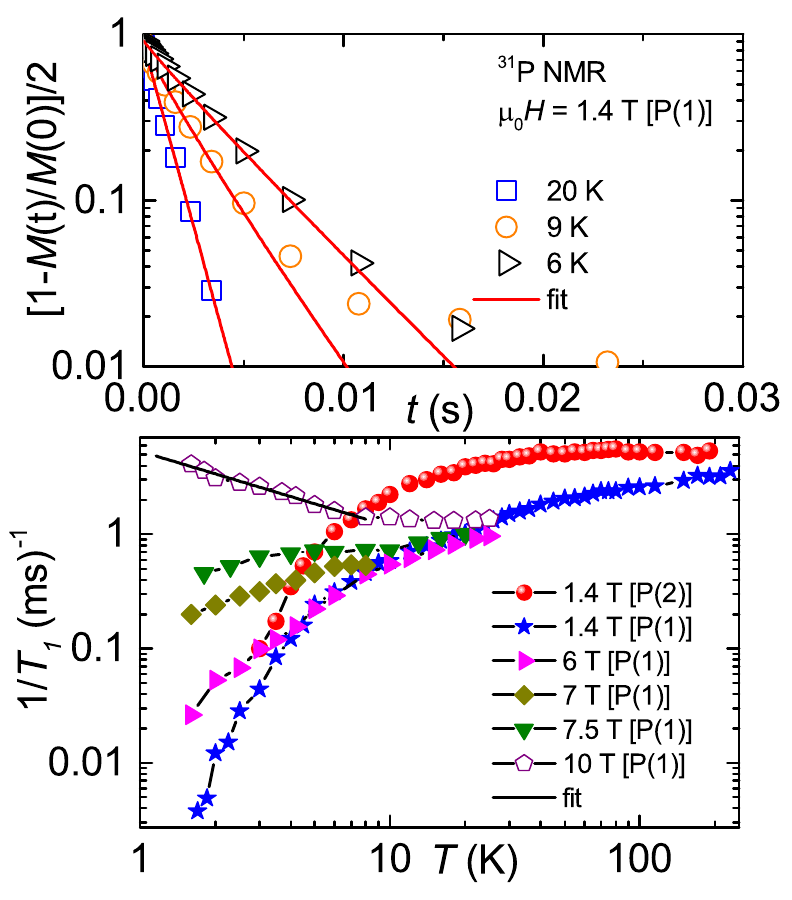}
\caption{Upper panel: recovery of the longitudinal magnetization as a function of waiting time $t$ at three different temperatures for the P(1) site. Solid lines are the fits using Eq.~\eqref{T1}. Lower panel: temperature variation of $1/T_1$ measured in different magnetic fields. For $\mu_{0}H = 1.4$~T, measurements are done on both P(1) and P(2) sites while for other fields, only P(1) site is probed. The solid line represents the fit using $1/T_1 \propto T^{\alpha}$ to the 10~T data at low temperatures with $\alpha \simeq -0.6$.}
\label{Fig11}
\end{figure}
The spin-lattice relaxation rate $1/T_1$ measures the dynamic susceptibility, which provides direct access to the low-energy spin excitations or spin-spin correlation function~\cite{Moriya23}. $1/T_1$ was measured at the central peak position of the spectra at each temperature using an inversion pulse sequence down to $T = 1.6$~K. Since $^{31}$P has the nuclear spin $I=1/2$, the value of $T_{1}$ at each temperature was estimated by fitting the recovery curve of the longitudinal magnetization to a single exponential function
\begin{equation}
\frac{1}{2}\left[\frac{M(0)-M(t)}{M(0)}\right]= Ae^{-(t/T_{1})}.
\label{T1}
\end{equation}
Here, $M(t)$ is the nuclear magnetization at a time $t$ after the inversion pulse and $M(0)$ is the equilibrium magnetization.
The upper panel of Fig.~\ref{Fig11} shows the recovery curves at three different temperatures probed for the P(1) site at $\mu_{0}H = 1.4$~T. The recovery curves show linearity over one and half decades when the $y$ axis is plotted in log scale. $1/T_1$ was estimated by fitting Eq.~\eqref{T1} in this linear regime.

$1/T_1$ estimated from the above fit is shown in the lower panel of Fig.~\ref{Fig11}. Our measurements are done at different field values ranging from 1.4~T to 10~T, above and below $H_{\rm c1}$. For $\mu_{0}H = 1.4$~T, the measurements are done at both P(1) and P(2) sites and over the whole temperature range, while for other fields only the P(1) site is probed and the measurements are restricted to low temperatures ($T < 30$~K). Since there is a large difference in the magnitude of $A_{\rm hf}$ for the P(1) and P(2) sites, they experience different local magnetic fields induced by the V$^{4+}$ spins. Therefore, it is expected that the resulting temperature-dependent $1/T_1$ will have different values accordingly. Indeed, for $\mu_{0}H = 1.4$~T, $1/T_1$ of the P(2) site has larger magnitude than that of the P(1) site, as $A_{\rm hf}$ of P(2) is larger than that of P(1). For both the P-sites, $1/T_1$ follows a temperature-independent behavior due to the random fluctuation of the paramagnetic moments at high temperatures~\cite{Moriya23}. At lower temperatures, $1/T_{1}$ starts to decrease and below about 10~K it drops rapidly towards zero. The 1.6~K value is almost two orders of magnitude lower than the room-temperature one, indicating the opening of a spin gap. In higher fields, the low-temperature values of $1/T_1$ increase and show an upward curvature.

The spin gap should be closed at $H_{\rm c1}$ and thereafter an AFM LRO sets in. Therefore, we measured $1/T_1$ at different fields above $H_{\rm c1}$. The increase in the low-temperature values of $1/T_1$ confirms the closing of the spin gap and the growth of 3D AFM correlations due to the field-induced LRO~\cite{Klanjsek137207}. Since our measurements are limited down to 1.6~K, we are unable to detect the field-induced LRO from the $1/T_1$ data. Nevertheless, the systematic increase of $1/T_1$ with field implies that $T_{\rm N}$ shifts toward higher temperatures with increasing $H$, in good agreement with our $C_{\rm p}(T)$ data, where field-induced LRO is detected above $H_{\rm c1}$.


The data sets for the P(1) site at various field values exhibit a fanlike pattern in the low-temperature regime. Similar behavior has been reported for spin-$1/2$ ladder compound (C$_5$H$_{12}$N)$_2$CuBr$_4$ (BPCB) and spin-1 chain compound NiCl$_2$-4SC(NH$_2$)$_2$ (DTN)~\cite{Mukhopadhyay177206}. It is expected that $1/T_1(T)$ data for different fields, around the quantum critical point (QCP) (i.e., around $H_{\rm c1}$) should follow a power-law behavior, $1/T_1 \propto T^{\alpha}$. The exponent $\alpha$ should vary across $H_{\rm c1}$, resulting in a fanlike pattern of the $1/T_1$ data. For instance, in the gapless TLL region ($H>H_{\rm c1}$), the power law has the form $1/T_1 \propto T^{1/(2K)-1}$, with $K$ being the TLL exponent~\cite{Klanjsek137207,Giamarchi11398}. The value of $K$ increases gradually as one approaches QCP from the TLL regime and reaches the maximum value 1 that corresponds to $\alpha [= 1/(2K)-1] = -0.5$. In order to test this formalism, we fitted the $1/T_1$ data below 5~K measured at 10~T by the power law. As shown in the lower panel of Fig.~\ref{Fig11}, our fit yields $\alpha \simeq -0.6$, which is very close to the predicted value ($\alpha = -0.5$) in the TLL regime. This indicates the pronounced 1D character of the compound.

\begin{figure}
	\includegraphics[width = \linewidth]{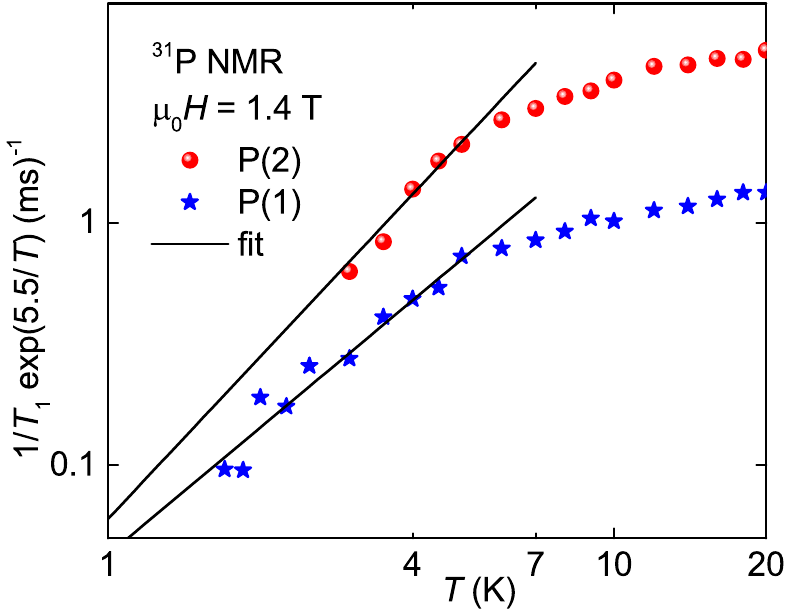}
	\caption{$1/T_{1}e^{5.5/T}$ vs $T$ in the log-log plot. The solid lines are the fits, as described in the text with $d \simeq 2.74$ and 3.2, for the P(1) and P(2) sites, respectively.}
	\label{Fig12}
\end{figure}

On the other hand, from analyzing the $K(T)$ data in magnetic fields below $H_{c1}$ we already inferred that 3D spin-spin correlations caused by the interchain coupling may be relevant. According to Mukhopadhyay \textit{et~al.}~\cite{Mukhopadhyay177206}, in the gapped region ($H \leq H_{c1}$) with 3D correlations $1/T_1$ follows an activated behavior of the form,
\begin{equation}\label{3Dmagnongap}
1/T_{1}\propto T^{\alpha_{\rm 0}} \exp\left[\frac{g\mu_{B}(H-H_{\rm c1})}{k_{\rm B}T}\right].
\end{equation}
The exponent $\alpha_0$ in the above equation depends on the effective dimension of the magnon dispersion relation as set by the thermal fluctuations $k_{\rm B}T$. With increasing temperature, $\alpha_0$ slowly decreases from 2 for $k_{\rm B}T< J_{\rm 3D}$ (3D regime) to 0 for $J_{3D} < k_{\rm B}T < J_{1D}$ (1D regime). In order to detect the effective dimensionality of the spin-spin correlations, Eq.~\eqref{3Dmagnongap} was further simplified as $1/T_1 \propto T^{d-1} e^{-\Delta_{\rm 1.4\,T}/k_{\rm B}T}$ by putting $\alpha_{\rm 0} = d-1$ and $\Delta_{\rm 1.4~T}/k_{\rm B} = g \mu_{\rm B}(H_{\rm c1}-H)/k_{\rm B}$ and fitted to the low temperature $1/T_1$ data. The fit for $T \leq 5$~K, taking $\Delta_{\rm 1.4\,T}/k_{\rm B} \simeq 5.5$~K results in $d \simeq 2.74$ and 3.2 for the P(1) and P(2) sites, respectively. The average value of $d\simeq 2.97$ is consistent with the value obtained from the $K(T)$ analysis.
This further confirms the importance of interchain couplings at low temperatures, where activated behavior is observed. Figure~\ref{Fig12} presents the $1/T_1 e^{5.5/T}$ vs $T$ plot for the data measured at $\mu_{0} H = 1.4$~T along with the fit using Eq.~\eqref{3Dmagnongap}. Both the $x$ and $y$ axes are shown in log scale to highlight the power-law prefactor to the activated behavior, at low temperatures.

 
\subsection{Microscopic magnetic model}
\label{sec:model}
Similar to Refs.~\cite{Tsirlin144412,Arjun014421,Mukharjee144433}, we use two complementary computational methods to derive exchange couplings in $\varepsilon$-LiVOPO$_4$. For a single magnetic orbital of V$^{4+}$, superexchange theory yields antiferromagnetic exchange couplings $J_i^{\rm AFM}=4t_i^2/U_{\rm eff}$, where $t_i$ are V--V hoppings extracted from the uncorrelated (PBE) band structure, and $U_{\rm eff}$ is an effective Coulomb repulsion in the V $3d$ bands. On the other hand, exchange couplings $J_i$ can be obtained from DFT+$U$ by a mapping procedure, where both ferromagnetic and antiferromagnetic contributions are taken into account. 

\begin{table}
\caption{\label{tab:couplings}
Exchange couplings in $\varepsilon$-LiVOPO$_4$. The $t_i$ values are the V--V hoppings extracted from the uncorrelated band structure, and show relative strengths of the AFM contributions to the exchange couplings $J_i^{\rm AFM}\sim t_i^2$. The $J_i$ are exchange interactions obtained by the mapping procedure within DFT+$U$.
}
\begin{ruledtabular}
\begin{tabular}{c@{\hspace{1em}}c@{\hspace{0em}}c@{\hspace{2em}}rr}\smallskip
      & $d_{\rm V-V}$ (\r A) & & $t_i$ (meV) & $J_i$ (K)  \\
$J_1$     & 5.250 & V1--V1 & $-72$  & 33 \\
$J_1'$    & 5.101 & V1--V1 & $-55$  & 23 \\
$J_2$     & 5.275 & V2--V2 & $-117$ & 63 \\\smallskip
$J_2'$    & 5.303 & V2--V2 & $-78$  & 22 \\
$J_{c1}$  & 3.599 & V1--V2 & 0      & $-15$ \\
$J_{c2}$  & 3.629 & V1--V2 & 0      & $-15$ \\
$J_{a1}$  & 6.018 & V1--V2 & $-21$  & 12 \\
$J_{a2}$  & 6.070 & V1--V2 & $-32$  & 7  \\
\end{tabular}
\end{ruledtabular}
\end{table}

In Table~\ref{tab:couplings}, we list the $t_i$ values for the uncorrelated band structure and the exchange couplings $J_i$ obtained from DFT+$U$. The two methods are in excellent agreement and consistently show the stronger couplings within chain 2. Moreover, we find that within each spin chain the stronger couplings involve the P(2) bridges and the weaker couplings involve the P(1) bridges. On the structural level, this difference should be traced back to the lateral displacements $d_i$ of the VO$_6$ octahedra within the spin chain [Fig.~\ref{Fig1}(b)], where smaller displacement leads to a stronger coupling~\cite{Mukharjee144433,Roca3167}. Indeed, we find $d_1=0.71$\,\r A for $J_1$ vs $d_1'=0.97$\,\r A for $J_1'$ and $d_2=0.08$\,\r A for $J_2$ vs $d_2'=0.23$\,\r A for $J_2'$. The smaller lateral displacements $d_2$ and $d_2'$ could also explain the overall stronger couplings in chain 2, although in this case other geometrical parameters~\cite{Roca3167} are relevant as well, because $J_1$ is about as strong as $J_2'$, despite the fact that $d_1>d_2'$. 

Regarding the interchain couplings, the microscopic scenario is very similar to that of monoclinic $A$VO$X$O$_4$ with $A$ = Ag, Na and $X$ = P, As~\cite{Tsirlin144412,Arjun014421,Mukharjee144433}. Shorter V--O--V bridges render $J_{c1}$ and $J_{c2}$ ferromagnetic, whereas the long-range couplings $J_{a1}$ and $J_{a2}$ are weakly antiferromagnetic. These ferromagnetic and antiferromagnetic interactions compete and make the spin lattice of $\varepsilon$-LiVOPO$_4$ frustrated. 

\begin{table}
\caption{\label{tab:fit}
Exchange couplings (in K) extracted from the $\chi(T)$ and $M(H)$ data using the fits shown in Fig.~\ref{fig:fit}. The susceptibility fit using Eq.~\eqref{eq:chi} returns $\chi_0=2.7\times 10^{-5}$\,cm$^3$/mol, $C_{\rm imp}=0.013$\,cm$^3$K/mol (3.5\% of paramagnetic impurities), $\theta_{\rm imp}=0.9$\,K, and $g=2.03$. This $g$ value is slightly higher than 1.98 obtained from the Curie-Weiss fit, probably because the interchain couplings were not taken into account. For magnetization data, $g=1.98$ has been used as a fixed parameter.
}
\begin{ruledtabular}
\begin{tabular}{c@{\hspace{3em}}cccc}\smallskip
 & $J_1$ & $J_1'$ & $J_2$ & $J_2'$ \\
$\chi(T)$ & 20 & 12 & 70 & 20 \\
$M(H)$    & 19 & 12 & 63 & 22 \\
\end{tabular}
\end{ruledtabular}
\end{table}

\begin{figure}
\includegraphics{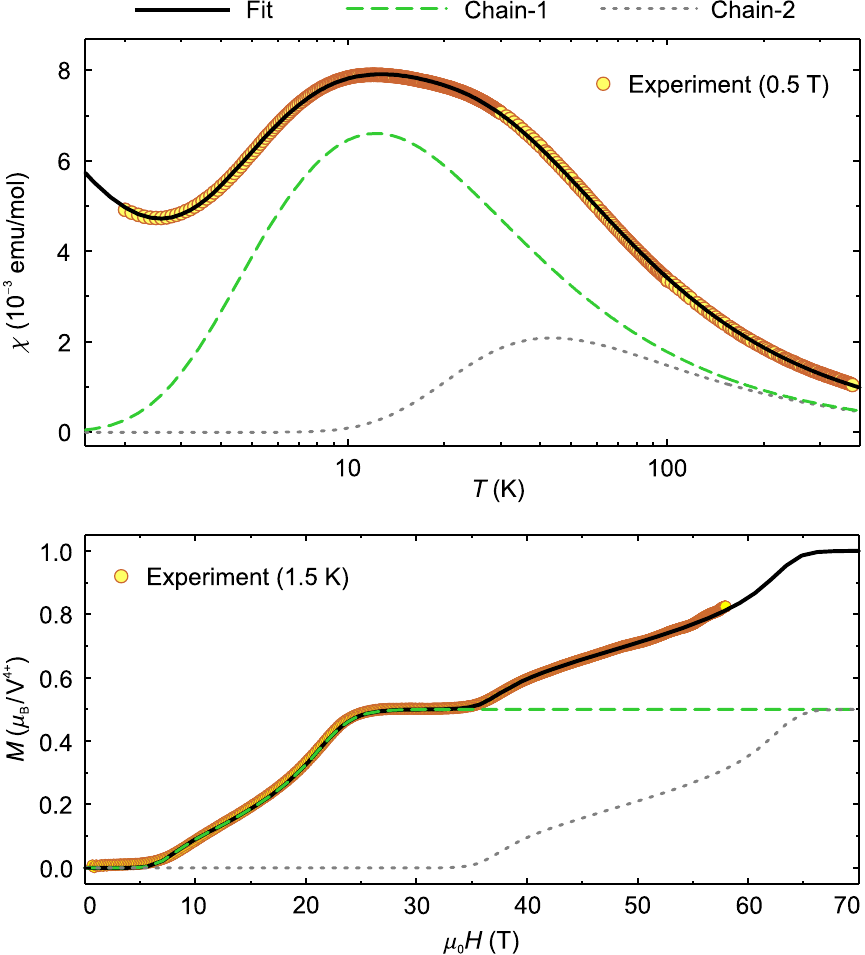}
\caption{\label{fig:fit}
Temperature-dependent susceptibility (top) and field-dependent magnetization (bottom) of $\varepsilon$-LiVOPO$_4$ fitted using the model of two non-equivalent alternating spin-$\frac12$ chains, as explained in the text. See Table~\ref{tab:fit} for the fitting parameters.}
\end{figure}

Our DFT results suggest that chain 1 shows only a moderate degree of alternation ($\alpha_1=J_1'/J_1\simeq 0.6$) that, together with the lower energy scale of the couplings, leads to a relatively small spin gap closed at $H_{\rm c1}$. In contrast, the alternation ratio of $\alpha_2=J_2'/J_2\simeq 0.3$ renders chain 2 strongly dimerized with a larger spin gap that is closed at the much higher field $H_{\rm c3}$. The model of two alternating spin-$\frac12$ chains was further used to calculate temperature-dependent magnetic susceptibility and field-dependent magnetization of $\varepsilon$-LiVOPO$_4$ (Fig.~\ref{fig:fit}). For the susceptibility, we used the analytical expression from Ref.~\cite{Johnston9558} augmented by additional terms that account for the temperature-independent ($\chi_0$) and Curie-type impurity contributions,
\begin{equation}
 \chi=\chi_0+\frac{C_{\rm imp}}{T+\theta_{\rm imp}}+\chi_{\rm ch1}+\chi_{\rm ch2},
\label{eq:chi}\end{equation}
where $\chi_{\rm ch1}$ and $\chi_{\rm ch2}$ are susceptibilities of two nonequivalent alternating spin-$\frac12$ chains with the interaction parameters $J_1$, $J_1'$ and $J_2$, $J_2'$, respectively (same $g$ factor is used for both chains). Magnetization curve of $\varepsilon$-LiVOPO$_4$ was modeled by the sum of simulated magnetization curves for the V(1) and V(2) sublattices obtained by the method described in Ref.~\cite{Tsirlin144412}.

The fitting results listed in Table~\ref{tab:fit} show good agreement between the fits to the susceptibility and magnetization. They are also consistent with the exchange parameters calculated by DFT (Table~\ref{tab:couplings}). We chose not to include interchain couplings into the susceptibility fit, which became ambiguous when more than four exchange parameters were involved. The effect of the interchain couplings can be seen from the fact that for an isolated chain 1 one finds, using $J_1$ and $J_1'$ from the susceptibility fit, the zero-field spin gap of 11.3\,K, which is somewhat larger than 7.3\,K obtained experimentally~\footnote{Similar to Ref.~\cite{Tsirlin144412}, magnetization curve was simulated by including a weak interchain coupling of $J_{\perp}/J_1=-0.05$ in order to reproduce the experimental data around $H_{\rm c1}$. This interchain coupling is much lower than estimated by DFT (Table~\ref{tab:couplings}), possibly because $J_{\perp}$ reflects a cumulative effect of the frustrated couplings $J_{a1}$, $J_{a2}$ and $J_{c1}$, $J_{c2}$. }. 

\section{Discussion and summary}
			
			
			
			
			

$\varepsilon$-LiVOPO$_4$ features two types of alternating spin-$\frac12$ chains that manifest themselves in the double maxima of the susceptibility and magnetic specific heat and in the two-step magnetization process. This unusual microscopic scenario is reminiscent of the ambient-pressure polymorph of (VO)$_2$P$_2$O$_7$~\cite{Johnston134403}, where two spin gaps corresponding to two types of spin chains were directly observed by NMR~\cite{Yamauchi3729} and inelastic neutron scattering~\cite{Garrett745}. On the other hand, large size of these gaps (35\,K and 68\,K, respectively) and high critical fields associated with them preclude experimental access to field-induced transitions, where triplon excitations of the spin chains consecutively condense leading to a long-range magnetic order.

\begin{figure}
\includegraphics[width = \linewidth]{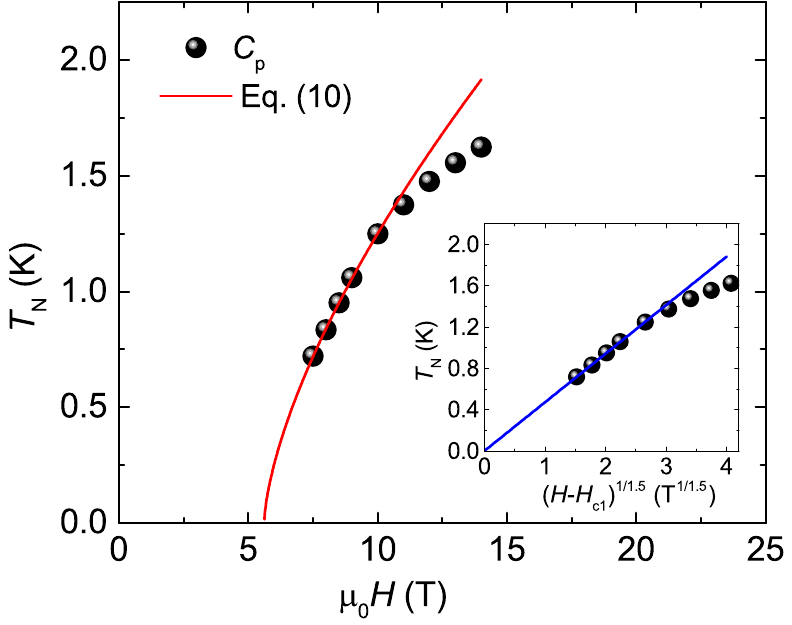}
\caption{$H-T$ phase diagram of $\varepsilon$-LiVOPO$_{4}$ obtained using the data points from $C_{\rm p}(T)$ measurements. The solid line corresponds to Eq.~\eqref{Power-law}. Inset: $T_{\rm N}$ vs ($H-H_{\rm c1})^{1/1.5}$ to highlight the low-temperature linear regime and the agreement of the simulated curve with the experimental data points.}
\label{Fig14}
\end{figure}
With its lower critical fields, $\varepsilon$-LiVOPO$_4$ offers a much better platform for studying these transitions experimentally. Indeed, we observed field-induced magnetic order already above $\mu_{0}H_{\rm c1}\simeq 5.6$\,T. Transition temperature systematically increases with field and tracks the $H$-$T$ phase boundary shown in Fig.~\ref{Fig14}. 

The field-induced transition in gapped quantum magnets is often understood as triplon BEC. In this case, the phase boundary close to $H_{\rm c1}$ should follow the universal power law~\cite{Zapf563,Giamarchi11398,Nohadani220402},
\begin{equation}\label{Power-law}
T_{\rm N} \propto (H - H_{\rm c1})^{\frac{1}{\phi}},
\end{equation}
where $\phi=d/2$ is the critical exponent reflecting the universality class of the quantum phase transition at $H_{\rm c1}$, and $d$ is the lattice dimensionality. In the absence of the low-temperature data immediately above $H_{\rm c1}$, we simulate the anticipated critical behavior by choosing $d = 3$, $\phi=\frac32$, and $\mu_{0}H_{\rm c1}\simeq 5.6$~T in Eq.~\eqref{Power-law}. The resulting curves match our data up to 1.2~K (Fig.~\ref{Fig14}), although this agreement may be partly accidental, because the temperature of 1.2~K is rather high compared to the temperature scale of the BEC transition given by $T_N^{\max}$ at the tip of the BEC dome. Since 14~T is roughly the midpoint between $H_{\rm c1}$ and $H_{\rm c2}$, we expect $T_N^{\max}\simeq 1.7$~K, and 1.2~K is more than half of this temperature. Further measurements of the phase boundary around $H_{\rm c1}$ and below 0.7~K would be clearly interesting to confirm the tentative BEC critical behavior in $\varepsilon$-LiVOPO$_4$. Impurity effects witnessed by the low-temperature upturn in the magnetic susceptibility (Fig.~\ref{Fig4}) may also be important.

The fate of the ordered phase in fields above 14~T is of significant interest too. The two-step increase in the magnetization suggests that the transition at $H_{\rm c1}$ corresponds to chain 1 and will lead to a BEC dome between $H_{\rm c1}$ and $H_{\rm c2}$, while chain 2 remains in the singlet state up to $H_{\rm c3}$, where another BEC dome should appear. Phenomenologically, this behavior would be similar to the two-dome $H$-$T$ phase diagrams of spin-1 dimer magnets~\cite{Samulon047202}, although in $\varepsilon$-LiVOPO$_4$ with local \mbox{spin $\frac12$} it should have a very different microscopic origin. It does in fact arise from the coexistence of the nonequivalent magnetic sublattices. We also point out that the $\frac12$-magnetization plateau in this compound is not caused by Wigner crystallization and is thus dissimilar to the magnetization plateaus in SrCu$_2$(BO$_3)_2$. Because $H_{\rm c3}$ lies above $H_{\rm c2}$, magnetic orders related to chain 1 and chain 2 should be mostly decoupled. This is different from other systems with multiple spin gaps, where intertwined BEC transitions lead to an unusual critical behavior, as in the dimer magnet BaCuSi$_2$O$_6$~\cite{Mazurenko107202,Allenspach2020}. 

In summary, we have shown that $\varepsilon$-LiVOPO$_4$ is a gapped quantum magnet that features singlet ground state in zero field. With two non-equivalent alternating spin-$\frac12$ chains, it shows double maxima in the susceptibility and magnetic specific heat and a two-step increase in the magnetization. Chain 1 features weaker couplings and a weaker alternation ($J_1\simeq 20$~K, $\alpha_1\simeq 0.6$), whereas chain 2 reveals stronger couplings and lies closer to the dimer limit ($J_2\simeq 60$~K, $\alpha_2\simeq 0.3$). The zero-field spin gap of $\Delta_0/k_B\simeq 7.3$~K is closed at $\mu_{0}H_{\rm c1}\simeq 5.6$~T. The magnetization increases up to $\mu_{0}H_{\rm c2}\simeq 25$~T, flattens out within the $\frac12$ plateau, and increases again above $\mu_{0}H_{\rm c3}\simeq 35$~T. The gap closing above $H_{\rm c1}$ leads to a field-induced LRO that can be understood as Bose-Einstein condensation of triplons.

\acknowledgments
P.K.M. and R.N. acknowledge BRNS, India for financial support bearing sanction No.37(3)/14/26/2017-BRNS. We also thank C. Klausnitzer (MPI-CPfS) for the technical support. A.A.T was funded by the Federal Ministry for Education and Research through the Sofja Kovalevskaya Award of Alexander von Humboldt Foundation. We also acknowledge the support of the HLD at HZDR, member of European Magnetic Field Laboratory (EMFL).


\begin{thebibliography}{78}%
	\makeatletter
	\providecommand \@ifxundefined [1]{%
		\@ifx{#1\undefined}
	}%
	\providecommand \@ifnum [1]{%
		\ifnum #1\expandafter \@firstoftwo
		\else \expandafter \@secondoftwo
		\fi
	}%
	\providecommand \@ifx [1]{%
		\ifx #1\expandafter \@firstoftwo
		\else \expandafter \@secondoftwo
		\fi
	}%
	\providecommand \natexlab [1]{#1}%
	\providecommand \enquote  [1]{``#1''}%
	\providecommand \bibnamefont  [1]{#1}%
	\providecommand \bibfnamefont [1]{#1}%
	\providecommand \citenamefont [1]{#1}%
	\providecommand \href@noop [0]{\@secondoftwo}%
	\providecommand \href [0]{\begingroup \@sanitize@url \@href}%
	\providecommand \@href[1]{\@@startlink{#1}\@@href}%
	\providecommand \@@href[1]{\endgroup#1\@@endlink}%
	\providecommand \@sanitize@url [0]{\catcode `\\12\catcode `\$12\catcode
		`\&12\catcode `\#12\catcode `\^12\catcode `\_12\catcode `\%12\relax}%
	\providecommand \@@startlink[1]{}%
	\providecommand \@@endlink[0]{}%
	\providecommand \url  [0]{\begingroup\@sanitize@url \@url }%
	\providecommand \@url [1]{\endgroup\@href {#1}{\urlprefix }}%
	\providecommand \urlprefix  [0]{URL }%
	\providecommand \Eprint [0]{\href }%
	\providecommand \doibase [0]{http://dx.doi.org/}%
	\providecommand \selectlanguage [0]{\@gobble}%
	\providecommand \bibinfo  [0]{\@secondoftwo}%
	\providecommand \bibfield  [0]{\@secondoftwo}%
	\providecommand \translation [1]{[#1]}%
	\providecommand \BibitemOpen [0]{}%
	\providecommand \bibitemStop [0]{}%
	\providecommand \bibitemNoStop [0]{.\EOS\space}%
	\providecommand \EOS [0]{\spacefactor3000\relax}%
	\providecommand \BibitemShut  [1]{\csname bibitem#1\endcsname}%
	\let\auto@bib@innerbib\@empty
	\bibitem [{\citenamefont {Matsubara}\ and\ \citenamefont
		{Matsuda}(1956)}]{Matsubara569}%
	\BibitemOpen
	\bibfield  {author} {\bibinfo {author} {\bibfnamefont {T.}~\bibnamefont
			{Matsubara}}\ and\ \bibinfo {author} {\bibfnamefont {H.}~\bibnamefont
			{Matsuda}},\ }\bibfield  {title} {\enquote {\bibinfo {title} {{A Lattice
					Model of Liquid Helium, I}},}\ }\href {\doibase 10.1143/PTP.16.569}
	{\bibfield  {journal} {\bibinfo  {journal} {Prog. Theor. Phys.}\ }\textbf
		{\bibinfo {volume} {16}},\ \bibinfo {pages} {569} (\bibinfo {year}
		{1956})}\BibitemShut {NoStop}%
	\bibitem [{\citenamefont {Giamarchi}\ \emph {et~al.}(2008)\citenamefont
		{Giamarchi}, \citenamefont {R{\"u}egg},\ and\ \citenamefont
		{Tchernyshyov}}]{Giamarchi198}%
	\BibitemOpen
	\bibfield  {author} {\bibinfo {author} {\bibfnamefont {T.}~\bibnamefont
			{Giamarchi}}, \bibinfo {author} {\bibfnamefont {Ch.}\ \bibnamefont
			{R{\"u}egg}}, \ and\ \bibinfo {author} {\bibfnamefont {O.}~\bibnamefont
			{Tchernyshyov}},\ }\bibfield  {title} {\enquote {\bibinfo {title}
			{Bose-{E}instein condensation in magnetic insulators},}\ }\href@noop {}
	{\bibfield  {journal} {\bibinfo  {journal} {Nat. Phys.}\ }\textbf {\bibinfo
			{volume} {4}},\ \bibinfo {pages} {198} (\bibinfo {year} {2008})}\BibitemShut
	{NoStop}%
	\bibitem [{\citenamefont {Zapf}\ \emph {et~al.}(2014)\citenamefont {Zapf},
		\citenamefont {Jaime},\ and\ \citenamefont {Batista}}]{Zapf563}%
	\BibitemOpen
	\bibfield  {author} {\bibinfo {author} {\bibfnamefont {V.}~\bibnamefont
			{Zapf}}, \bibinfo {author} {\bibfnamefont {M.}~\bibnamefont {Jaime}}, \ and\
		\bibinfo {author} {\bibfnamefont {C.~D.}\ \bibnamefont {Batista}},\
	}\bibfield  {title} {\enquote {\bibinfo {title} {Bose-{E}instein condensation
				in quantum magnets},}\ }\href {\doibase 10.1103/RevModPhys.86.563} {\bibfield
		{journal} {\bibinfo  {journal} {Rev. Mod. Phys.}\ }\textbf {\bibinfo
			{volume} {86}},\ \bibinfo {pages} {563} (\bibinfo {year} {2014})}\BibitemShut
	{NoStop}%
	\bibitem [{\citenamefont {Mukhopadhyay}\ \emph {et~al.}(2012)\citenamefont
		{Mukhopadhyay}, \citenamefont {Klanj\ifmmode~\check{s}\else \v{s}\fi{}ek},
		\citenamefont {Grbi\ifmmode~\acute{c}\else \'{c}\fi{}}, \citenamefont
		{Blinder}, \citenamefont {Mayaffre}, \citenamefont {Berthier}, \citenamefont
		{Horvati\ifmmode~\acute{c}\else \'{c}\fi{}}, \citenamefont {Continentino},
		\citenamefont {Paduan-Filho}, \citenamefont {Chiari},\ and\ \citenamefont
		{Piovesana}}]{Mukhopadhyay177206}%
	\BibitemOpen
	\bibfield  {author} {\bibinfo {author} {\bibfnamefont {S.}~\bibnamefont
			{Mukhopadhyay}}, \bibinfo {author} {\bibfnamefont {M.}~\bibnamefont
			{Klanj\ifmmode~\check{s}\else \v{s}\fi{}ek}}, \bibinfo {author}
		{\bibfnamefont {M.~S.}\ \bibnamefont {Grbi\ifmmode~\acute{c}\else
				\'{c}\fi{}}}, \bibinfo {author} {\bibfnamefont {R.}~\bibnamefont {Blinder}},
		\bibinfo {author} {\bibfnamefont {H.}~\bibnamefont {Mayaffre}}, \bibinfo
		{author} {\bibfnamefont {C.}~\bibnamefont {Berthier}}, \bibinfo {author}
		{\bibfnamefont {M.}~\bibnamefont {Horvati\ifmmode~\acute{c}\else
				\'{c}\fi{}}}, \bibinfo {author} {\bibfnamefont {M.~A.}\ \bibnamefont
			{Continentino}}, \bibinfo {author} {\bibfnamefont {A.}~\bibnamefont
			{Paduan-Filho}}, \bibinfo {author} {\bibfnamefont {B.}~\bibnamefont
			{Chiari}}, \ and\ \bibinfo {author} {\bibfnamefont {O.}~\bibnamefont
			{Piovesana}},\ }\bibfield  {title} {\enquote {\bibinfo {title}
			{Quantum-critical spin dynamics in quasi-one-dimensional antiferromagnets},}\
	}\href {\doibase 10.1103/PhysRevLett.109.177206} {\bibfield  {journal}
		{\bibinfo  {journal} {Phys. Rev. Lett.}\ }\textbf {\bibinfo {volume} {109}},\
		\bibinfo {pages} {177206} (\bibinfo {year} {2012})}\BibitemShut {NoStop}%
	\bibitem [{\citenamefont {Sebastian}\ \emph {et~al.}(2006)\citenamefont
		{Sebastian}, \citenamefont {Harrison}, \citenamefont {Batista}, \citenamefont
		{Balicas}, \citenamefont {Jaime}, \citenamefont {Sharma}, \citenamefont
		{Kawashima},\ and\ \citenamefont {Fisher}}]{Sebastian617}%
	\BibitemOpen
	\bibfield  {author} {\bibinfo {author} {\bibfnamefont {S.~E.}\ \bibnamefont
			{Sebastian}}, \bibinfo {author} {\bibfnamefont {N.}~\bibnamefont {Harrison}},
		\bibinfo {author} {\bibfnamefont {C.~D.}\ \bibnamefont {Batista}}, \bibinfo
		{author} {\bibfnamefont {L.}~\bibnamefont {Balicas}}, \bibinfo {author}
		{\bibfnamefont {M.}~\bibnamefont {Jaime}}, \bibinfo {author} {\bibfnamefont
			{P.~A.}\ \bibnamefont {Sharma}}, \bibinfo {author} {\bibfnamefont
			{N.}~\bibnamefont {Kawashima}}, \ and\ \bibinfo {author} {\bibfnamefont
			{I.~R.}\ \bibnamefont {Fisher}},\ }\bibfield  {title} {\enquote {\bibinfo
			{title} {Dimensional reduction at a quantum critical point},}\ }\href
	{\doibase 10.1038/nature04732} {\bibfield  {journal} {\bibinfo  {journal}
			{Nature}\ }\textbf {\bibinfo {volume} {441}},\ \bibinfo {pages} {617}
		(\bibinfo {year} {2006})}\BibitemShut {NoStop}%
	\bibitem [{\citenamefont {Nikuni}\ \emph {et~al.}(2000)\citenamefont {Nikuni},
		\citenamefont {Oshikawa}, \citenamefont {Oosawa},\ and\ \citenamefont
		{Tanaka}}]{Nikuni5868}%
	\BibitemOpen
	\bibfield  {author} {\bibinfo {author} {\bibfnamefont {T.}~\bibnamefont
			{Nikuni}}, \bibinfo {author} {\bibfnamefont {M.}~\bibnamefont {Oshikawa}},
		\bibinfo {author} {\bibfnamefont {A.}~\bibnamefont {Oosawa}}, \ and\ \bibinfo
		{author} {\bibfnamefont {H.}~\bibnamefont {Tanaka}},\ }\bibfield  {title}
	{\enquote {\bibinfo {title} {Bose-{E}instein condensation of dilute magnons
				in {TlCuCl$_3$}},}\ }\href {\doibase 10.1103/PhysRevLett.84.5868} {\bibfield
		{journal} {\bibinfo  {journal} {Phys. Rev. Lett.}\ }\textbf {\bibinfo
			{volume} {84}},\ \bibinfo {pages} {5868} (\bibinfo {year}
		{2000})}\BibitemShut {NoStop}%
	\bibitem [{\citenamefont {Klanj\ifmmode~\check{s}\else \v{s}\fi{}ek}\ \emph
		{et~al.}(2008)\citenamefont {Klanj\ifmmode~\check{s}\else \v{s}\fi{}ek},
		\citenamefont {Mayaffre}, \citenamefont {Berthier}, \citenamefont
		{Horvati\ifmmode~\acute{c}\else \'{c}\fi{}}, \citenamefont {Chiari},
		\citenamefont {Piovesana}, \citenamefont {Bouillot}, \citenamefont {Kollath},
		\citenamefont {Orignac}, \citenamefont {Citro},\ and\ \citenamefont
		{Giamarchi}}]{Klanjsek137207}%
	\BibitemOpen
	\bibfield  {author} {\bibinfo {author} {\bibfnamefont {M.}~\bibnamefont
			{Klanj\ifmmode~\check{s}\else \v{s}\fi{}ek}}, \bibinfo {author}
		{\bibfnamefont {H.}~\bibnamefont {Mayaffre}}, \bibinfo {author}
		{\bibfnamefont {C.}~\bibnamefont {Berthier}}, \bibinfo {author}
		{\bibfnamefont {M.}~\bibnamefont {Horvati\ifmmode~\acute{c}\else
				\'{c}\fi{}}}, \bibinfo {author} {\bibfnamefont {B.}~\bibnamefont {Chiari}},
		\bibinfo {author} {\bibfnamefont {O.}~\bibnamefont {Piovesana}}, \bibinfo
		{author} {\bibfnamefont {P.}~\bibnamefont {Bouillot}}, \bibinfo {author}
		{\bibfnamefont {C.}~\bibnamefont {Kollath}}, \bibinfo {author} {\bibfnamefont
			{E.}~\bibnamefont {Orignac}}, \bibinfo {author} {\bibfnamefont
			{R.}~\bibnamefont {Citro}}, \ and\ \bibinfo {author} {\bibfnamefont
			{T.}~\bibnamefont {Giamarchi}},\ }\bibfield  {title} {\enquote {\bibinfo
			{title} {Controlling luttinger liquid physics in spin ladders under a
				magnetic field},}\ }\href {\doibase 10.1103/PhysRevLett.101.137207}
	{\bibfield  {journal} {\bibinfo  {journal} {Phys. Rev. Lett.}\ }\textbf
		{\bibinfo {volume} {101}},\ \bibinfo {pages} {137207} (\bibinfo {year}
		{2008})}\BibitemShut {NoStop}%
	\bibitem [{\citenamefont {Matsushita}\ \emph {et~al.}(2017)\citenamefont
		{Matsushita}, \citenamefont {Hori}, \citenamefont {Takata}, \citenamefont
		{Wada}, \citenamefont {Amaya},\ and\ \citenamefont
		{Hosokoshi}}]{Matsushita020408}%
	\BibitemOpen
	\bibfield  {author} {\bibinfo {author} {\bibfnamefont {T.}~\bibnamefont
			{Matsushita}}, \bibinfo {author} {\bibfnamefont {N.}~\bibnamefont {Hori}},
		\bibinfo {author} {\bibfnamefont {S.}~\bibnamefont {Takata}}, \bibinfo
		{author} {\bibfnamefont {N.}~\bibnamefont {Wada}}, \bibinfo {author}
		{\bibfnamefont {N.}~\bibnamefont {Amaya}}, \ and\ \bibinfo {author}
		{\bibfnamefont {Y.}~\bibnamefont {Hosokoshi}},\ }\bibfield  {title} {\enquote
		{\bibinfo {title} {Direct three-dimensional ordering of quasi-one-dimensional
				quantum dimer system near critical fields},}\ }\href {\doibase
		10.1103/PhysRevB.95.020408} {\bibfield  {journal} {\bibinfo  {journal} {Phys.
				Rev. B}\ }\textbf {\bibinfo {volume} {95}},\ \bibinfo {pages} {020408(R)}
		(\bibinfo {year} {2017})}\BibitemShut {NoStop}%
	\bibitem [{\citenamefont {Willenberg}\ \emph {et~al.}(2015)\citenamefont
		{Willenberg}, \citenamefont {Ryll}, \citenamefont {Kiefer}, \citenamefont
		{Tennant}, \citenamefont {Groitl}, \citenamefont {Rolfs}, \citenamefont
		{Manuel}, \citenamefont {Khalyavin}, \citenamefont {Rule}, \citenamefont
		{Wolter},\ and\ \citenamefont {S\"ullow}}]{Willenberg060407}%
	\BibitemOpen
	\bibfield  {author} {\bibinfo {author} {\bibfnamefont {B.}~\bibnamefont
			{Willenberg}}, \bibinfo {author} {\bibfnamefont {H.}~\bibnamefont {Ryll}},
		\bibinfo {author} {\bibfnamefont {K.}~\bibnamefont {Kiefer}}, \bibinfo
		{author} {\bibfnamefont {D.~A.}\ \bibnamefont {Tennant}}, \bibinfo {author}
		{\bibfnamefont {F.}~\bibnamefont {Groitl}}, \bibinfo {author} {\bibfnamefont
			{K.}~\bibnamefont {Rolfs}}, \bibinfo {author} {\bibfnamefont
			{P.}~\bibnamefont {Manuel}}, \bibinfo {author} {\bibfnamefont
			{D.}~\bibnamefont {Khalyavin}}, \bibinfo {author} {\bibfnamefont {K.~C.}\
			\bibnamefont {Rule}}, \bibinfo {author} {\bibfnamefont {A.~U.~B.}\
			\bibnamefont {Wolter}}, \ and\ \bibinfo {author} {\bibfnamefont
			{S.}~\bibnamefont {S\"ullow}},\ }\bibfield  {title} {\enquote {\bibinfo
			{title} {Luttinger liquid behavior in the alternating spin-chain system
				copper nitrate},}\ }\href {\doibase 10.1103/PhysRevB.91.060407} {\bibfield
		{journal} {\bibinfo  {journal} {Phys. Rev. B}\ }\textbf {\bibinfo {volume}
			{91}},\ \bibinfo {pages} {060407(R)} (\bibinfo {year} {2015})}\BibitemShut
	{NoStop}%
	\bibitem [{\citenamefont {Thielemann}\ \emph {et~al.}(2009)\citenamefont
		{Thielemann}, \citenamefont {R\"uegg}, \citenamefont {Kiefer}, \citenamefont
		{R\o{}nnow}, \citenamefont {Normand}, \citenamefont {Bouillot}, \citenamefont
		{Kollath}, \citenamefont {Orignac}, \citenamefont {Citro}, \citenamefont
		{Giamarchi}, \citenamefont {L\"auchli}, \citenamefont {Biner}, \citenamefont
		{Kr\"amer}, \citenamefont {Wolff-Fabris}, \citenamefont {Zapf}, \citenamefont
		{Jaime}, \citenamefont {Stahn}, \citenamefont {Christensen}, \citenamefont
		{Grenier}, \citenamefont {McMorrow},\ and\ \citenamefont
		{Mesot}}]{Thielemann020408}%
	\BibitemOpen
	\bibfield  {author} {\bibinfo {author} {\bibfnamefont {B.}~\bibnamefont
			{Thielemann}}, \bibinfo {author} {\bibfnamefont {Ch.}\ \bibnamefont
			{R\"uegg}}, \bibinfo {author} {\bibfnamefont {K.}~\bibnamefont {Kiefer}},
		\bibinfo {author} {\bibfnamefont {H.~M.}\ \bibnamefont {R\o{}nnow}}, \bibinfo
		{author} {\bibfnamefont {B.}~\bibnamefont {Normand}}, \bibinfo {author}
		{\bibfnamefont {P.}~\bibnamefont {Bouillot}}, \bibinfo {author}
		{\bibfnamefont {C.}~\bibnamefont {Kollath}}, \bibinfo {author} {\bibfnamefont
			{E.}~\bibnamefont {Orignac}}, \bibinfo {author} {\bibfnamefont
			{R.}~\bibnamefont {Citro}}, \bibinfo {author} {\bibfnamefont
			{T.}~\bibnamefont {Giamarchi}}, \bibinfo {author} {\bibfnamefont {A.~M.}\
			\bibnamefont {L\"auchli}}, \bibinfo {author} {\bibfnamefont {D.}~\bibnamefont
			{Biner}}, \bibinfo {author} {\bibfnamefont {K.~W.}\ \bibnamefont {Kr\"amer}},
		\bibinfo {author} {\bibfnamefont {F.}~\bibnamefont {Wolff-Fabris}}, \bibinfo
		{author} {\bibfnamefont {V.~S.}\ \bibnamefont {Zapf}}, \bibinfo {author}
		{\bibfnamefont {M.}~\bibnamefont {Jaime}}, \bibinfo {author} {\bibfnamefont
			{J.}~\bibnamefont {Stahn}}, \bibinfo {author} {\bibfnamefont {N.~B.}\
			\bibnamefont {Christensen}}, \bibinfo {author} {\bibfnamefont
			{B.}~\bibnamefont {Grenier}}, \bibinfo {author} {\bibfnamefont {D.~F.}\
			\bibnamefont {McMorrow}}, \ and\ \bibinfo {author} {\bibfnamefont
			{J.}~\bibnamefont {Mesot}},\ }\bibfield  {title} {\enquote {\bibinfo {title}
			{Field-controlled magnetic order in the quantum spin-ladder system
				{(Hpip)$_2$CuBr$_4$}},}\ }\href {\doibase 10.1103/PhysRevB.79.020408}
	{\bibfield  {journal} {\bibinfo  {journal} {Phys. Rev. B}\ }\textbf {\bibinfo
			{volume} {79}},\ \bibinfo {pages} {020408(R)} (\bibinfo {year}
		{2009})}\BibitemShut {NoStop}%
	\bibitem [{\citenamefont {Rice}(2002)}]{Rice760}%
	\BibitemOpen
	\bibfield  {author} {\bibinfo {author} {\bibfnamefont {T.~M.}\ \bibnamefont
			{Rice}},\ }\bibfield  {title} {\enquote {\bibinfo {title} {To condense or not
				to condense},}\ }\href {\doibase 10.1126/science.1078819} {\bibfield
		{journal} {\bibinfo  {journal} {Science}\ }\textbf {\bibinfo {volume}
			{298}},\ \bibinfo {pages} {760} (\bibinfo {year} {2002})}\BibitemShut
	{NoStop}%
	\bibitem [{\citenamefont {Narumi}\ \emph {et~al.}(1998)\citenamefont {Narumi},
		\citenamefont {Hagiwara}, \citenamefont {Sato}, \citenamefont {Kindo},
		\citenamefont {Nakano},\ and\ \citenamefont {Takahashi}}]{Narumi509}%
	\BibitemOpen
	\bibfield  {author} {\bibinfo {author} {\bibfnamefont {Y.}~\bibnamefont
			{Narumi}}, \bibinfo {author} {\bibfnamefont {M.}~\bibnamefont {Hagiwara}},
		\bibinfo {author} {\bibfnamefont {R.}~\bibnamefont {Sato}}, \bibinfo {author}
		{\bibfnamefont {K.}~\bibnamefont {Kindo}}, \bibinfo {author} {\bibfnamefont
			{H.}~\bibnamefont {Nakano}}, \ and\ \bibinfo {author} {\bibfnamefont
			{M.}~\bibnamefont {Takahashi}},\ }\bibfield  {title} {\enquote {\bibinfo
			{title} {High field magnetization in a {$S=1$} antiferromagnetic chain with
				bond alternation},}\ }\href {\doibase
		https://doi.org/10.1016/S0921-4526(97)00974-5} {\bibfield  {journal}
		{\bibinfo  {journal} {Physica B: Cond. Mat.}\ }\textbf {\bibinfo {volume}
			{246}},\ \bibinfo {pages} {509} (\bibinfo {year} {1998})}\BibitemShut
	{NoStop}%
	\bibitem [{\citenamefont {Shiramura}\ \emph {et~al.}(1998)\citenamefont
		{Shiramura}, \citenamefont {Takatsu}, \citenamefont {Kurniawan},
		\citenamefont {Tanaka}, \citenamefont {Uekusa}, \citenamefont {Ohashi},
		\citenamefont {Takizawa}, \citenamefont {Mitamura},\ and\ \citenamefont
		{Goto}}]{Shiramura1548}%
	\BibitemOpen
	\bibfield  {author} {\bibinfo {author} {\bibfnamefont {B.}~\bibnamefont
			{Shiramura}}, \bibinfo {author} {\bibfnamefont {K.}~\bibnamefont {Takatsu}},
		\bibinfo {author} {\bibfnamefont {B.}~\bibnamefont {Kurniawan}}, \bibinfo
		{author} {\bibfnamefont {H.}~\bibnamefont {Tanaka}}, \bibinfo {author}
		{\bibfnamefont {H.}~\bibnamefont {Uekusa}}, \bibinfo {author} {\bibfnamefont
			{Y.}~\bibnamefont {Ohashi}}, \bibinfo {author} {\bibfnamefont
			{K.}~\bibnamefont {Takizawa}}, \bibinfo {author} {\bibfnamefont
			{H.}~\bibnamefont {Mitamura}}, \ and\ \bibinfo {author} {\bibfnamefont
			{T.}~\bibnamefont {Goto}},\ }\bibfield  {title} {\enquote {\bibinfo {title}
			{Magnetization plateaus in {NH$_4$CuCl$_3$}},}\ }\href {\doibase
		10.1143/jpsj.67.1548} {\bibfield  {journal} {\bibinfo  {journal} {J. Phys.
				Soc. Jpn.}\ }\textbf {\bibinfo {volume} {67}},\ \bibinfo {pages} {1548}
		(\bibinfo {year} {1998})}\BibitemShut {NoStop}%
	\bibitem [{\citenamefont {Kodama}\ \emph {et~al.}(2002)\citenamefont {Kodama},
		\citenamefont {Takigawa}, \citenamefont {Horvati{\'c}}, \citenamefont
		{Berthier}, \citenamefont {Kageyama}, \citenamefont {Ueda}, \citenamefont
		{Miyahara}, \citenamefont {Becca},\ and\ \citenamefont {Mila}}]{Kodama395}%
	\BibitemOpen
	\bibfield  {author} {\bibinfo {author} {\bibfnamefont {K.}~\bibnamefont
			{Kodama}}, \bibinfo {author} {\bibfnamefont {M.}~\bibnamefont {Takigawa}},
		\bibinfo {author} {\bibfnamefont {M.}~\bibnamefont {Horvati{\'c}}}, \bibinfo
		{author} {\bibfnamefont {C.}~\bibnamefont {Berthier}}, \bibinfo {author}
		{\bibfnamefont {H.}~\bibnamefont {Kageyama}}, \bibinfo {author}
		{\bibfnamefont {Y.}~\bibnamefont {Ueda}}, \bibinfo {author} {\bibfnamefont
			{S.}~\bibnamefont {Miyahara}}, \bibinfo {author} {\bibfnamefont
			{F.}~\bibnamefont {Becca}}, \ and\ \bibinfo {author} {\bibfnamefont
			{F.}~\bibnamefont {Mila}},\ }\bibfield  {title} {\enquote {\bibinfo {title}
			{Magnetic superstructure in the two-dimensional quantum antiferromagnet
				{SrCu$_2$(BO$_3$)$_2$}},}\ }\href {\doibase 10.1126/science.1075045}
	{\bibfield  {journal} {\bibinfo  {journal} {Science}\ }\textbf {\bibinfo
			{volume} {298}},\ \bibinfo {pages} {395} (\bibinfo {year}
		{2002})}\BibitemShut {NoStop}%
	\bibitem [{\citenamefont {Kageyama}\ \emph {et~al.}(1999)\citenamefont
		{Kageyama}, \citenamefont {Yoshimura}, \citenamefont {Stern}, \citenamefont
		{Mushnikov}, \citenamefont {Onizuka}, \citenamefont {Kato}, \citenamefont
		{Kosuge}, \citenamefont {Slichter}, \citenamefont {Goto},\ and\ \citenamefont
		{Ueda}}]{Kageyama3168}%
	\BibitemOpen
	\bibfield  {author} {\bibinfo {author} {\bibfnamefont {H.}~\bibnamefont
			{Kageyama}}, \bibinfo {author} {\bibfnamefont {K.}~\bibnamefont {Yoshimura}},
		\bibinfo {author} {\bibfnamefont {R.}~\bibnamefont {Stern}}, \bibinfo
		{author} {\bibfnamefont {N.~V.}\ \bibnamefont {Mushnikov}}, \bibinfo {author}
		{\bibfnamefont {K.}~\bibnamefont {Onizuka}}, \bibinfo {author} {\bibfnamefont
			{M.}~\bibnamefont {Kato}}, \bibinfo {author} {\bibfnamefont {K.}~\bibnamefont
			{Kosuge}}, \bibinfo {author} {\bibfnamefont {C.~P.}\ \bibnamefont
			{Slichter}}, \bibinfo {author} {\bibfnamefont {T.}~\bibnamefont {Goto}}, \
		and\ \bibinfo {author} {\bibfnamefont {Y.}~\bibnamefont {Ueda}},\ }\bibfield
	{title} {\enquote {\bibinfo {title} {Exact dimer ground state and quantized
				magnetization plateaus in the two-dimensional spin system
				{SrCu$_2$(BO$_3$)$_2$}},}\ }\href {\doibase 10.1103/PhysRevLett.82.3168}
	{\bibfield  {journal} {\bibinfo  {journal} {Phys. Rev. Lett.}\ }\textbf
		{\bibinfo {volume} {82}},\ \bibinfo {pages} {3168} (\bibinfo {year}
		{1999})}\BibitemShut {NoStop}%
	\bibitem [{\citenamefont {R{\"u}egg}\ \emph {et~al.}(2003)\citenamefont
		{R{\"u}egg}, \citenamefont {Cavadini}, \citenamefont {Furrer}, \citenamefont
		{G{\"u}del}, \citenamefont {Kr{\"a}mer}, \citenamefont {Mutka}, \citenamefont
		{Wildes}, \citenamefont {Habicht},\ and\ \citenamefont
		{Vorderwisch}}]{Ruegg62}%
	\BibitemOpen
	\bibfield  {author} {\bibinfo {author} {\bibfnamefont {C.}~\bibnamefont
			{R{\"u}egg}}, \bibinfo {author} {\bibfnamefont {N.}~\bibnamefont {Cavadini}},
		\bibinfo {author} {\bibfnamefont {A.}~\bibnamefont {Furrer}}, \bibinfo
		{author} {\bibfnamefont {H.~U.}\ \bibnamefont {G{\"u}del}}, \bibinfo {author}
		{\bibfnamefont {K.}~\bibnamefont {Kr{\"a}mer}}, \bibinfo {author}
		{\bibfnamefont {H.}~\bibnamefont {Mutka}}, \bibinfo {author} {\bibfnamefont
			{A.}~\bibnamefont {Wildes}}, \bibinfo {author} {\bibfnamefont
			{K.}~\bibnamefont {Habicht}}, \ and\ \bibinfo {author} {\bibfnamefont
			{P.}~\bibnamefont {Vorderwisch}},\ }\bibfield  {title} {\enquote {\bibinfo
			{title} {Bose-{E}instein condensation of the triplet states in the magnetic
				insulator {TlCuCl$_3$}},}\ }\href {https://doi.org/10.1038/nature01617}
	{\bibfield  {journal} {\bibinfo  {journal} {Nature}\ }\textbf {\bibinfo
			{volume} {423}},\ \bibinfo {pages} {62} (\bibinfo {year} {2003})}\BibitemShut
	{NoStop}%
	\bibitem [{\citenamefont {Jaime}\ \emph {et~al.}(2004)\citenamefont {Jaime},
		\citenamefont {Correa}, \citenamefont {Harrison}, \citenamefont {Batista},
		\citenamefont {Kawashima}, \citenamefont {Kazuma}, \citenamefont {Jorge},
		\citenamefont {Stern}, \citenamefont {Heinmaa}, \citenamefont {Zvyagin},
		\citenamefont {Sasago},\ and\ \citenamefont {Uchinokura}}]{Jaime087203}%
	\BibitemOpen
	\bibfield  {author} {\bibinfo {author} {\bibfnamefont {M.}~\bibnamefont
			{Jaime}}, \bibinfo {author} {\bibfnamefont {V.~F.}\ \bibnamefont {Correa}},
		\bibinfo {author} {\bibfnamefont {N.}~\bibnamefont {Harrison}}, \bibinfo
		{author} {\bibfnamefont {C.~D.}\ \bibnamefont {Batista}}, \bibinfo {author}
		{\bibfnamefont {N.}~\bibnamefont {Kawashima}}, \bibinfo {author}
		{\bibfnamefont {Y.}~\bibnamefont {Kazuma}}, \bibinfo {author} {\bibfnamefont
			{G.~A.}\ \bibnamefont {Jorge}}, \bibinfo {author} {\bibfnamefont
			{R.}~\bibnamefont {Stern}}, \bibinfo {author} {\bibfnamefont
			{I.}~\bibnamefont {Heinmaa}}, \bibinfo {author} {\bibfnamefont {S.~A.}\
			\bibnamefont {Zvyagin}}, \bibinfo {author} {\bibfnamefont {Y.}~\bibnamefont
			{Sasago}}, \ and\ \bibinfo {author} {\bibfnamefont {K.}~\bibnamefont
			{Uchinokura}},\ }\bibfield  {title} {\enquote {\bibinfo {title}
			{Magnetic-field-induced condensation of triplons in han purple pigment
				{BaCuSi$_2$O$_6$}},}\ }\href {\doibase 10.1103/PhysRevLett.93.087203}
	{\bibfield  {journal} {\bibinfo  {journal} {Phys. Rev. Lett.}\ }\textbf
		{\bibinfo {volume} {93}},\ \bibinfo {pages} {087203} (\bibinfo {year}
		{2004})}\BibitemShut {NoStop}%
	\bibitem [{\citenamefont {Aczel}\ \emph
		{et~al.}(2009{\natexlab{a}})\citenamefont {Aczel}, \citenamefont {Kohama},
		\citenamefont {Marcenat}, \citenamefont {Weickert}, \citenamefont {Jaime},
		\citenamefont {Ayala-Valenzuela}, \citenamefont {McDonald}, \citenamefont
		{Selesnic}, \citenamefont {Dabkowska},\ and\ \citenamefont
		{Luke}}]{Aczel207203}%
	\BibitemOpen
	\bibfield  {author} {\bibinfo {author} {\bibfnamefont {A.~A.}\ \bibnamefont
			{Aczel}}, \bibinfo {author} {\bibfnamefont {Y.}~\bibnamefont {Kohama}},
		\bibinfo {author} {\bibfnamefont {C.}~\bibnamefont {Marcenat}}, \bibinfo
		{author} {\bibfnamefont {F.}~\bibnamefont {Weickert}}, \bibinfo {author}
		{\bibfnamefont {M.}~\bibnamefont {Jaime}}, \bibinfo {author} {\bibfnamefont
			{O.~E.}\ \bibnamefont {Ayala-Valenzuela}}, \bibinfo {author} {\bibfnamefont
			{R.~D.}\ \bibnamefont {McDonald}}, \bibinfo {author} {\bibfnamefont {S.~D.}\
			\bibnamefont {Selesnic}}, \bibinfo {author} {\bibfnamefont {H.~A.}\
			\bibnamefont {Dabkowska}}, \ and\ \bibinfo {author} {\bibfnamefont {G.~M.}\
			\bibnamefont {Luke}},\ }\bibfield  {title} {\enquote {\bibinfo {title}
			{Field-induced {B}ose-{E}instein condensation of triplons up to 8 {K} in
				{Sr$_3$Cr$_2$O$_8$}},}\ }\href {\doibase 10.1103/PhysRevLett.103.207203}
	{\bibfield  {journal} {\bibinfo  {journal} {Phys. Rev. Lett.}\ }\textbf
		{\bibinfo {volume} {103}},\ \bibinfo {pages} {207203} (\bibinfo {year}
		{2009}{\natexlab{a}})}\BibitemShut {NoStop}%
	\bibitem [{\citenamefont {Aczel}\ \emph
		{et~al.}(2009{\natexlab{b}})\citenamefont {Aczel}, \citenamefont {Kohama},
		\citenamefont {Jaime}, \citenamefont {Ninios}, \citenamefont {Chan},
		\citenamefont {Balicas}, \citenamefont {Dabkowska},\ and\ \citenamefont
		{Luke}}]{Aczel100409}%
	\BibitemOpen
	\bibfield  {author} {\bibinfo {author} {\bibfnamefont {A.~A.}\ \bibnamefont
			{Aczel}}, \bibinfo {author} {\bibfnamefont {Y.}~\bibnamefont {Kohama}},
		\bibinfo {author} {\bibfnamefont {M.}~\bibnamefont {Jaime}}, \bibinfo
		{author} {\bibfnamefont {K.}~\bibnamefont {Ninios}}, \bibinfo {author}
		{\bibfnamefont {H.~B.}\ \bibnamefont {Chan}}, \bibinfo {author}
		{\bibfnamefont {L.}~\bibnamefont {Balicas}}, \bibinfo {author} {\bibfnamefont
			{H.~A.}\ \bibnamefont {Dabkowska}}, \ and\ \bibinfo {author} {\bibfnamefont
			{G.~M.}\ \bibnamefont {Luke}},\ }\bibfield  {title} {\enquote {\bibinfo
			{title} {Bose-{E}instein condensation of triplons in {Ba$_3$Cr$_2$O$_8$}},}\
	}\href {\doibase 10.1103/PhysRevB.79.100409} {\bibfield  {journal} {\bibinfo
			{journal} {Phys. Rev. B}\ }\textbf {\bibinfo {volume} {79}},\ \bibinfo
		{pages} {100409(R)} (\bibinfo {year} {2009}{\natexlab{b}})}\BibitemShut
	{NoStop}%
	\bibitem [{\citenamefont {Jeong}\ \emph {et~al.}(2013)\citenamefont {Jeong},
		\citenamefont {Mayaffre}, \citenamefont {Berthier}, \citenamefont
		{Schmidiger}, \citenamefont {Zheludev},\ and\ \citenamefont
		{Horvati\ifmmode~\acute{c}\else \'{c}\fi{}}}]{Jeong106404}%
	\BibitemOpen
	\bibfield  {author} {\bibinfo {author} {\bibfnamefont {M.}~\bibnamefont
			{Jeong}}, \bibinfo {author} {\bibfnamefont {H.}~\bibnamefont {Mayaffre}},
		\bibinfo {author} {\bibfnamefont {C.}~\bibnamefont {Berthier}}, \bibinfo
		{author} {\bibfnamefont {D.}~\bibnamefont {Schmidiger}}, \bibinfo {author}
		{\bibfnamefont {A.}~\bibnamefont {Zheludev}}, \ and\ \bibinfo {author}
		{\bibfnamefont {M.}~\bibnamefont {Horvati\ifmmode~\acute{c}\else
				\'{c}\fi{}}},\ }\bibfield  {title} {\enquote {\bibinfo {title} {Attractive
				{Tomonaga-Luttinger} liquid in a quantum spin ladder},}\ }\href {\doibase
		10.1103/PhysRevLett.111.106404} {\bibfield  {journal} {\bibinfo  {journal}
			{Phys. Rev. Lett.}\ }\textbf {\bibinfo {volume} {111}},\ \bibinfo {pages}
		{106404} (\bibinfo {year} {2013})}\BibitemShut {NoStop}%
	\bibitem [{\citenamefont {Jeong}\ \emph {et~al.}(2017)\citenamefont {Jeong},
		\citenamefont {Mayaffre}, \citenamefont {Berthier}, \citenamefont
		{Schmidiger}, \citenamefont {Zheludev},\ and\ \citenamefont
		{Horvati\ifmmode~\acute{c}\else \'{c}\fi{}}}]{Jeong167206}%
	\BibitemOpen
	\bibfield  {author} {\bibinfo {author} {\bibfnamefont {M.}~\bibnamefont
			{Jeong}}, \bibinfo {author} {\bibfnamefont {H.}~\bibnamefont {Mayaffre}},
		\bibinfo {author} {\bibfnamefont {C.}~\bibnamefont {Berthier}}, \bibinfo
		{author} {\bibfnamefont {D.}~\bibnamefont {Schmidiger}}, \bibinfo {author}
		{\bibfnamefont {A.}~\bibnamefont {Zheludev}}, \ and\ \bibinfo {author}
		{\bibfnamefont {M.}~\bibnamefont {Horvati\ifmmode~\acute{c}\else
				\'{c}\fi{}}},\ }\bibfield  {title} {\enquote {\bibinfo {title}
			{Magnetic-order crossover in coupled spin ladders},}\ }\href {\doibase
		10.1103/PhysRevLett.118.167206} {\bibfield  {journal} {\bibinfo  {journal}
			{Phys. Rev. Lett.}\ }\textbf {\bibinfo {volume} {118}},\ \bibinfo {pages}
		{167206} (\bibinfo {year} {2017})}\BibitemShut {NoStop}%
	\bibitem [{\citenamefont {M\"oller}\ \emph {et~al.}(2017)\citenamefont
		{M\"oller}, \citenamefont {Lancaster}, \citenamefont {Blundell},
		\citenamefont {Pratt}, \citenamefont {Baker}, \citenamefont {Xiao},
		\citenamefont {Williams}, \citenamefont {Hayes}, \citenamefont {Turnbull},\
		and\ \citenamefont {Landee}}]{Moller020402}%
	\BibitemOpen
	\bibfield  {author} {\bibinfo {author} {\bibfnamefont {J.~S.}\ \bibnamefont
			{M\"oller}}, \bibinfo {author} {\bibfnamefont {T.}~\bibnamefont {Lancaster}},
		\bibinfo {author} {\bibfnamefont {S.~J.}\ \bibnamefont {Blundell}}, \bibinfo
		{author} {\bibfnamefont {F.~L.}\ \bibnamefont {Pratt}}, \bibinfo {author}
		{\bibfnamefont {P.~J.}\ \bibnamefont {Baker}}, \bibinfo {author}
		{\bibfnamefont {F.}~\bibnamefont {Xiao}}, \bibinfo {author} {\bibfnamefont
			{R.~C.}\ \bibnamefont {Williams}}, \bibinfo {author} {\bibfnamefont
			{W.}~\bibnamefont {Hayes}}, \bibinfo {author} {\bibfnamefont {M.~M.}\
			\bibnamefont {Turnbull}}, \ and\ \bibinfo {author} {\bibfnamefont {C.~P.}\
			\bibnamefont {Landee}},\ }\bibfield  {title} {\enquote {\bibinfo {title}
			{Quantum-critical spin dynamics in a {Tomonaga-Luttinger} liquid studied with
				muon-spin relaxation},}\ }\href {\doibase 10.1103/PhysRevB.95.020402}
	{\bibfield  {journal} {\bibinfo  {journal} {Phys. Rev. B}\ }\textbf {\bibinfo
			{volume} {95}},\ \bibinfo {pages} {020402(R)} (\bibinfo {year}
		{2017})}\BibitemShut {NoStop}%
	\bibitem [{\citenamefont {R\"uegg}\ \emph {et~al.}(2008)\citenamefont
		{R\"uegg}, \citenamefont {Kiefer}, \citenamefont {Thielemann}, \citenamefont
		{McMorrow}, \citenamefont {Zapf}, \citenamefont {Normand}, \citenamefont
		{Zvonarev}, \citenamefont {Bouillot}, \citenamefont {Kollath}, \citenamefont
		{Giamarchi}, \citenamefont {Capponi}, \citenamefont {Poilblanc},
		\citenamefont {Biner},\ and\ \citenamefont {Kr\"amer}}]{Ruegg247202}%
	\BibitemOpen
	\bibfield  {author} {\bibinfo {author} {\bibfnamefont {Ch.}\ \bibnamefont
			{R\"uegg}}, \bibinfo {author} {\bibfnamefont {K.}~\bibnamefont {Kiefer}},
		\bibinfo {author} {\bibfnamefont {B.}~\bibnamefont {Thielemann}}, \bibinfo
		{author} {\bibfnamefont {D.~F.}\ \bibnamefont {McMorrow}}, \bibinfo {author}
		{\bibfnamefont {V.}~\bibnamefont {Zapf}}, \bibinfo {author} {\bibfnamefont
			{B.}~\bibnamefont {Normand}}, \bibinfo {author} {\bibfnamefont {M.~B.}\
			\bibnamefont {Zvonarev}}, \bibinfo {author} {\bibfnamefont {P.}~\bibnamefont
			{Bouillot}}, \bibinfo {author} {\bibfnamefont {C.}~\bibnamefont {Kollath}},
		\bibinfo {author} {\bibfnamefont {T.}~\bibnamefont {Giamarchi}}, \bibinfo
		{author} {\bibfnamefont {S.}~\bibnamefont {Capponi}}, \bibinfo {author}
		{\bibfnamefont {D.}~\bibnamefont {Poilblanc}}, \bibinfo {author}
		{\bibfnamefont {D.}~\bibnamefont {Biner}}, \ and\ \bibinfo {author}
		{\bibfnamefont {K.~W.}\ \bibnamefont {Kr\"amer}},\ }\bibfield  {title}
	{\enquote {\bibinfo {title} {Thermodynamics of the spin {Luttinger} liquid in
				a model ladder material},}\ }\href {\doibase 10.1103/PhysRevLett.101.247202}
	{\bibfield  {journal} {\bibinfo  {journal} {Phys. Rev. Lett.}\ }\textbf
		{\bibinfo {volume} {101}},\ \bibinfo {pages} {247202} (\bibinfo {year}
		{2008})}\BibitemShut {NoStop}%
	\bibitem [{\citenamefont {Ueda}(1998)}]{Ueda2653}%
	\BibitemOpen
	\bibfield  {author} {\bibinfo {author} {\bibfnamefont {Y.}~\bibnamefont
			{Ueda}},\ }\bibfield  {title} {\enquote {\bibinfo {title} {Vanadate family as
				spin-gap systems},}\ }\href {\doibase 10.1021/cm980215w} {\bibfield
		{journal} {\bibinfo  {journal} {Chem. Mater.}\ }\textbf {\bibinfo {volume}
			{10}},\ \bibinfo {pages} {2653} (\bibinfo {year} {1998})}\BibitemShut
	{NoStop}%
	\bibitem [{\citenamefont {Yamauchi}\ \emph {et~al.}(1999)\citenamefont
		{Yamauchi}, \citenamefont {Narumi}, \citenamefont {Kikuchi}, \citenamefont
		{Ueda}, \citenamefont {Tatani}, \citenamefont {Kobayashi}, \citenamefont
		{Kindo},\ and\ \citenamefont {Motoya}}]{Yamauchi3729}%
	\BibitemOpen
	\bibfield  {author} {\bibinfo {author} {\bibfnamefont {T.}~\bibnamefont
			{Yamauchi}}, \bibinfo {author} {\bibfnamefont {Y.}~\bibnamefont {Narumi}},
		\bibinfo {author} {\bibfnamefont {J.}~\bibnamefont {Kikuchi}}, \bibinfo
		{author} {\bibfnamefont {Y.}~\bibnamefont {Ueda}}, \bibinfo {author}
		{\bibfnamefont {K.}~\bibnamefont {Tatani}}, \bibinfo {author} {\bibfnamefont
			{T.~C.}\ \bibnamefont {Kobayashi}}, \bibinfo {author} {\bibfnamefont
			{K.}~\bibnamefont {Kindo}}, \ and\ \bibinfo {author} {\bibfnamefont
			{K.}~\bibnamefont {Motoya}},\ }\bibfield  {title} {\enquote {\bibinfo {title}
			{Two gaps in {(VO)$_2$P$_2$O$_7$}: Observation using high-field magnetization
				and {NMR}},}\ }\href {\doibase 10.1103/PhysRevLett.83.3729} {\bibfield
		{journal} {\bibinfo  {journal} {Phys. Rev. Lett.}\ }\textbf {\bibinfo
			{volume} {83}},\ \bibinfo {pages} {3729} (\bibinfo {year}
		{1999})}\BibitemShut {NoStop}%
	\bibitem [{\citenamefont {Johnston}\ \emph {et~al.}(1987)\citenamefont
		{Johnston}, \citenamefont {Johnson}, \citenamefont {Goshorn},\ and\
		\citenamefont {Jacobson}}]{Johnston219}%
	\BibitemOpen
	\bibfield  {author} {\bibinfo {author} {\bibfnamefont {D.~C.}\ \bibnamefont
			{Johnston}}, \bibinfo {author} {\bibfnamefont {J.~W.}\ \bibnamefont
			{Johnson}}, \bibinfo {author} {\bibfnamefont {D.~P.}\ \bibnamefont
			{Goshorn}}, \ and\ \bibinfo {author} {\bibfnamefont {A.~J.}\ \bibnamefont
			{Jacobson}},\ }\bibfield  {title} {\enquote {\bibinfo {title} {Magnetic
				susceptibility of {(VO)$_2$P$_2$O$_7$}: A one-dimensional spin-1/2
				{Heisenberg} antiferromagnet with a ladder spin configuration and a singlet
				ground state},}\ }\href {\doibase 10.1103/PhysRevB.35.219} {\bibfield
		{journal} {\bibinfo  {journal} {Phys. Rev. B}\ }\textbf {\bibinfo {volume}
			{35}},\ \bibinfo {pages} {219} (\bibinfo {year} {1987})}\BibitemShut
	{NoStop}%
	\bibitem [{\citenamefont {Ghoshray}\ \emph {et~al.}(2005)\citenamefont
		{Ghoshray}, \citenamefont {Pahari}, \citenamefont {Bandyopadhyay},
		\citenamefont {Sarkar},\ and\ \citenamefont {Ghoshray}}]{Ghoshray214401}%
	\BibitemOpen
	\bibfield  {author} {\bibinfo {author} {\bibfnamefont {K.}~\bibnamefont
			{Ghoshray}}, \bibinfo {author} {\bibfnamefont {B.}~\bibnamefont {Pahari}},
		\bibinfo {author} {\bibfnamefont {B.}~\bibnamefont {Bandyopadhyay}}, \bibinfo
		{author} {\bibfnamefont {R.}~\bibnamefont {Sarkar}}, \ and\ \bibinfo {author}
		{\bibfnamefont {A.}~\bibnamefont {Ghoshray}},\ }\bibfield  {title} {\enquote
		{\bibinfo {title} {{$^{51}\rm{V}$} {NMR} study of the quasi-one-dimensional
				alternating chain compound
				{${\mathrm{BaCu}}_{2}{\mathrm{V}}_{2}{\mathrm{O}}_{8}$}},}\ }\href {\doibase
		10.1103/PhysRevB.71.214401} {\bibfield  {journal} {\bibinfo  {journal} {Phys.
				Rev. B}\ }\textbf {\bibinfo {volume} {71}},\ \bibinfo {pages} {214401}
		(\bibinfo {year} {2005})}\BibitemShut {NoStop}%
	\bibitem [{\citenamefont {Mukharjee}\ \emph {et~al.}(2019)\citenamefont
		{Mukharjee}, \citenamefont {Ranjith}, \citenamefont {Koo}, \citenamefont
		{Sichelschmidt}, \citenamefont {Baenitz}, \citenamefont {Skourski},
		\citenamefont {Inagaki}, \citenamefont {Furukawa}, \citenamefont {Tsirlin},\
		and\ \citenamefont {Nath}}]{Mukharjee144433}%
	\BibitemOpen
	\bibfield  {author} {\bibinfo {author} {\bibfnamefont {P.~K.}\ \bibnamefont
			{Mukharjee}}, \bibinfo {author} {\bibfnamefont {K.~M.}\ \bibnamefont
			{Ranjith}}, \bibinfo {author} {\bibfnamefont {B.}~\bibnamefont {Koo}},
		\bibinfo {author} {\bibfnamefont {J.}~\bibnamefont {Sichelschmidt}}, \bibinfo
		{author} {\bibfnamefont {M.}~\bibnamefont {Baenitz}}, \bibinfo {author}
		{\bibfnamefont {Y.}~\bibnamefont {Skourski}}, \bibinfo {author}
		{\bibfnamefont {Y.}~\bibnamefont {Inagaki}}, \bibinfo {author} {\bibfnamefont
			{Y.}~\bibnamefont {Furukawa}}, \bibinfo {author} {\bibfnamefont {A.~A.}\
			\bibnamefont {Tsirlin}}, \ and\ \bibinfo {author} {\bibfnamefont
			{R.}~\bibnamefont {Nath}},\ }\bibfield  {title} {\enquote {\bibinfo {title}
			{Bose-{E}instein condensation of triplons close to the quantum critical point
				in the quasi-one-dimensional spin-$\frac{1}{2}$ antiferromagnet
				{NaVOPO$_4$}},}\ }\href {\doibase 10.1103/PhysRevB.100.144433} {\bibfield
		{journal} {\bibinfo  {journal} {Phys. Rev. B}\ }\textbf {\bibinfo {volume}
			{100}},\ \bibinfo {pages} {144433} (\bibinfo {year} {2019})}\BibitemShut
	{NoStop}%
	\bibitem [{\citenamefont {Arjun}\ \emph {et~al.}(2019)\citenamefont {Arjun},
		\citenamefont {Ranjith}, \citenamefont {Koo}, \citenamefont {Sichelschmidt},
		\citenamefont {Skourski}, \citenamefont {Baenitz}, \citenamefont {Tsirlin},\
		and\ \citenamefont {Nath}}]{Arjun014421}%
	\BibitemOpen
	\bibfield  {author} {\bibinfo {author} {\bibfnamefont {U.}~\bibnamefont
			{Arjun}}, \bibinfo {author} {\bibfnamefont {K.~M.}\ \bibnamefont {Ranjith}},
		\bibinfo {author} {\bibfnamefont {B.}~\bibnamefont {Koo}}, \bibinfo {author}
		{\bibfnamefont {J.}~\bibnamefont {Sichelschmidt}}, \bibinfo {author}
		{\bibfnamefont {Y.}~\bibnamefont {Skourski}}, \bibinfo {author}
		{\bibfnamefont {M.}~\bibnamefont {Baenitz}}, \bibinfo {author} {\bibfnamefont
			{A.~A.}\ \bibnamefont {Tsirlin}}, \ and\ \bibinfo {author} {\bibfnamefont
			{R.}~\bibnamefont {Nath}},\ }\bibfield  {title} {\enquote {\bibinfo {title}
			{Singlet ground state in the alternating spin-$\frac{1}{2}$ chain compound
				{NaVOAsO$_4$}},}\ }\href {\doibase 10.1103/PhysRevB.99.014421} {\bibfield
		{journal} {\bibinfo  {journal} {Phys. Rev. B}\ }\textbf {\bibinfo {volume}
			{99}},\ \bibinfo {pages} {014421} (\bibinfo {year} {2019})}\BibitemShut
	{NoStop}%
	\bibitem [{\citenamefont {Ahmed}\ \emph {et~al.}(2017)\citenamefont {Ahmed},
		\citenamefont {Khuntia}, \citenamefont {Ranjith}, \citenamefont {Rosner},
		\citenamefont {Baenitz}, \citenamefont {Tsirlin},\ and\ \citenamefont
		{Nath}}]{Ahmed224423}%
	\BibitemOpen
	\bibfield  {author} {\bibinfo {author} {\bibfnamefont {N.}~\bibnamefont
			{Ahmed}}, \bibinfo {author} {\bibfnamefont {P.}~\bibnamefont {Khuntia}},
		\bibinfo {author} {\bibfnamefont {K.~M.}\ \bibnamefont {Ranjith}}, \bibinfo
		{author} {\bibfnamefont {H.}~\bibnamefont {Rosner}}, \bibinfo {author}
		{\bibfnamefont {M.}~\bibnamefont {Baenitz}}, \bibinfo {author} {\bibfnamefont
			{A.~A.}\ \bibnamefont {Tsirlin}}, \ and\ \bibinfo {author} {\bibfnamefont
			{R.}~\bibnamefont {Nath}},\ }\bibfield  {title} {\enquote {\bibinfo {title}
			{Alternating spin chain compound {AgVOAsO$_4$} probed by $^{75}\mathrm{As}$
				{NMR}},}\ }\href {\doibase 10.1103/PhysRevB.96.224423} {\bibfield  {journal}
		{\bibinfo  {journal} {Phys. Rev. B}\ }\textbf {\bibinfo {volume} {96}},\
		\bibinfo {pages} {224423} (\bibinfo {year} {2017})}\BibitemShut {NoStop}%
	\bibitem [{\citenamefont {Tsirlin}\ \emph {et~al.}(2011)\citenamefont
		{Tsirlin}, \citenamefont {Nath}, \citenamefont {Sichelschmidt}, \citenamefont
		{Skourski}, \citenamefont {Geibel},\ and\ \citenamefont
		{Rosner}}]{Tsirlin144412}%
	\BibitemOpen
	\bibfield  {author} {\bibinfo {author} {\bibfnamefont {A.~A.}\ \bibnamefont
			{Tsirlin}}, \bibinfo {author} {\bibfnamefont {R.}~\bibnamefont {Nath}},
		\bibinfo {author} {\bibfnamefont {J.}~\bibnamefont {Sichelschmidt}}, \bibinfo
		{author} {\bibfnamefont {Y.}~\bibnamefont {Skourski}}, \bibinfo {author}
		{\bibfnamefont {C.}~\bibnamefont {Geibel}}, \ and\ \bibinfo {author}
		{\bibfnamefont {H.}~\bibnamefont {Rosner}},\ }\bibfield  {title} {\enquote
		{\bibinfo {title} {Frustrated couplings between alternating
				spin-$\frac{1}{2}$ chains in {AgVOAsO$_4$}},}\ }\href {\doibase
		10.1103/PhysRevB.83.144412} {\bibfield  {journal} {\bibinfo  {journal} {Phys.
				Rev. B}\ }\textbf {\bibinfo {volume} {83}},\ \bibinfo {pages} {144412}
		(\bibinfo {year} {2011})}\BibitemShut {NoStop}%
	\bibitem [{\citenamefont {Weickert}\ \emph {et~al.}(2019)\citenamefont
		{Weickert}, \citenamefont {Aczel}, \citenamefont {Stone}, \citenamefont
		{Garlea}, \citenamefont {Dong}, \citenamefont {Kohama}, \citenamefont
		{Movshovich}, \citenamefont {Demuer}, \citenamefont {Harrison}, \citenamefont
		{Gam\ifmmode~\dot{z}\else \.{z}\fi{}a}, \citenamefont {Steppke},
		\citenamefont {Brando}, \citenamefont {Rosner},\ and\ \citenamefont
		{Tsirlin}}]{Weickert104422}%
	\BibitemOpen
	\bibfield  {author} {\bibinfo {author} {\bibfnamefont {F.}~\bibnamefont
			{Weickert}}, \bibinfo {author} {\bibfnamefont {A.~A.}\ \bibnamefont {Aczel}},
		\bibinfo {author} {\bibfnamefont {M.~B.}\ \bibnamefont {Stone}}, \bibinfo
		{author} {\bibfnamefont {V.~O.}\ \bibnamefont {Garlea}}, \bibinfo {author}
		{\bibfnamefont {C.}~\bibnamefont {Dong}}, \bibinfo {author} {\bibfnamefont
			{Y.}~\bibnamefont {Kohama}}, \bibinfo {author} {\bibfnamefont
			{R.}~\bibnamefont {Movshovich}}, \bibinfo {author} {\bibfnamefont
			{A.}~\bibnamefont {Demuer}}, \bibinfo {author} {\bibfnamefont
			{N.}~\bibnamefont {Harrison}}, \bibinfo {author} {\bibfnamefont {M.~B.}\
			\bibnamefont {Gam\ifmmode~\dot{z}\else \.{z}\fi{}a}}, \bibinfo {author}
		{\bibfnamefont {A.}~\bibnamefont {Steppke}}, \bibinfo {author} {\bibfnamefont
			{M.}~\bibnamefont {Brando}}, \bibinfo {author} {\bibfnamefont
			{H.}~\bibnamefont {Rosner}}, \ and\ \bibinfo {author} {\bibfnamefont {A.~A.}\
			\bibnamefont {Tsirlin}},\ }\bibfield  {title} {\enquote {\bibinfo {title}
			{Field-induced double dome and {Bose-Einstein} condensation in the crossing
				quantum spin chain system {AgVOAsO$_4$}},}\ }\href {\doibase
		10.1103/PhysRevB.100.104422} {\bibfield  {journal} {\bibinfo  {journal}
			{Phys. Rev. B}\ }\textbf {\bibinfo {volume} {100}},\ \bibinfo {pages}
		{104422} (\bibinfo {year} {2019})}\BibitemShut {NoStop}%
	\bibitem [{\citenamefont {Harrison}\ and\ \citenamefont
		{Manthiram}(2013)}]{Harrison1751}%
	\BibitemOpen
	\bibfield  {author} {\bibinfo {author} {\bibfnamefont {K.L.}\ \bibnamefont
			{Harrison}}\ and\ \bibinfo {author} {\bibfnamefont {A.}~\bibnamefont
			{Manthiram}},\ }\bibfield  {title} {\enquote {\bibinfo {title}
			{Microwave-assisted solvothermal synthesis and characterization of various
				polymorphs of {LiVOPO$_4$}},}\ }\href {\doibase 10.1021/cm400227j} {\bibfield
		{journal} {\bibinfo  {journal} {Chem. Mater.}\ }\textbf {\bibinfo {volume}
			{25}},\ \bibinfo {pages} {1751} (\bibinfo {year} {2013})}\BibitemShut
	{NoStop}%
	\bibitem [{\citenamefont {Hidalgo}\ \emph {et~al.}(2019)\citenamefont
		{Hidalgo}, \citenamefont {Lin}, \citenamefont {Grenier}, \citenamefont
		{Xiao}, \citenamefont {Rana}, \citenamefont {Tran}, \citenamefont {Xin},
		\citenamefont {Zuba}, \citenamefont {Donohue}, \citenamefont {Omenya},
		\citenamefont {Chu}, \citenamefont {Wang}, \citenamefont {Li}, \citenamefont
		{Chernova}, \citenamefont {Chapman}, \citenamefont {Zhou}, \citenamefont
		{Piper}, \citenamefont {Ong},\ and\ \citenamefont
		{Whittingham}}]{Hidalgo8423}%
	\BibitemOpen
	\bibfield  {author} {\bibinfo {author} {\bibfnamefont {M.F.V.}\ \bibnamefont
			{Hidalgo}}, \bibinfo {author} {\bibfnamefont {Y.-C.}\ \bibnamefont {Lin}},
		\bibinfo {author} {\bibfnamefont {A.}~\bibnamefont {Grenier}}, \bibinfo
		{author} {\bibfnamefont {D.}~\bibnamefont {Xiao}}, \bibinfo {author}
		{\bibfnamefont {J.}~\bibnamefont {Rana}}, \bibinfo {author} {\bibfnamefont
			{R.}~\bibnamefont {Tran}}, \bibinfo {author} {\bibfnamefont {H.}~\bibnamefont
			{Xin}}, \bibinfo {author} {\bibfnamefont {M.}~\bibnamefont {Zuba}}, \bibinfo
		{author} {\bibfnamefont {J.}~\bibnamefont {Donohue}}, \bibinfo {author}
		{\bibfnamefont {F.O.}\ \bibnamefont {Omenya}}, \bibinfo {author}
		{\bibfnamefont {I.-H.}\ \bibnamefont {Chu}}, \bibinfo {author} {\bibfnamefont
			{Z.}~\bibnamefont {Wang}}, \bibinfo {author} {\bibfnamefont {X.G.}\
			\bibnamefont {Li}}, \bibinfo {author} {\bibfnamefont {N.A.}\ \bibnamefont
			{Chernova}}, \bibinfo {author} {\bibfnamefont {K.W.}\ \bibnamefont
			{Chapman}}, \bibinfo {author} {\bibfnamefont {G.}~\bibnamefont {Zhou}},
		\bibinfo {author} {\bibfnamefont {L.}~\bibnamefont {Piper}}, \bibinfo
		{author} {\bibfnamefont {S.P.}\ \bibnamefont {Ong}}, \ and\ \bibinfo {author}
		{\bibfnamefont {M.S.}\ \bibnamefont {Whittingham}},\ }\bibfield  {title}
	{\enquote {\bibinfo {title} {Rational synthesis and electrochemical
				performance of {LiVOPO$_4$} polymorphs},}\ }\href {\doibase
		10.1039/C8TA12531G} {\bibfield  {journal} {\bibinfo  {journal} {J. Mater.
				Chem. A}\ }\textbf {\bibinfo {volume} {7}},\ \bibinfo {pages} {8423}
		(\bibinfo {year} {2019})}\BibitemShut {NoStop}%
	\bibitem [{Note1()}]{Note1}%
	\BibitemOpen
	\bibinfo {note} {Triclinic crystal structure is derived from $\varepsilon
		$-VOPO$_4$. Alternatively, triclinic LiVOPO$_4$ is sometimes referred to as
		the $\alpha $-phase, because it was the earliest discovered LiVOPO$_4$
		polymorph.}\BibitemShut {Stop}%
	\bibitem [{\citenamefont {Yang}\ \emph {et~al.}(2008)\citenamefont {Yang},
		\citenamefont {Fang}, \citenamefont {Zheng}, \citenamefont {Li},
		\citenamefont {Li},\ and\ \citenamefont {Yan}}]{Yang2008}%
	\BibitemOpen
	\bibfield  {author} {\bibinfo {author} {\bibfnamefont {Y.}~\bibnamefont
			{Yang}}, \bibinfo {author} {\bibfnamefont {H.}~\bibnamefont {Fang}}, \bibinfo
		{author} {\bibfnamefont {J.}~\bibnamefont {Zheng}}, \bibinfo {author}
		{\bibfnamefont {L.}~\bibnamefont {Li}}, \bibinfo {author} {\bibfnamefont
			{G.}~\bibnamefont {Li}}, \ and\ \bibinfo {author} {\bibfnamefont
			{G.}~\bibnamefont {Yan}},\ }\bibfield  {title} {\enquote {\bibinfo {title}
			{Towards the understanding of poor electrochemical activity of triclinic
				{LiVOPO$_4$}: Experimental characterization and theoretical
				investigations},}\ }\href {\doibase 10.1016/j.solidstatesciences.2008.01.028}
	{\bibfield  {journal} {\bibinfo  {journal} {Solid State Sci.}\ }\textbf
		{\bibinfo {volume} {10}},\ \bibinfo {pages} {1292} (\bibinfo {year}
		{2008})}\BibitemShut {NoStop}%
	\bibitem [{\citenamefont {Quackenbush}\ \emph {et~al.}(2015)\citenamefont
		{Quackenbush}, \citenamefont {Wangoh}, \citenamefont {Scanlon}, \citenamefont
		{Zhang}, \citenamefont {Chung}, \citenamefont {Chen}, \citenamefont {Wen},
		\citenamefont {Lin}, \citenamefont {Woicik}, \citenamefont {Chernova},
		\citenamefont {Ong}, \citenamefont {Whittingham},\ and\ \citenamefont
		{Piper}}]{Quackenbush2015}%
	\BibitemOpen
	\bibfield  {author} {\bibinfo {author} {\bibfnamefont {N.~F.}\ \bibnamefont
			{Quackenbush}}, \bibinfo {author} {\bibfnamefont {L.}~\bibnamefont {Wangoh}},
		\bibinfo {author} {\bibfnamefont {D.~O.}\ \bibnamefont {Scanlon}}, \bibinfo
		{author} {\bibfnamefont {R.}~\bibnamefont {Zhang}}, \bibinfo {author}
		{\bibfnamefont {Y.}~\bibnamefont {Chung}}, \bibinfo {author} {\bibfnamefont
			{Z.}~\bibnamefont {Chen}}, \bibinfo {author} {\bibfnamefont {B.}~\bibnamefont
			{Wen}}, \bibinfo {author} {\bibfnamefont {Y.}~\bibnamefont {Lin}}, \bibinfo
		{author} {\bibfnamefont {J.~C.}\ \bibnamefont {Woicik}}, \bibinfo {author}
		{\bibfnamefont {N.~A.}\ \bibnamefont {Chernova}}, \bibinfo {author}
		{\bibfnamefont {S.~P.}\ \bibnamefont {Ong}}, \bibinfo {author} {\bibfnamefont
			{M.S.}\ \bibnamefont {Whittingham}}, \ and\ \bibinfo {author} {\bibfnamefont
			{L.~F.~J.}\ \bibnamefont {Piper}},\ }\bibfield  {title} {\enquote {\bibinfo
			{title} {Interfacial effects in {$\varepsilon$-Li$_x$VOPO$_4$} and evolution
				of the electronic structure},}\ }\href {\doibase
		10.1021/acs.chemmater.5b02145} {\bibfield  {journal} {\bibinfo  {journal}
			{Chem. Mater.}\ }\textbf {\bibinfo {volume} {27}},\ \bibinfo {pages} {8211}
		(\bibinfo {year} {2015})}\BibitemShut {NoStop}%
	\bibitem [{\citenamefont {Lin}\ \emph {et~al.}(2016)\citenamefont {Lin},
		\citenamefont {Wen}, \citenamefont {Wiaderek}, \citenamefont {Sallis},
		\citenamefont {Liu}, \citenamefont {Lapidus}, \citenamefont {Borkiewicz},
		\citenamefont {Quackenbush}, \citenamefont {Chernova}, \citenamefont {Karki},
		\citenamefont {Omenya}, \citenamefont {Chupas}, \citenamefont {Piper},
		\citenamefont {Whittingham}, \citenamefont {Chapman},\ and\ \citenamefont
		{Ong}}]{Lin2016}%
	\BibitemOpen
	\bibfield  {author} {\bibinfo {author} {\bibfnamefont {Y.-C.}\ \bibnamefont
			{Lin}}, \bibinfo {author} {\bibfnamefont {B.}~\bibnamefont {Wen}}, \bibinfo
		{author} {\bibfnamefont {K.~M.}\ \bibnamefont {Wiaderek}}, \bibinfo {author}
		{\bibfnamefont {S.}~\bibnamefont {Sallis}}, \bibinfo {author} {\bibfnamefont
			{H.}~\bibnamefont {Liu}}, \bibinfo {author} {\bibfnamefont {S.~H.}\
			\bibnamefont {Lapidus}}, \bibinfo {author} {\bibfnamefont {O.~J.}\
			\bibnamefont {Borkiewicz}}, \bibinfo {author} {\bibfnamefont {N.~F.}\
			\bibnamefont {Quackenbush}}, \bibinfo {author} {\bibfnamefont {N.~A.}\
			\bibnamefont {Chernova}}, \bibinfo {author} {\bibfnamefont {K.}~\bibnamefont
			{Karki}}, \bibinfo {author} {\bibfnamefont {F.}~\bibnamefont {Omenya}},
		\bibinfo {author} {\bibfnamefont {P.~J.}\ \bibnamefont {Chupas}}, \bibinfo
		{author} {\bibfnamefont {L.~F.~J.}\ \bibnamefont {Piper}}, \bibinfo {author}
		{\bibfnamefont {M.~S.}\ \bibnamefont {Whittingham}}, \bibinfo {author}
		{\bibfnamefont {K.~W.}\ \bibnamefont {Chapman}}, \ and\ \bibinfo {author}
		{\bibfnamefont {S.~P.}\ \bibnamefont {Ong}},\ }\bibfield  {title} {\enquote
		{\bibinfo {title} {Thermodynamics, kinetics and structural evolution of
				{$\epsilon$-LiVOPO$_4$} over multiple lithium intercalation},}\ }\href
	{\doibase 10.1021/acs.chemmater.5b04880} {\bibfield  {journal} {\bibinfo
			{journal} {Chem. Mater.}\ }\textbf {\bibinfo {volume} {28}},\ \bibinfo
		{pages} {1794} (\bibinfo {year} {2016})}\BibitemShut {NoStop}%
	\bibitem [{\citenamefont {Shi}\ \emph {et~al.}(2018)\citenamefont {Shi},
		\citenamefont {Zhou}, \citenamefont {Seymour}, \citenamefont {Britto},
		\citenamefont {Rana}, \citenamefont {Wangoh}, \citenamefont {Huang},
		\citenamefont {Yin}, \citenamefont {Reeves}, \citenamefont {Zuba},
		\citenamefont {Chung}, \citenamefont {Omenya}, \citenamefont {Chernova},
		\citenamefont {Zhou}, \citenamefont {Piper}, \citenamefont {Grey},\ and\
		\citenamefont {Whittingham}}]{Shi2018}%
	\BibitemOpen
	\bibfield  {author} {\bibinfo {author} {\bibfnamefont {Y.}~\bibnamefont
			{Shi}}, \bibinfo {author} {\bibfnamefont {H.}~\bibnamefont {Zhou}}, \bibinfo
		{author} {\bibfnamefont {I.D.}\ \bibnamefont {Seymour}}, \bibinfo {author}
		{\bibfnamefont {S.}~\bibnamefont {Britto}}, \bibinfo {author} {\bibfnamefont
			{J.}~\bibnamefont {Rana}}, \bibinfo {author} {\bibfnamefont {L.W.}\
			\bibnamefont {Wangoh}}, \bibinfo {author} {\bibfnamefont {Y.}~\bibnamefont
			{Huang}}, \bibinfo {author} {\bibfnamefont {Q.}~\bibnamefont {Yin}}, \bibinfo
		{author} {\bibfnamefont {P.J.}\ \bibnamefont {Reeves}}, \bibinfo {author}
		{\bibfnamefont {M.}~\bibnamefont {Zuba}}, \bibinfo {author} {\bibfnamefont
			{Y.}~\bibnamefont {Chung}}, \bibinfo {author} {\bibfnamefont
			{F.}~\bibnamefont {Omenya}}, \bibinfo {author} {\bibfnamefont {N.A.}\
			\bibnamefont {Chernova}}, \bibinfo {author} {\bibfnamefont {G.}~\bibnamefont
			{Zhou}}, \bibinfo {author} {\bibfnamefont {L.F.J.}\ \bibnamefont {Piper}},
		\bibinfo {author} {\bibfnamefont {C.P.}\ \bibnamefont {Grey}}, \ and\
		\bibinfo {author} {\bibfnamefont {M.S.}\ \bibnamefont {Whittingham}},\
	}\bibfield  {title} {\enquote {\bibinfo {title} {Electrochemical performance
				of nanosized disordered {LiVOPO$_4$}},}\ }\href {\doibase
		10.1021/acsomega.8b00763} {\bibfield  {journal} {\bibinfo  {journal} {ACS
				Omega}\ }\textbf {\bibinfo {volume} {3}},\ \bibinfo {pages} {7310} (\bibinfo
		{year} {2018})}\BibitemShut {NoStop}%
	\bibitem [{\citenamefont {Chung}\ \emph {et~al.}(2019)\citenamefont {Chung},
		\citenamefont {Cassidy}, \citenamefont {Lee}, \citenamefont {Siu},
		\citenamefont {Huang}, \citenamefont {Omenya}, \citenamefont {Rana},
		\citenamefont {Wiaderek}, \citenamefont {Chernova}, \citenamefont {Chapman},
		\citenamefont {Piper},\ and\ \citenamefont {Whittingham}}]{Chung2019}%
	\BibitemOpen
	\bibfield  {author} {\bibinfo {author} {\bibfnamefont {Y.}~\bibnamefont
			{Chung}}, \bibinfo {author} {\bibfnamefont {E.}~\bibnamefont {Cassidy}},
		\bibinfo {author} {\bibfnamefont {K.}~\bibnamefont {Lee}}, \bibinfo {author}
		{\bibfnamefont {C.}~\bibnamefont {Siu}}, \bibinfo {author} {\bibfnamefont
			{Y.}~\bibnamefont {Huang}}, \bibinfo {author} {\bibfnamefont
			{F.}~\bibnamefont {Omenya}}, \bibinfo {author} {\bibfnamefont
			{J.}~\bibnamefont {Rana}}, \bibinfo {author} {\bibfnamefont {K.M.}\
			\bibnamefont {Wiaderek}}, \bibinfo {author} {\bibfnamefont {N.A.}\
			\bibnamefont {Chernova}}, \bibinfo {author} {\bibfnamefont {K.W.}\
			\bibnamefont {Chapman}}, \bibinfo {author} {\bibfnamefont {L.F.J.}\
			\bibnamefont {Piper}}, \ and\ \bibinfo {author} {\bibfnamefont {M.S.}\
			\bibnamefont {Whittingham}},\ }\bibfield  {title} {\enquote {\bibinfo {title}
			{Nonstoichiometry and defects in hydrothermally synthesized
				{$\varepsilon$-LiVOPO$_4$}},}\ }\href {\doibase 10.1021/acsaem.9b00448}
	{\bibfield  {journal} {\bibinfo  {journal} {ACS Appl. Energy Mater.}\
		}\textbf {\bibinfo {volume} {2}},\ \bibinfo {pages} {4792} (\bibinfo {year}
		{2019})}\BibitemShut {NoStop}%
	\bibitem [{\citenamefont {Onoda}\ and\ \citenamefont
		{Ikeda}(2013)}]{Onoda053801}%
	\BibitemOpen
	\bibfield  {author} {\bibinfo {author} {\bibfnamefont {M.}~\bibnamefont
			{Onoda}}\ and\ \bibinfo {author} {\bibfnamefont {S.}~\bibnamefont {Ikeda}},\
	}\bibfield  {title} {\enquote {\bibinfo {title} {Crystal structure and
				spin-singlet state of the {Li$_x$VOPO$_4$} insertion electrode system with
				alternating-bond chain},}\ }\href {\doibase 10.7566/JPSJ.82.053801}
	{\bibfield  {journal} {\bibinfo  {journal} {J. Phys. Soc. Jpn.}\ }\textbf
		{\bibinfo {volume} {82}},\ \bibinfo {pages} {053801} (\bibinfo {year}
		{2013})}\BibitemShut {NoStop}%
	\bibitem [{\citenamefont {Carvajal}(1993)}]{Carvajal55}%
	\BibitemOpen
	\bibfield  {author} {\bibinfo {author} {\bibfnamefont {J.~R.}\ \bibnamefont
			{Carvajal}},\ }\bibfield  {title} {\enquote {\bibinfo {title} {Recent
				advances in magnetic structure determination by neutron powder
				diffraction},}\ }\href {\doibase
		https://doi.org/10.1016/0921-4526(93)90108-I} {\bibfield  {journal} {\bibinfo
			{journal} {Physica B: Cond. Mat.}\ }\textbf {\bibinfo {volume} {192}},\
		\bibinfo {pages} {55} (\bibinfo {year} {1993})}\BibitemShut {NoStop}%
	\bibitem [{\citenamefont {A.~Mba}\ \emph {et~al.}(2012)\citenamefont {A.~Mba},
		\citenamefont {Masquelier}, \citenamefont {Suard},\ and\ \citenamefont
		{Croguennec}}]{Ateba1223}%
	\BibitemOpen
	\bibfield  {author} {\bibinfo {author} {\bibfnamefont {J.}~\bibnamefont
			{A.~Mba}}, \bibinfo {author} {\bibfnamefont {C.}~\bibnamefont {Masquelier}},
		\bibinfo {author} {\bibfnamefont {E.}~\bibnamefont {Suard}}, \ and\ \bibinfo
		{author} {\bibfnamefont {L.}~\bibnamefont {Croguennec}},\ }\bibfield  {title}
	{\enquote {\bibinfo {title} {Synthesis and crystallographic study of
				homeotypic {LiVPO$_4$F} and {LiVOPO$_4$}},}\ }\href {\doibase
		10.1021/cm3003996} {\bibfield  {journal} {\bibinfo  {journal} {Chem. Mater.}\
		}\textbf {\bibinfo {volume} {24}},\ \bibinfo {pages} {1223} (\bibinfo {year}
		{2012})}\BibitemShut {NoStop}%
	\bibitem [{\citenamefont {Tsirlin}\ \emph {et~al.}(2009)\citenamefont
		{Tsirlin}, \citenamefont {Schmidt}, \citenamefont {Skourski}, \citenamefont
		{Nath}, \citenamefont {Geibel},\ and\ \citenamefont
		{Rosner}}]{Tsirlin132407}%
	\BibitemOpen
	\bibfield  {author} {\bibinfo {author} {\bibfnamefont {A.~A.}\ \bibnamefont
			{Tsirlin}}, \bibinfo {author} {\bibfnamefont {B.}~\bibnamefont {Schmidt}},
		\bibinfo {author} {\bibfnamefont {Y.}~\bibnamefont {Skourski}}, \bibinfo
		{author} {\bibfnamefont {R.}~\bibnamefont {Nath}}, \bibinfo {author}
		{\bibfnamefont {C.}~\bibnamefont {Geibel}}, \ and\ \bibinfo {author}
		{\bibfnamefont {H.}~\bibnamefont {Rosner}},\ }\bibfield  {title} {\enquote
		{\bibinfo {title} {Exploring the spin-{$\frac{1}{2}$} frustrated square
				lattice model with high-field magnetization studies},}\ }\href {\doibase
		10.1103/PhysRevB.80.132407} {\bibfield  {journal} {\bibinfo  {journal} {Phys.
				Rev. B}\ }\textbf {\bibinfo {volume} {80}},\ \bibinfo {pages} {132407}
		(\bibinfo {year} {2009})}\BibitemShut {NoStop}%
	\bibitem [{\citenamefont {Skourski}\ \emph {et~al.}(2011)\citenamefont
		{Skourski}, \citenamefont {Kuz'min}, \citenamefont {Skokov}, \citenamefont
		{Andreev},\ and\ \citenamefont {Wosnitza}}]{Skourski214420}%
	\BibitemOpen
	\bibfield  {author} {\bibinfo {author} {\bibfnamefont {Y.}~\bibnamefont
			{Skourski}}, \bibinfo {author} {\bibfnamefont {M.~D.}\ \bibnamefont
			{Kuz'min}}, \bibinfo {author} {\bibfnamefont {K.~P.}\ \bibnamefont {Skokov}},
		\bibinfo {author} {\bibfnamefont {A.~V.}\ \bibnamefont {Andreev}}, \ and\
		\bibinfo {author} {\bibfnamefont {J.}~\bibnamefont {Wosnitza}},\ }\bibfield
	{title} {\enquote {\bibinfo {title} {High-field magnetization of
				{Ho$_{2}$Fe$_{17}$}},}\ }\href {\doibase 10.1103/PhysRevB.83.214420}
	{\bibfield  {journal} {\bibinfo  {journal} {Phys. Rev. B}\ }\textbf {\bibinfo
			{volume} {83}},\ \bibinfo {pages} {214420} (\bibinfo {year}
		{2011})}\BibitemShut {NoStop}%
	\bibitem [{\citenamefont {Koepernik}\ and\ \citenamefont
		{Eschrig}(1999)}]{fplo}%
	\BibitemOpen
	\bibfield  {author} {\bibinfo {author} {\bibfnamefont {K.}~\bibnamefont
			{Koepernik}}\ and\ \bibinfo {author} {\bibfnamefont {H.}~\bibnamefont
			{Eschrig}},\ }\bibfield  {title} {\enquote {\bibinfo {title} {Full-potential
				nonorthogonal local-orbital minimum-basis band-structure scheme},}\ }\href
	{\doibase 10.1103/PhysRevB.59.1743} {\bibfield  {journal} {\bibinfo
			{journal} {Phys. Rev. B}\ }\textbf {\bibinfo {volume} {59}},\ \bibinfo
		{pages} {1743} (\bibinfo {year} {1999})}\BibitemShut {NoStop}%
	\bibitem [{\citenamefont {Lavrov}\ \emph {et~al.}(1982)\citenamefont {Lavrov},
		\citenamefont {Nikolaev}, \citenamefont {Sadikov},\ and\ \citenamefont
		{Poray-Koshits}}]{lavrov1982}%
	\BibitemOpen
	\bibfield  {author} {\bibinfo {author} {\bibfnamefont {A.V.}\ \bibnamefont
			{Lavrov}}, \bibinfo {author} {\bibfnamefont {V.P.}\ \bibnamefont {Nikolaev}},
		\bibinfo {author} {\bibfnamefont {G.G.}\ \bibnamefont {Sadikov}}, \ and\
		\bibinfo {author} {\bibfnamefont {M.A.}\ \bibnamefont {Poray-Koshits}},\
	}\bibfield  {title} {\enquote {\bibinfo {title} {Synthesis and crystal
				structure of mixed vanadyl and lithium orthophosphate {LiVOPO$_4$}},}\ }\href
	{http://mi.mathnet.ru/dan45588} {\bibfield  {journal} {\bibinfo  {journal}
			{Dokl. Akad. Nauk SSSR}\ }\textbf {\bibinfo {volume} {266}},\ \bibinfo
		{pages} {343} (\bibinfo {year} {1982})}\BibitemShut {NoStop}%
	\bibitem [{\citenamefont {Perdew}\ \emph {et~al.}(1996)\citenamefont {Perdew},
		\citenamefont {Burke},\ and\ \citenamefont {Ernzerhof}}]{pbe96}%
	\BibitemOpen
	\bibfield  {author} {\bibinfo {author} {\bibfnamefont {J.~P.}\ \bibnamefont
			{Perdew}}, \bibinfo {author} {\bibfnamefont {K.}~\bibnamefont {Burke}}, \
		and\ \bibinfo {author} {\bibfnamefont {M.}~\bibnamefont {Ernzerhof}},\
	}\bibfield  {title} {\enquote {\bibinfo {title} {Generalized gradient
				approximation made simple},}\ }\href {\doibase 10.1103/PhysRevLett.77.3865}
	{\bibfield  {journal} {\bibinfo  {journal} {Phys. Rev. Lett.}\ }\textbf
		{\bibinfo {volume} {77}},\ \bibinfo {pages} {3865} (\bibinfo {year}
		{1996})}\BibitemShut {NoStop}%
	\bibitem [{\citenamefont {Tsirlin}(2014)}]{Tsirlin014405}%
	\BibitemOpen
	\bibfield  {author} {\bibinfo {author} {\bibfnamefont {A.A.}\ \bibnamefont
			{Tsirlin}},\ }\bibfield  {title} {\enquote {\bibinfo {title} {Spin-chain
				magnetism and uniform {Dzyaloshinsky-Moriya} anisotropy in {BaV$_3$O$_8$}},}\
	}\href {\doibase 10.1103/PhysRevB.89.014405} {\bibfield  {journal} {\bibinfo
			{journal} {Phys. Rev. B}\ }\textbf {\bibinfo {volume} {89}},\ \bibinfo
		{pages} {014405} (\bibinfo {year} {2014})}\BibitemShut {NoStop}%
	\bibitem [{\citenamefont {Nath}\ \emph
		{et~al.}(2008{\natexlab{a}})\citenamefont {Nath}, \citenamefont {Tsirlin},
		\citenamefont {Kaul}, \citenamefont {Baenitz}, \citenamefont {B\"uttgen},
		\citenamefont {Geibel},\ and\ \citenamefont {Rosner}}]{Nath024418}%
	\BibitemOpen
	\bibfield  {author} {\bibinfo {author} {\bibfnamefont {R.}~\bibnamefont
			{Nath}}, \bibinfo {author} {\bibfnamefont {A.A.}\ \bibnamefont {Tsirlin}},
		\bibinfo {author} {\bibfnamefont {E.E.}\ \bibnamefont {Kaul}}, \bibinfo
		{author} {\bibfnamefont {M.}~\bibnamefont {Baenitz}}, \bibinfo {author}
		{\bibfnamefont {N.}~\bibnamefont {B\"uttgen}}, \bibinfo {author}
		{\bibfnamefont {C.}~\bibnamefont {Geibel}}, \ and\ \bibinfo {author}
		{\bibfnamefont {H.}~\bibnamefont {Rosner}},\ }\bibfield  {title} {\enquote
		{\bibinfo {title} {Strong frustration due to competing ferromagnetic and
				antiferromagnetic interactions: Magnetic properties of {M(VO)$_2$(PO$_4)_2$
					(M = Ca and Sr)}},}\ }\href {\doibase 10.1103/PhysRevB.78.024418} {\bibfield
		{journal} {\bibinfo  {journal} {Phys. Rev. B}\ }\textbf {\bibinfo {volume}
			{78}},\ \bibinfo {pages} {024418} (\bibinfo {year}
		{2008}{\natexlab{a}})}\BibitemShut {NoStop}%
	\bibitem [{\citenamefont {Tsirlin}\ \emph {et~al.}(2008)\citenamefont
		{Tsirlin}, \citenamefont {Nath}, \citenamefont {Geibel},\ and\ \citenamefont
		{Rosner}}]{Tsirlin104436}%
	\BibitemOpen
	\bibfield  {author} {\bibinfo {author} {\bibfnamefont {A.A.}\ \bibnamefont
			{Tsirlin}}, \bibinfo {author} {\bibfnamefont {R.}~\bibnamefont {Nath}},
		\bibinfo {author} {\bibfnamefont {C.}~\bibnamefont {Geibel}}, \ and\ \bibinfo
		{author} {\bibfnamefont {H.}~\bibnamefont {Rosner}},\ }\bibfield  {title}
	{\enquote {\bibinfo {title} {Magnetic properties of {Ag$_2$VOP$_2$O$_7$}: An
				unexpected spin dimer system},}\ }\href {\doibase 10.1103/PhysRevB.77.104436}
	{\bibfield  {journal} {\bibinfo  {journal} {Phys. Rev. B}\ }\textbf {\bibinfo
			{volume} {77}},\ \bibinfo {pages} {104436} (\bibinfo {year}
		{2008})}\BibitemShut {NoStop}%
	\bibitem [{\citenamefont {Isobe}\ \emph {et~al.}(2002)\citenamefont {Isobe},
		\citenamefont {Ninomiya}, \citenamefont {Vasil'ev},\ and\ \citenamefont
		{Yutaka}}]{Isobe1423}%
	\BibitemOpen
	\bibfield  {author} {\bibinfo {author} {\bibfnamefont {M.}~\bibnamefont
			{Isobe}}, \bibinfo {author} {\bibfnamefont {E.}~\bibnamefont {Ninomiya}},
		\bibinfo {author} {\bibfnamefont {A.~N.}\ \bibnamefont {Vasil'ev}}, \ and\
		\bibinfo {author} {\bibfnamefont {U.}~\bibnamefont {Yutaka}},\ }\bibfield
	{title} {\enquote {\bibinfo {title} {Novel phase transition in
				spin-$\frac{1}{2}$ linear chain systems: {NaTiSi$_2$O$_6$} and
				{LiTiSi$_2$O$_6$}},}\ }\href@noop {} {\bibfield  {journal} {\bibinfo
			{journal} {J. Phys. Soc. Jpn.}\ }\textbf {\bibinfo {volume} {71}},\ \bibinfo
		{pages} {1423} (\bibinfo {year} {2002})}\BibitemShut {NoStop}%
	\bibitem [{\citenamefont {Popovi\ifmmode~\acute{c}\else \'{c}\fi{}}\ \emph
		{et~al.}(2004)\citenamefont {Popovi\ifmmode~\acute{c}\else \'{c}\fi{}},
		\citenamefont {\ifmmode \check{S}\else
			\v{S}\fi{}ljivan\ifmmode~\check{c}\else \v{c}\fi{}anin},\ and\ \citenamefont
		{Vukajlovi\ifmmode~\acute{c}\else \'{c}\fi{}}}]{Popovic036401}%
	\BibitemOpen
	\bibfield  {author} {\bibinfo {author} {\bibfnamefont {Z.~S.}\ \bibnamefont
			{Popovi\ifmmode~\acute{c}\else \'{c}\fi{}}}, \bibinfo {author} {\bibfnamefont
			{Z.~V.}\ \bibnamefont {\ifmmode \check{S}\else
				\v{S}\fi{}ljivan\ifmmode~\check{c}\else \v{c}\fi{}anin}}, \ and\ \bibinfo
		{author} {\bibfnamefont {F.~R.}\ \bibnamefont
			{Vukajlovi\ifmmode~\acute{c}\else \'{c}\fi{}}},\ }\bibfield  {title}
	{\enquote {\bibinfo {title} {Sodium pyroxene {NaTiSi$_2$O$_6$}: Possible
				{Haldane} spin-1 chain system},}\ }\href {\doibase
		10.1103/PhysRevLett.93.036401} {\bibfield  {journal} {\bibinfo  {journal}
			{Phys. Rev. Lett.}\ }\textbf {\bibinfo {volume} {93}},\ \bibinfo {pages}
		{036401} (\bibinfo {year} {2004})}\BibitemShut {NoStop}%
	\bibitem [{\citenamefont {Hirota}\ \emph {et~al.}(1994)\citenamefont {Hirota},
		\citenamefont {Cox}, \citenamefont {Lorenzo}, \citenamefont {Shirane},
		\citenamefont {Tranquada}, \citenamefont {Hase}, \citenamefont {Uchinokura},
		\citenamefont {Kojima}, \citenamefont {Shibuya},\ and\ \citenamefont
		{Tanaka}}]{Hirota736}%
	\BibitemOpen
	\bibfield  {author} {\bibinfo {author} {\bibfnamefont {K.}~\bibnamefont
			{Hirota}}, \bibinfo {author} {\bibfnamefont {D.~E.}\ \bibnamefont {Cox}},
		\bibinfo {author} {\bibfnamefont {J.~E.}\ \bibnamefont {Lorenzo}}, \bibinfo
		{author} {\bibfnamefont {G.}~\bibnamefont {Shirane}}, \bibinfo {author}
		{\bibfnamefont {J.~M.}\ \bibnamefont {Tranquada}}, \bibinfo {author}
		{\bibfnamefont {M.}~\bibnamefont {Hase}}, \bibinfo {author} {\bibfnamefont
			{K.}~\bibnamefont {Uchinokura}}, \bibinfo {author} {\bibfnamefont
			{H.}~\bibnamefont {Kojima}}, \bibinfo {author} {\bibfnamefont
			{Y.}~\bibnamefont {Shibuya}}, \ and\ \bibinfo {author} {\bibfnamefont
			{I.}~\bibnamefont {Tanaka}},\ }\bibfield  {title} {\enquote {\bibinfo {title}
			{Dimerization of {CuGeO$_3$} in the spin-{Peierls} state},}\ }\href {\doibase
		10.1103/PhysRevLett.73.736} {\bibfield  {journal} {\bibinfo  {journal} {Phys.
				Rev. Lett.}\ }\textbf {\bibinfo {volume} {73}},\ \bibinfo {pages} {736}
		(\bibinfo {year} {1994})}\BibitemShut {NoStop}%
	\bibitem [{\citenamefont {Fujii}\ \emph {et~al.}(1997)\citenamefont {Fujii},
		\citenamefont {Nakao}, \citenamefont {Yosihama}, \citenamefont {Nishi},
		\citenamefont {Nakajima}, \citenamefont {Kakurai}, \citenamefont {Isobe},
		\citenamefont {Ueda},\ and\ \citenamefont {Sawa}}]{Fujii326}%
	\BibitemOpen
	\bibfield  {author} {\bibinfo {author} {\bibfnamefont {Y.}~\bibnamefont
			{Fujii}}, \bibinfo {author} {\bibfnamefont {H.}~\bibnamefont {Nakao}},
		\bibinfo {author} {\bibfnamefont {T.}~\bibnamefont {Yosihama}}, \bibinfo
		{author} {\bibfnamefont {M.}~\bibnamefont {Nishi}}, \bibinfo {author}
		{\bibfnamefont {K.}~\bibnamefont {Nakajima}}, \bibinfo {author}
		{\bibfnamefont {K.}~\bibnamefont {Kakurai}}, \bibinfo {author} {\bibfnamefont
			{M.}~\bibnamefont {Isobe}}, \bibinfo {author} {\bibfnamefont
			{Y.}~\bibnamefont {Ueda}}, \ and\ \bibinfo {author} {\bibfnamefont
			{H.}~\bibnamefont {Sawa}},\ }\bibfield  {title} {\enquote {\bibinfo {title}
			{New inorganic spin-{Peierls} compound {NaV$_2$O$_5$} evidenced by x-ray and
				neutron scattering},}\ }\href@noop {} {\bibfield  {journal} {\bibinfo
			{journal} {J. Phys. Soc. Jpn.}\ }\textbf {\bibinfo {volume} {66}},\ \bibinfo
		{pages} {326} (\bibinfo {year} {1997})}\BibitemShut {NoStop}%
	\bibitem [{\citenamefont {L\'epine}\ \emph {et~al.}(1978)\citenamefont
		{L\'epine}, \citenamefont {Caill\'e},\ and\ \citenamefont
		{Larochelle}}]{Lepine3585}%
	\BibitemOpen
	\bibfield  {author} {\bibinfo {author} {\bibfnamefont {Y.}~\bibnamefont
			{L\'epine}}, \bibinfo {author} {\bibfnamefont {A.}~\bibnamefont {Caill\'e}},
		\ and\ \bibinfo {author} {\bibfnamefont {V.}~\bibnamefont {Larochelle}},\
	}\bibfield  {title} {\enquote {\bibinfo {title}
			{Potassium-tetracyanoquinodimethane {(K-TCNQ)}: A spin-{P}eierls system},}\
	}\href {\doibase 10.1103/PhysRevB.18.3585} {\bibfield  {journal} {\bibinfo
			{journal} {Phys. Rev. B}\ }\textbf {\bibinfo {volume} {18}},\ \bibinfo
		{pages} {3585} (\bibinfo {year} {1978})}\BibitemShut {NoStop}%
	\bibitem [{\citenamefont {Kittel}(1986)}]{Kittel1986}%
	\BibitemOpen
	\bibfield  {author} {\bibinfo {author} {\bibfnamefont {Charles}\ \bibnamefont
			{Kittel}},\ }\href@noop {} {\emph {\bibinfo {title} {{Introduction to Solid
					State Physics}}}},\ \bibinfo {edition} {6th}\ ed.\ (\bibinfo  {publisher}
	{John Wiley \& Sons, Inc.},\ \bibinfo {address} {New York},\ \bibinfo {year}
	{1986})\BibitemShut {NoStop}%
	\bibitem [{\citenamefont {Bag}\ \emph {et~al.}(2018)\citenamefont {Bag},
		\citenamefont {Baral},\ and\ \citenamefont {Nath}}]{Bag144436}%
	\BibitemOpen
	\bibfield  {author} {\bibinfo {author} {\bibfnamefont {P.}~\bibnamefont
			{Bag}}, \bibinfo {author} {\bibfnamefont {P.~R.}\ \bibnamefont {Baral}}, \
		and\ \bibinfo {author} {\bibfnamefont {R.}~\bibnamefont {Nath}},\ }\bibfield
	{title} {\enquote {\bibinfo {title} {Cluster spin-glass behavior and memory
				effect in {Cr$_{0.5}$Fe$_{0.5}$Ga}},}\ }\href {\doibase
		10.1103/PhysRevB.98.144436} {\bibfield  {journal} {\bibinfo  {journal} {Phys.
				Rev. B}\ }\textbf {\bibinfo {volume} {98}},\ \bibinfo {pages} {144436}
		(\bibinfo {year} {2018})}\BibitemShut {NoStop}%
	\bibitem [{\citenamefont {Johnston}\ \emph {et~al.}(2000)\citenamefont
		{Johnston}, \citenamefont {Kremer}, \citenamefont {Troyer}, \citenamefont
		{Wang}, \citenamefont {Kl{\"u}mper}, \citenamefont {Budko}, \citenamefont
		{Panchula},\ and\ \citenamefont {Canfield}}]{Johnston9558}%
	\BibitemOpen
	\bibfield  {author} {\bibinfo {author} {\bibfnamefont {D.C.}\ \bibnamefont
			{Johnston}}, \bibinfo {author} {\bibfnamefont {R.~K.}\ \bibnamefont
			{Kremer}}, \bibinfo {author} {\bibfnamefont {M.}~\bibnamefont {Troyer}},
		\bibinfo {author} {\bibfnamefont {X.}~\bibnamefont {Wang}}, \bibinfo {author}
		{\bibfnamefont {A.}~\bibnamefont {Kl{\"u}mper}}, \bibinfo {author}
		{\bibfnamefont {S.~L.}\ \bibnamefont {Budko}}, \bibinfo {author}
		{\bibfnamefont {A.~F.}\ \bibnamefont {Panchula}}, \ and\ \bibinfo {author}
		{\bibfnamefont {P.~C.}\ \bibnamefont {Canfield}},\ }\bibfield  {title}
	{\enquote {\bibinfo {title} {Thermodynamics of spin {$S$}= $\frac{1}{2}$
				antiferromagnetic uniform and alternating-exchange {H}eisenberg chains},}\
	}\href {\doibase 10.1103/PhysRevB.61.9558} {\bibfield  {journal} {\bibinfo
			{journal} {Phys. Rev. B}\ }\textbf {\bibinfo {volume} {61}},\ \bibinfo
		{pages} {9558} (\bibinfo {year} {2000})}\BibitemShut {NoStop}%
	\bibitem [{\citenamefont {Eggert}\ and\ \citenamefont
		{Affleck}(1995)}]{Eggert934}%
	\BibitemOpen
	\bibfield  {author} {\bibinfo {author} {\bibfnamefont {S.}~\bibnamefont
			{Eggert}}\ and\ \bibinfo {author} {\bibfnamefont {I.}~\bibnamefont
			{Affleck}},\ }\bibfield  {title} {\enquote {\bibinfo {title} {Impurities in
				{$S = \frac{1}{2}$} {Heisenberg} antiferromagnetic chains: Consequences for
				neutron scattering and knight shift},}\ }\href {\doibase
		10.1103/PhysRevLett.75.934} {\bibfield  {journal} {\bibinfo  {journal} {Phys.
				Rev. Lett.}\ }\textbf {\bibinfo {volume} {75}},\ \bibinfo {pages} {934}
		(\bibinfo {year} {1995})}\BibitemShut {NoStop}%
	\bibitem [{\citenamefont {Yogi}\ \emph {et~al.}(2015)\citenamefont {Yogi},
		\citenamefont {Ahmed}, \citenamefont {Nath}, \citenamefont {Tsirlin},
		\citenamefont {Kundu}, \citenamefont {Mahajan}, \citenamefont
		{Sichelschmidt}, \citenamefont {Roy},\ and\ \citenamefont
		{Furukawa}}]{Yogi024413}%
	\BibitemOpen
	\bibfield  {author} {\bibinfo {author} {\bibfnamefont {A.}~\bibnamefont
			{Yogi}}, \bibinfo {author} {\bibfnamefont {N.}~\bibnamefont {Ahmed}},
		\bibinfo {author} {\bibfnamefont {R.}~\bibnamefont {Nath}}, \bibinfo {author}
		{\bibfnamefont {A.~A.}\ \bibnamefont {Tsirlin}}, \bibinfo {author}
		{\bibfnamefont {S.}~\bibnamefont {Kundu}}, \bibinfo {author} {\bibfnamefont
			{A.~V.}\ \bibnamefont {Mahajan}}, \bibinfo {author} {\bibfnamefont
			{J.}~\bibnamefont {Sichelschmidt}}, \bibinfo {author} {\bibfnamefont
			{B.}~\bibnamefont {Roy}}, \ and\ \bibinfo {author} {\bibfnamefont
			{Y.}~\bibnamefont {Furukawa}},\ }\bibfield  {title} {\enquote {\bibinfo
			{title} {Antiferromagnetism of {Zn$_2$VO(PO$_4$)$_2$} and the dilution with
				{Ti$^{4+}$}},}\ }\href {\doibase 10.1103/PhysRevB.91.024413} {\bibfield
		{journal} {\bibinfo  {journal} {Phys. Rev. B}\ }\textbf {\bibinfo {volume}
			{91}},\ \bibinfo {pages} {024413} (\bibinfo {year} {2015})}\BibitemShut
	{NoStop}%
	\bibitem [{\citenamefont {Selwood}(2013)}]{Selwood2013}%
	\BibitemOpen
	\bibfield  {author} {\bibinfo {author} {\bibfnamefont {P.~W.}\ \bibnamefont
			{Selwood}},\ }\href@noop {} {\emph {\bibinfo {title} {Magnetochemistry}}}\
	(\bibinfo  {publisher} {Read Books Ltd},\ \bibinfo {year} {2013})\BibitemShut
	{NoStop}%
	\bibitem [{\citenamefont {Samulon}\ \emph {et~al.}(2009)\citenamefont
		{Samulon}, \citenamefont {Kohama}, \citenamefont {McDonald}, \citenamefont
		{Shapiro}, \citenamefont {Al-Hassanieh}, \citenamefont {Batista},
		\citenamefont {Jaime},\ and\ \citenamefont {Fisher}}]{Samulon047202}%
	\BibitemOpen
	\bibfield  {author} {\bibinfo {author} {\bibfnamefont {E.C.}\ \bibnamefont
			{Samulon}}, \bibinfo {author} {\bibfnamefont {Y.}~\bibnamefont {Kohama}},
		\bibinfo {author} {\bibfnamefont {R.D.}\ \bibnamefont {McDonald}}, \bibinfo
		{author} {\bibfnamefont {M.C.}\ \bibnamefont {Shapiro}}, \bibinfo {author}
		{\bibfnamefont {K.A.}\ \bibnamefont {Al-Hassanieh}}, \bibinfo {author}
		{\bibfnamefont {C.D.}\ \bibnamefont {Batista}}, \bibinfo {author}
		{\bibfnamefont {M.}~\bibnamefont {Jaime}}, \ and\ \bibinfo {author}
		{\bibfnamefont {I.R.}\ \bibnamefont {Fisher}},\ }\bibfield  {title} {\enquote
		{\bibinfo {title} {Asymmetric quintuplet condensation in the frustrated
				{$S=1$} spin-dimer compound {Ba$_3$Mn$_2$O$_8$}},}\ }\href {\doibase
		10.1103/PhysRevLett.103.047202} {\bibfield  {journal} {\bibinfo  {journal}
			{Phys. Rev. Lett.}\ }\textbf {\bibinfo {volume} {103}},\ \bibinfo {pages}
		{047202} (\bibinfo {year} {2009})}\BibitemShut {NoStop}%
	\bibitem [{\citenamefont {Nath}\ \emph
		{et~al.}(2008{\natexlab{b}})\citenamefont {Nath}, \citenamefont {Tsirlin},
		\citenamefont {Rosner},\ and\ \citenamefont {Geibel}}]{Nath064422}%
	\BibitemOpen
	\bibfield  {author} {\bibinfo {author} {\bibfnamefont {R.}~\bibnamefont
			{Nath}}, \bibinfo {author} {\bibfnamefont {A.~A.}\ \bibnamefont {Tsirlin}},
		\bibinfo {author} {\bibfnamefont {H.}~\bibnamefont {Rosner}}, \ and\ \bibinfo
		{author} {\bibfnamefont {C.}~\bibnamefont {Geibel}},\ }\bibfield  {title}
	{\enquote {\bibinfo {title} {Magnetic properties of
				$\text{BaCdVO}{({\text{PO}}_{4})}_{2}$: A strongly frustrated
				spin-$\frac{1}{2}$ square lattice close to the quantum critical regime},}\
	}\href {\doibase 10.1103/PhysRevB.78.064422} {\bibfield  {journal} {\bibinfo
			{journal} {Phys. Rev. B}\ }\textbf {\bibinfo {volume} {78}},\ \bibinfo
		{pages} {064422} (\bibinfo {year} {2008}{\natexlab{b}})}\BibitemShut
	{NoStop}%
	\bibitem [{\citenamefont {Walstedt}\ and\ \citenamefont
		{Walker}(1974)}]{Walstedt1974}%
	\BibitemOpen
	\bibfield  {author} {\bibinfo {author} {\bibfnamefont {R.~E.}\ \bibnamefont
			{Walstedt}}\ and\ \bibinfo {author} {\bibfnamefont {L.~R.}\ \bibnamefont
			{Walker}},\ }\bibfield  {title} {\enquote {\bibinfo {title}
			{Nuclear-resonance line shapes due to magnetic impurities in metals},}\
	}\href {\doibase 10.1103/PhysRevB.9.4857} {\bibfield  {journal} {\bibinfo
			{journal} {Phys. Rev. B}\ }\textbf {\bibinfo {volume} {9}},\ \bibinfo {pages}
		{4857} (\bibinfo {year} {1974})}\BibitemShut {NoStop}%
	\bibitem [{\citenamefont {Kikuchi}\ \emph {et~al.}(1999)\citenamefont
		{Kikuchi}, \citenamefont {Motoya}, \citenamefont {Yamauchi},\ and\
		\citenamefont {Ueda}}]{Kikuchi6731}%
	\BibitemOpen
	\bibfield  {author} {\bibinfo {author} {\bibfnamefont {J.}~\bibnamefont
			{Kikuchi}}, \bibinfo {author} {\bibfnamefont {K.}~\bibnamefont {Motoya}},
		\bibinfo {author} {\bibfnamefont {T.}~\bibnamefont {Yamauchi}}, \ and\
		\bibinfo {author} {\bibfnamefont {Y.}~\bibnamefont {Ueda}},\ }\bibfield
	{title} {\enquote {\bibinfo {title} {Coexistence of double alternating
				antiferromagnetic chains in {(VO)$_2$P$_2$O$_7$}:{NMR} study},}\ }\href
	{\doibase 10.1103/PhysRevB.60.6731} {\bibfield  {journal} {\bibinfo
			{journal} {Phys. Rev. B}\ }\textbf {\bibinfo {volume} {60}},\ \bibinfo
		{pages} {6731} (\bibinfo {year} {1999})}\BibitemShut {NoStop}%
	\bibitem [{\citenamefont {Nath}\ \emph
		{et~al.}(2008{\natexlab{c}})\citenamefont {Nath}, \citenamefont {Kasinathan},
		\citenamefont {Rosner}, \citenamefont {Baenitz},\ and\ \citenamefont
		{Geibel}}]{Nath134451}%
	\BibitemOpen
	\bibfield  {author} {\bibinfo {author} {\bibfnamefont {R.}~\bibnamefont
			{Nath}}, \bibinfo {author} {\bibfnamefont {D.}~\bibnamefont {Kasinathan}},
		\bibinfo {author} {\bibfnamefont {H.}~\bibnamefont {Rosner}}, \bibinfo
		{author} {\bibfnamefont {M.}~\bibnamefont {Baenitz}}, \ and\ \bibinfo
		{author} {\bibfnamefont {C.}~\bibnamefont {Geibel}},\ }\bibfield  {title}
	{\enquote {\bibinfo {title} {Electronic and magnetic properties of
				{K$_2$CuP$_2$O$_7$}: A model {$S=\frac{1}{2}$} {H}eisenberg chain system},}\
	}\href {\doibase 10.1103/PhysRevB.77.134451} {\bibfield  {journal} {\bibinfo
			{journal} {Phys. Rev. B}\ }\textbf {\bibinfo {volume} {77}},\ \bibinfo
		{pages} {134451} (\bibinfo {year} {2008}{\natexlab{c}})}\BibitemShut
	{NoStop}%
	\bibitem [{\citenamefont {Nath}\ \emph {et~al.}(2005)\citenamefont {Nath},
		\citenamefont {Mahajan}, \citenamefont {B\"uttgen}, \citenamefont {Kegler},
		\citenamefont {Loidl},\ and\ \citenamefont {Bobroff}}]{Nath174436}%
	\BibitemOpen
	\bibfield  {author} {\bibinfo {author} {\bibfnamefont {R.}~\bibnamefont
			{Nath}}, \bibinfo {author} {\bibfnamefont {A.~V.}\ \bibnamefont {Mahajan}},
		\bibinfo {author} {\bibfnamefont {N.}~\bibnamefont {B\"uttgen}}, \bibinfo
		{author} {\bibfnamefont {C.}~\bibnamefont {Kegler}}, \bibinfo {author}
		{\bibfnamefont {A.}~\bibnamefont {Loidl}}, \ and\ \bibinfo {author}
		{\bibfnamefont {J.}~\bibnamefont {Bobroff}},\ }\bibfield  {title} {\enquote
		{\bibinfo {title} {Study of one-dimensional nature of {$S=\frac{1}{2}$} of
				{(Sr,Ba)$_2$Cu(PO$_4$)$_2$} and {BaCuP$_2$O$_7$} via {$^{31}$P} {NMR}},}\
	}\href {\doibase 10.1103/PhysRevB.71.174436} {\bibfield  {journal} {\bibinfo
			{journal} {Phys. Rev. B}\ }\textbf {\bibinfo {volume} {71}},\ \bibinfo
		{pages} {174436} (\bibinfo {year} {2005})}\BibitemShut {NoStop}%
	\bibitem [{\citenamefont {Taniguchi}\ \emph {et~al.}(1995)\citenamefont
		{Taniguchi}, \citenamefont {Nishikawa}, \citenamefont {Yasui}, \citenamefont
		{Kobayashi}, \citenamefont {Sato}, \citenamefont {Nishioka}, \citenamefont
		{Kontani},\ and\ \citenamefont {Sano}}]{Taniguchi2758}%
	\BibitemOpen
	\bibfield  {author} {\bibinfo {author} {\bibfnamefont {S.}~\bibnamefont
			{Taniguchi}}, \bibinfo {author} {\bibfnamefont {T.}~\bibnamefont
			{Nishikawa}}, \bibinfo {author} {\bibfnamefont {Y.}~\bibnamefont {Yasui}},
		\bibinfo {author} {\bibfnamefont {Y.}~\bibnamefont {Kobayashi}}, \bibinfo
		{author} {\bibfnamefont {M.}~\bibnamefont {Sato}}, \bibinfo {author}
		{\bibfnamefont {T.}~\bibnamefont {Nishioka}}, \bibinfo {author}
		{\bibfnamefont {M.}~\bibnamefont {Kontani}}, \ and\ \bibinfo {author}
		{\bibfnamefont {K.}~\bibnamefont {Sano}},\ }\bibfield  {title} {\enquote
		{\bibinfo {title} {Spin gap behavior of {$S$}= $\frac{1}{2}$
				quasi-two-dimensional system {CaV$_4$O$_9$}},}\ }\href {\doibase
		10.1143/JPSJ.64.2758} {\bibfield  {journal} {\bibinfo  {journal} {J. Phys.
				Soc. Jpn.}\ }\textbf {\bibinfo {volume} {64}},\ \bibinfo {pages} {2758}
		(\bibinfo {year} {1995})}\BibitemShut {NoStop}%
	\bibitem [{\citenamefont {Moriya}(1956)}]{Moriya23}%
	\BibitemOpen
	\bibfield  {author} {\bibinfo {author} {\bibfnamefont {T.}~\bibnamefont
			{Moriya}},\ }\bibfield  {title} {\enquote {\bibinfo {title} {{Nuclear
					magnetic relaxation in antiferromagnetics}},}\ }\href {\doibase
		10.1143/PTP.16.23} {\bibfield  {journal} {\bibinfo  {journal} {Prog. Theor.
				Phys.}\ }\textbf {\bibinfo {volume} {16}},\ \bibinfo {pages} {23} (\bibinfo
		{year} {1956})}\BibitemShut {NoStop}%
	\bibitem [{\citenamefont {Giamarchi}\ and\ \citenamefont
		{Tsvelik}(1999)}]{Giamarchi11398}%
	\BibitemOpen
	\bibfield  {author} {\bibinfo {author} {\bibfnamefont {T.}~\bibnamefont
			{Giamarchi}}\ and\ \bibinfo {author} {\bibfnamefont {A.~M.}\ \bibnamefont
			{Tsvelik}},\ }\bibfield  {title} {\enquote {\bibinfo {title} {Coupled ladders
				in a magnetic field},}\ }\href {\doibase 10.1103/PhysRevB.59.11398}
	{\bibfield  {journal} {\bibinfo  {journal} {Phys. Rev. B}\ }\textbf {\bibinfo
			{volume} {59}},\ \bibinfo {pages} {11398} (\bibinfo {year}
		{1999})}\BibitemShut {NoStop}%
	\bibitem [{\citenamefont {Roca}\ \emph {et~al.}(1998)\citenamefont {Roca},
		\citenamefont {Amor\'os}, \citenamefont {Cano}, \citenamefont {{Dolores
				Marcos}}, \citenamefont {Alamo}, \citenamefont {Beltr\'an-Porter},\ and\
		\citenamefont {Beltr\'an-Porter}}]{Roca3167}%
	\BibitemOpen
	\bibfield  {author} {\bibinfo {author} {\bibfnamefont {M.}~\bibnamefont
			{Roca}}, \bibinfo {author} {\bibfnamefont {P.}~\bibnamefont {Amor\'os}},
		\bibinfo {author} {\bibfnamefont {J.}~\bibnamefont {Cano}}, \bibinfo {author}
		{\bibfnamefont {M.}~\bibnamefont {{Dolores Marcos}}}, \bibinfo {author}
		{\bibfnamefont {J.}~\bibnamefont {Alamo}}, \bibinfo {author} {\bibfnamefont
			{A.}~\bibnamefont {Beltr\'an-Porter}}, \ and\ \bibinfo {author}
		{\bibfnamefont {D.}~\bibnamefont {Beltr\'an-Porter}},\ }\bibfield  {title}
	{\enquote {\bibinfo {title} {Prediction of magnetic properties in
				{oxovanadium(IV)} phosphates: The role of the bridging {PO$_4$} anions},}\
	}\href {\doibase 10.1021/ic971210o} {\bibfield  {journal} {\bibinfo
			{journal} {Inorg. Chem.}\ }\textbf {\bibinfo {volume} {37}},\ \bibinfo
		{pages} {3167} (\bibinfo {year} {1998})}\BibitemShut {NoStop}%
	\bibitem [{Note2()}]{Note2}%
	\BibitemOpen
	\bibinfo {note} {Similar to Ref.~\cite {Tsirlin144412}, magnetization curve
		was simulated by including a weak interchain coupling of $J_{\perp
		}/J_1=-0.05$ in order to reproduce the experimental data around $H_{\protect
			\rm c1}$. This interchain coupling is much lower than estimated by DFT
		(Table~\ref {tab:couplings}), possibly because $J_{\perp }$ reflects a
		cumulative effect of the frustrated couplings $J_{a1}$, $J_{a2}$ and
		$J_{c1}$, $J_{c2}$.}\BibitemShut {Stop}%
	\bibitem [{\citenamefont {Johnston}\ \emph {et~al.}(2001)\citenamefont
		{Johnston}, \citenamefont {Saito}, \citenamefont {Azuma}, \citenamefont
		{Takano}, \citenamefont {Yamauchi},\ and\ \citenamefont
		{Ueda}}]{Johnston134403}%
	\BibitemOpen
	\bibfield  {author} {\bibinfo {author} {\bibfnamefont {D.C.}\ \bibnamefont
			{Johnston}}, \bibinfo {author} {\bibfnamefont {T.}~\bibnamefont {Saito}},
		\bibinfo {author} {\bibfnamefont {M.}~\bibnamefont {Azuma}}, \bibinfo
		{author} {\bibfnamefont {M.}~\bibnamefont {Takano}}, \bibinfo {author}
		{\bibfnamefont {T.}~\bibnamefont {Yamauchi}}, \ and\ \bibinfo {author}
		{\bibfnamefont {Y.}~\bibnamefont {Ueda}},\ }\bibfield  {title} {\enquote
		{\bibinfo {title} {Modeling of the magnetic susceptibilities of the ambient-
				and high-pressure phases of {(VO)$_2$P$_2$O$_7$}},}\ }\href {\doibase
		10.1103/PhysRevB.64.134403} {\bibfield  {journal} {\bibinfo  {journal} {Phys.
				Rev. B}\ }\textbf {\bibinfo {volume} {64}},\ \bibinfo {pages} {134403}
		(\bibinfo {year} {2001})}\BibitemShut {NoStop}%
	\bibitem [{\citenamefont {Garrett}\ \emph {et~al.}(1997)\citenamefont
		{Garrett}, \citenamefont {Nagler}, \citenamefont {Tennant}, \citenamefont
		{Sales},\ and\ \citenamefont {Barnes}}]{Garrett745}%
	\BibitemOpen
	\bibfield  {author} {\bibinfo {author} {\bibfnamefont {A.~W.}\ \bibnamefont
			{Garrett}}, \bibinfo {author} {\bibfnamefont {S.~E.}\ \bibnamefont {Nagler}},
		\bibinfo {author} {\bibfnamefont {D.~A.}\ \bibnamefont {Tennant}}, \bibinfo
		{author} {\bibfnamefont {B.~C.}\ \bibnamefont {Sales}}, \ and\ \bibinfo
		{author} {\bibfnamefont {T.}~\bibnamefont {Barnes}},\ }\bibfield  {title}
	{\enquote {\bibinfo {title} {Magnetic excitations in the {$S$} = $1/2$
				alternating chain compound {(VO)$_2$P$_2$O$_7$}},}\ }\href {\doibase
		10.1103/PhysRevLett.79.745} {\bibfield  {journal} {\bibinfo  {journal} {Phys.
				Rev. Lett.}\ }\textbf {\bibinfo {volume} {79}},\ \bibinfo {pages} {745}
		(\bibinfo {year} {1997})}\BibitemShut {NoStop}%
	\bibitem [{\citenamefont {Nohadani}\ \emph {et~al.}(2004)\citenamefont
		{Nohadani}, \citenamefont {Wessel}, \citenamefont {Normand},\ and\
		\citenamefont {Haas}}]{Nohadani220402}%
	\BibitemOpen
	\bibfield  {author} {\bibinfo {author} {\bibfnamefont {O.}~\bibnamefont
			{Nohadani}}, \bibinfo {author} {\bibfnamefont {S.}~\bibnamefont {Wessel}},
		\bibinfo {author} {\bibfnamefont {B.}~\bibnamefont {Normand}}, \ and\
		\bibinfo {author} {\bibfnamefont {S.}~\bibnamefont {Haas}},\ }\bibfield
	{title} {\enquote {\bibinfo {title} {Universal scaling at field-induced
				magnetic phase transitions},}\ }\href {\doibase 10.1103/PhysRevB.69.220402}
	{\bibfield  {journal} {\bibinfo  {journal} {Phys. Rev. B}\ }\textbf {\bibinfo
			{volume} {69}},\ \bibinfo {pages} {220402(R)} (\bibinfo {year}
		{2004})}\BibitemShut {NoStop}%
	\bibitem [{\citenamefont {Mazurenko}\ \emph {et~al.}(2014)\citenamefont
		{Mazurenko}, \citenamefont {Valentyuk}, \citenamefont {Stern},\ and\
		\citenamefont {Tsirlin}}]{Mazurenko107202}%
	\BibitemOpen
	\bibfield  {author} {\bibinfo {author} {\bibfnamefont {V.V.}\ \bibnamefont
			{Mazurenko}}, \bibinfo {author} {\bibfnamefont {M.V.}\ \bibnamefont
			{Valentyuk}}, \bibinfo {author} {\bibfnamefont {R.}~\bibnamefont {Stern}}, \
		and\ \bibinfo {author} {\bibfnamefont {A.A.}\ \bibnamefont {Tsirlin}},\
	}\bibfield  {title} {\enquote {\bibinfo {title} {Nonfrustrated interlayer
				order and its relevance to the {Bose-Einstein} condensation of magnons in
				{BaCuSi$_2$O$_6$}},}\ }\href {\doibase 10.1103/PhysRevLett.112.107202}
	{\bibfield  {journal} {\bibinfo  {journal} {Phys. Rev. Lett.}\ }\textbf
		{\bibinfo {volume} {112}},\ \bibinfo {pages} {107202} (\bibinfo {year}
		{2014})}\BibitemShut {NoStop}%
	\bibitem [{\citenamefont {Allenspach}\ \emph {et~al.}(2020)\citenamefont
		{Allenspach}, \citenamefont {Biffin}, \citenamefont {Stuhr}, \citenamefont
		{Tucker}, \citenamefont {Ohira-Kawamura}, \citenamefont {Kofu}, \citenamefont
		{Voneshen}, \citenamefont {Boehm}, \citenamefont {Normand}, \citenamefont
		{Laflorencie}, \citenamefont {Mila},\ and\ \citenamefont
		{R\"uegg}}]{Allenspach2020}%
	\BibitemOpen
	\bibfield  {author} {\bibinfo {author} {\bibfnamefont {S.}~\bibnamefont
			{Allenspach}}, \bibinfo {author} {\bibfnamefont {A.}~\bibnamefont {Biffin}},
		\bibinfo {author} {\bibfnamefont {U.}~\bibnamefont {Stuhr}}, \bibinfo
		{author} {\bibfnamefont {G.~S.}\ \bibnamefont {Tucker}}, \bibinfo {author}
		{\bibfnamefont {S.}~\bibnamefont {Ohira-Kawamura}}, \bibinfo {author}
		{\bibfnamefont {M.}~\bibnamefont {Kofu}}, \bibinfo {author} {\bibfnamefont
			{D.~J.}\ \bibnamefont {Voneshen}}, \bibinfo {author} {\bibfnamefont
			{M.}~\bibnamefont {Boehm}}, \bibinfo {author} {\bibfnamefont
			{B.}~\bibnamefont {Normand}}, \bibinfo {author} {\bibfnamefont
			{N.}~\bibnamefont {Laflorencie}}, \bibinfo {author} {\bibfnamefont
			{F.}~\bibnamefont {Mila}}, \ and\ \bibinfo {author} {\bibfnamefont {Ch.}\
			\bibnamefont {R\"uegg}},\ }\bibfield  {title} {\enquote {\bibinfo {title}
			{Multiple magnetic bilayers and unconventional criticality without
				frustration in {BaCuSi$_2$O$_6$}},}\ }\href {\doibase
		10.1103/PhysRevLett.124.177205} {\bibfield  {journal} {\bibinfo  {journal}
			{Phys. Rev. Lett.}\ }\textbf {\bibinfo {volume} {124}},\ \bibinfo {pages}
		{177205} (\bibinfo {year} {2020})}\BibitemShut {NoStop}%
\end{thebibliography}
%

\end{document}